\documentstyle[12pt,epsf]{article}
\newcommand{\bookfig}[5]{
\begin{figure}[htbp]\centering\mbox{\epsfysize=#5cm \epsfbox{#1}}
\caption[#2]{{\small #4}}\label{#3}
\end{figure}
}
\newcommand{\bookfigp}[5]{
\begin{figure}[p]\centering\mbox{\epsfysize=#5cm \epsfbox{#1}}
\caption[#2]{{\small #4}}\label{#3}
\end{figure}
}
\newcommand{\bookfigh}[5]{
\begin{figure}[ht]\centering\mbox{\epsfysize=#5cm \epsfbox{#1}}
\caption[#2]{{\small #4}}\label{#3}
\end{figure}
}
\newcommand{\bookfigb}[5]{
\begin{figure}[b]\centering\mbox{\epsfysize=#5cm \epsfbox{#1}}
\caption[#2]{{\small #4}}\label{#3}
\end{figure}
}

\newcommand{\nc}{\newcommand}
\nc{\rnc}{\renewcommand}
\rnc{\title}[1]{{\Large\bf\mbox{}\\\medskip#1\bigskip\medskip\\}}
\rnc{\author}[1]{{\large #1\smallskip\\}}
\nc{\address}[1]{{\em #1\medskip\\}}
\nc{\e}[1]{{\em #1\/}}
\nc{\comment}[1]{}
\nc{\itm}[2]{\\\noindent$\bullet$\ \ \e{#1}. \ #2}
\nc{\ru}[1]{\rule[-#1ex]{0ex}{#1ex}}
\rnc{\baselinestretch}{1.1}
\rnc{\arraystretch}{0.91}
\rnc{\thesection}{\arabic{section}\,.}
\rnc{\thesubsection}{\arabic{section}.\arabic{subsection}}
\rnc{\theequation}{\arabic{section}.\arabic{equation}}
\nc{\sect}[1]{\section{#1}\setcounter{equation}{0}}
\nc{\sub}[1]{\subsection{#1}}
\nc{\subsub}[1]{\subsubsection{#1}}

\nc{\beq}{\begin{equation}}
\nc{\eeq}{\end{equation}}
\nc{\be}{\begin{equation}}
\nc{\ee}{\end{equation}}
\nc{\beqa}{\begin{eqnarray}}
\nc{\eeqa}{\end{eqnarray}}
\nc{\bea}{\begin{eqnarray}}
\nc{\eea}{\end{eqnarray}}
\nc{\eql}[1]{\label{Eqn#1}}
\nc{\noeqno}{\nonumber\\}
\nc{\eqref}[1]{(\ref{Eqn#1})}
\nc{\disp}{\displaystyle}

\nc{\ade}{\mbox{$A$-$D$-$E$}}
\nc{\calA}{{\cal A}} \nc{\calB}{{\cal B}} \nc{\calC}{{\cal C}} 
\nc{\calD}{{\cal D}}
\nc{\calE}{{\cal E}} \nc{\calF}{{\cal F}} \nc{\calH}{{\cal H}}
\nc{\calI}{{\cal I}} \nc{\calM}{{\cal M}} \nc{\calN}{{\cal N}}
\nc{\calR}{{\cal R}} \nc{\calS}{{\cal S}} \nc{\calV}{{\cal V}}
\nc{\calU}{{\cal U}} \nc{\calW}{{\cal W}}
\nc{\GO}{\Omega}\nc\GTh{\Theta}
\nc{\phit}{\hat{\varphi}}
\nc{\chit}{\hat{\chi}}
\nc{\hcalN}{\hat{\calN}} \nc{\hcalS}{\hat{\calS}} 
\nc{\sigmad}{\sigma^\dagger}
\nc{\psid}{\psi^\dagger}

\def\hN{\hat{N}}
\def\hV{\hat{\n}} 
\nc{\hS}{\hat{S}}             

\nc\jb{\bar j} 
\def\a{a} 
\def\b{b}\def\c{c}
\def\n{n} 
\def\II{\relax{\rm I\kern-.18em I}}
\def\i{i}   

\def\llangle{\langle\!\langle}
\def\rrangle{\rangle\!\rangle}
\def\bra{\langle}\def\ket{\rangle}
\def\oh{{1\over 2}}
\def\tr{{\rm tr\,}}\def\mod{{\rm mod\,}}
\def\Exp{{\rm Exp}}
\def\psii#1#2{\psi_{#1}^{#2}} 
\def\s{m}
\def\slh{\widehat{sl}}

\long\def\omit#1{}
\def\chit{\hat{\chi}}

\nc{\nn}{\nonumber}

\def\za{\alpha} \def\zb{\beta} \def\zg{\gamma} \def\zd{\delta}
\def\ze{\varepsilon}   
   
  \def\zs{\sigma}

\def\un{{\bf 1}}

\font\tenmsb=msbm10 scaled \magstep1
\font\sevenmsb=msbm7 scaled \magstep1
\font\fivemsb=msbm5 scaled \magstep1
\newfam\msbfam
\textfont\msbfam=\tenmsb
\scriptfont\msbfam=\sevenmsb
\scriptscriptfont\msbfam=\fivemsb
\def\Bbb#1{{\fam\msbfam\relax#1}}

\begin{document}
\hfill\today
\begin{center}
\mbox{}\vspace{-.5in}
\title{Boundary Conditions \\
in Rational Conformal Field Theories } 
\author{Roger E. Behrend}
\address{C.N. Yang Institute for Theoretical Physics\\
State University of New York\\
Stony Brook, NY 11794-3840, USA}
\medskip
\author{Paul A. Pearce}
\address{Department of Mathematics and Statistics\\
University of Melbourne\\Parkville, Victoria 3052, Australia}
\medskip
\author{Valentina B. Petkova
\footnote{permanent address:
Institute for Nuclear Research and Nuclear Energy, 
Tzarigradsko Chaussee 72, 
1784 Sofia, Bulgaria}}
\address{Arnold Sommerfeld Institute for Mathematical Physics\\
TU Clausthal, Leibnizstr. 10\\ D-38678 Clausthal-Zellerfeld, Germany}
\medskip
\author{Jean-Bernard Zuber}
\address{Service de Physique Th\'eorique\\
CEA-Saclay \\ 91191 Gif-sur-Yvette Cedex, France}
\begin{abstract}
\noindent
We develop further the theory of Rational Conformal Field
Theories (RCFTs) on a cylinder with specified boundary
conditions emphasizing the role of a triplet of algebras: the Verlinde,
graph fusion and  Pasquier algebras. We show that solving Cardy's
equation, expressing consistency of a RCFT on a cylinder, is equivalent to
finding integer valued matrix representations of the Verlinde algebra.
These matrices allow us to naturally associate a
graph $G$ to each RCFT such that the conformal boundary
conditions are labelled by the nodes of $G$.
This approach is carried to completion for $sl(2)$ theories leading to
complete sets of conformal boundary conditions, their associated
cylinder partition functions and the \ade\ classification. We also review the
current status for WZW $sl(3)$ theories.
Finally,  a systematic generalization of the  
formalism  of Cardy-Lewellen 
is developed  to allow for multiplicities
arising from more general representations of the 
Verlinde algebra.
We obtain information on the bulk-boundary coefficients 
and reproduce the relevant algebraic structures
from the sewing constraints.

\end{abstract}
\end{center}


\sect{Introduction}

\subsection{History and motivation}

The subject of boundary conformal field theory has a fairly
long history. It was born more than ten years ago, in 
parallel work on open string theory~\cite{KLT,Ishi,IshO,Call,BD,Pol,PS,BSa}
and on conformal field theories (CFTs) describing
critical systems with boundaries~\cite{SaBa}.
The work of  Cardy~\cite{Ca89} was a landmark, leading to the unification
of methods, to the introduction of important concepts such as
boundary conformal fields and to the systematic investigation of their
properties and couplings~\cite{CL, L}.
The subject remained dormant for some time, in spite of some
 activity motivated again by string theory~\cite{PSS95, PSS, SS}
and of  beautiful applications to the  Kondo problem~\cite{AL,Affl}.
Lately, the subject has undergone a revival of interest
in connection with various problems. On the one hand, work on
 boundary conditions in integrable field theories and boundary
flows~\cite{Cher,GhZ,CDRS}
and on quantum impurities~\cite{OshAff, AfflOS98} motivated
 a closer  look at boundary CFT. On the other hand, within
statistical mechanics, integrable Boltzmann weights
satisfying the so-called Boundary Yang-Baxter Equation (BYBE)
were constructed  in lattice models~\cite{Skl,BehPeaObr96}.
Finally new progress in string theory was another reason to reconsider the
problem. Generalizations of D-branes as boundary conditions in CFT
have been studied by several
groups~\cite{RS1,RS2,FuchsS97,FuchsS98,FS98,SS1}.

In the present work, we want to reconsider  several issues in the
discussion of boundary conditions in (rational) conformal field
theories: what are the general boundary conditions that may be
imposed, what are the structure constants of the bulk and
boundary fields in the presence of these boundary conditions.
These  are the
basic questions that we want to address. The methods that we use
are not essentially new, but are based on the
systematic exploitation of the work of Cardy and Lewellen~\cite{Ca89,CL,L}.

Among the main results of this paper:
\begin{itemize}
\item We establish a
connection between the classification of boundary conditions
and the classification of integer valued representations of the
fusion algebra. A preliminary account of this result was 
given in \cite{BPZ,BPPZ}.  In the same vein, we show that it is
natural to associate graphs to these problems.  In particular, an
ADE classification of boundary conditions for Wess-Zumino-Witten
(WZW) and minimal $sl(2)$
theories emerges in a natural and simple way. A discussion of the state
of the art for $sl(3)$ models is also included.
\item We point out the deep connections between the features of
conformal field theory in the bulk and in the presence of boundaries.
The classification of the latter has some bearing on the classification
of the bulk properties (modular invariants etc).
 This is not
a new observation.   In particular, in string theory many connections 
are known to exist between open and closed string sectors, but
it seems that the point had  not been stressed enough.
A triplet of algebras,
specifically the graph fusion algebra and its dual, the Pasquier algebra,
appears naturally in our discussion, along with the Verlinde algebra.
\item We reanalyse in a systematic way the couplings (structure
constants) of fields in the presence of boundaries and the equations
they satisfy, generalising the formalism of Cardy-Lewellen to accomodate
the appearance of nontrivial multiplicities.  In the diagonal
cases  we find a direct relation between
the chiral duality identities of Moore and Seiberg
and the basic sewing relations of the boundary CFT. 
The main point is the observation that the
bulk boundary coefficients in the diagonal case
essentially coincide with the matrices $S(j)$ of modular
transformations of torus $1$-point functions. 
In this way the two basic bulk--boundary equations~\cite{CL,L} 
are shown to be equivalent to the torus duality identity of~\cite{MS1}.
\end{itemize}

Some more particular results include
the extension of Cardy's equation to non-specialized characters,
thus lifting an ambiguity in the original derivation,
the proof of uniqueness of boundary conditions for $\slh(2)$
WZW and minimal models and $\slh(N)_1$ models,
the clarification of the role of the graph  algebra and 
the recovery of this algebra along with its dual,
the Pasquier algebra, from the boundary sewing constraints.

\subsection{Background on bulk  CFT}
In this paper, we are only concerned with Rational Conformal Field Theories
(RCFTs). We
first establish notations etc. In the study of a RCFT, one first specifies a
chiral algebra $\calA$. It is the Virasoro algebra or one of its extensions:
current algebra, $\calW$ algebra etc. The generators of this algebra
will be denoted generically $W_n$ and include the Virasoro generators $L_n$.
At a given level, the theory is rational, i.e. $\calA$ has only a
finite set $\calI$ of irreducible representations $\calV_i$,
$i\in \calI$. The label $i^*$ indexes the representation conjugate
to $i$, and $i=1$ refers to the identity (or vacuum) representation.
We also suppose that the characters $\chi_i(q)=\tr_{\calV_i}
q^{L_0-{c\over 24}}$ of these representations,
the matrix $S$ of modular transformations of the $\chi$'s
and the fusion coefficients $N_{ij}{}^k$ of the $\calV$'s are all known.
The matrix $S_{ij}$ is symmetric and unitary and satisfies $S^2=C$, 
where $C$ is the conjugation matrix $C_{ij}=\delta_{j i^*}$. The
fusion coefficients are assumed to be given in terms of $S$ by the Verlinde
formula \cite{Verl}
\be
N_{ij}{}^k=  \sum_{l\in \calI} {S_{il}S_{jl}S_{kl}^*\over
S_{1l}}\ , \label{Verlinde} \ee
an assumption that rules out some cases of RCFTs.

A physical conformal theory is then defined by a collection of bulk
and boundary fields and their 3-point couplings 
(OPE Coefficients). In particular,  the spectrum of bulk fields
is described by the finite set Spec of pairs $(j, \jb)$
of representations, possibly appearing with some multiplicities $N_{j\jb}$,
of the left and right copies of $\calA$, such that the Hilbert
space of the theory on an infinitely long cylinder reads
\be
\calH=\oplus_{(j,\jb)\in {\rm Spec}} \calV_j\otimes
\overline{\calV}_{\jb} \ , \label{IIa}
\ee
with the same multiplicities $N_{j,\jb}$.
The modular invariant torus partition function
\be
Z_{{\rm torus}}=
\sum_{j,\jb} N_{j\jb} \chi_j(q) \left(\chi_{\jb}(q)\right)^*\  \label{IIaa}
\ee
is a convenient way to encode this information.
The finite subset $\calE$ of labels of
elements of the spectrum that are left-right symmetric
will play a central role in the following
\be
\calE=\{j | (j, \jb=j) \in {\rm Spec}\}\ , \label{IIab}
\ee
and will be called the set of {\it exponents} of the theory.
Recall that these exponents may come with some multiplicities.
To distinguish them as different elements of
the set $\calE$ a second index will be often added, i.e.,
$(j,\alpha)\in \calE\,,$ for $j\in \calI$.

In terms of all these data, one is in principle able to
compute exactly all correlation functions of the CFT on an arbitrary
2D surface, with or without boundaries~\cite{So,MS1,L}.
These data, however, are subject to consistency constraints:
 single-valuedness  of $n$-point functions on the plane,
modular invariance of the torus or annulus partition function, etc, all
rooted in the locality properties of the theory.
In this paper, we shall reexamine the conditions that stem from
surfaces with boundaries (half-plane or disk, cylinder or annulus) and
explore their consequences.

\medskip

For later reference, let us also recall that RCFTs fall in two classes.
In the first class (``type I''), the Hilbert space  (\ref{IIa})
is a diagonal sum of representations of a larger, ``extended'',
algebra $\calA'\supseteq \calA$. Accordingly, the partition function
(\ref{IIaa}) is a sum of squares of sums of characters
\be
Z=\sum_{{\rm blocks \ }B} |\sum_{i\in B}\chi_i |^2\ . \label{IIabc}
\ee
The second class (``type II'') is obtained from the first by
letting an automorphism $\zeta$ of the fusion rules of the extended
algebra $\calA'$ act on the right components, thus resulting in a
non-block-diagonal partition function
\be
Z= \sum_B \Big(\sum_{i\in B}\chi_i\Big)\Big(\sum_{j\in \zeta(B)}
\bar\chi_j\Big) \ . \label{IIabd}
\ee
For example,
in the classical case of $sl(2)$ theories, classified by \ade\ 
Dynkin diagrams, the $A$, $D_{2p}$, $E_6$ and $E_8$ cases are of the
first type, whereas the $D_{2p+1}$ and $E_7$ are obtained respectively
from the $A_{4p-1}$ and $D_{10}$ cases by a $\Bbb Z_2$ automorphism
of their fusion rules. We shall see below that the study of boundary conditions
on a cylinder has some bearing on these expressions of
torus partition functions.

\sect{Cardy Equation and Verlinde Algebra}

\subsection{Boundary states}

As discussed in~\cite{Ca84}, on the boundary of a domain such as  
the upper half plane or a semi-infinite cylinder,
one must impose a continuity condition of the form
\be
 T(z)=\bar T(\bar z)\big|_{z=\bar z} \qquad
W(z)=\overline{W}(\bar z)\big|_{z=\bar z}\ .
\ee
While the first of these conditions has the direct physical meaning
of the absence of energy-momentum flow across the boundary,  or
the preservation of the real boundary by diffeo\-mor\-phisms,  the
condition(s) on the other $W$ may be generalized to incorporate a possible
``gluing automorphism''~\cite{KO,RS1,FuchsS97}
\be
 W(z)=\Omega\overline{W}(\bar z)\big|_{z=\bar z}\ .
\ee
A semi-annular domain in the upper half-plane
may be conformally  mapped into an annulus in the complex plane by
$\zeta=\exp(- 2i\pi w/T)$, $w={L\over \pi} \log z$.  Then
as shown by Ishibashi~\cite{Ishi} and Cardy~\cite{Ca89},
the boundary condition becomes
\be
 \zeta^2 T(\zeta)= \overline{\zeta}^2 \overline{T}(\bar \zeta)
\qquad \zeta^{s_W} W(\zeta)= (-\overline{\zeta})^{s_W} \overline{W}(\bar
\zeta) \qquad {\rm for\ \  } |\zeta|=1 {\rm\ \ and\ \ } |\zeta|= e^{2\pi L/T}
\ee
where $s_W$ denotes the spin of $W$, or more generally
$$ \qquad \zeta^{s_W} W(\zeta)=
(-\bar\zeta)^{s_W} \Omega\overline{W}(\overline{\zeta})\ .$$
Through radial quantization, this translates into a condition on
{\it boundary states} $|\a\rangle_\GO$
\be
\left(W_n -(-1)^{s_W}\Omega({\overline{W}}_{-n})\right) |\a\rangle_\GO
=0 \ . \label{IIb}
\ee
This includes in particular the condition that
\be
(L_n-\bar L_{-n})|\a\ket_\GO=0\,, \label{IIbba}
\ee
assuming that the automorphism $\Omega$ keeps invariant the
Virasoro generators.
For the central charge operator  we have $(k
-\bar{k})|a\ket_\GO =0$.

Solutions to this linear system are spanned by special states
called Ishibashi states~\cite{Ishi}, labelled by the  finite set
$\calE_\GO =\{ j| (j, \jb=\omega(j)) \in {\rm Spec}\} $, where
$\omega$ depends in particular  on $\Omega$.   To see this, let us
consider first the simpler equation (\ref{IIbba}) in the case
when $\calA$ is the Virasoro algebra and $\GO$ is trivial. Then
observe that one may solve (\ref{IIbba}) in each component of
(\ref{IIa}) independently,  as these spaces are invariant under
the action of the two copies of $\calA$.  Now we recall that any
state  $A=\sum_{n,\bar n} a_{n,\bar n} |j,n\ket \otimes |\bar j, \bar n\ket $
in $\calV_j\otimes \calV_{\jb}$  is in one-to-one
correspondence with a homomorphism $X_A
=\sum_{n,\bar n} a_{n,\bar n} |j,n\ket  \bra\bar j, \bar n| $
 of $\calV_{\jb}$ into $\calV_{j}$. This uses  the scalar product in
 $\calV_{\jb}$. Since  $L_{-n}=L_n^\dagger$ for that
scalar product, (\ref{IIbba}) implies that $ L_n X_A=X_A L_n$,
i.e. that $X_A$ intertwines the action of $L_n$ in the two
representations $\calV_{j}$ and $\calV_{\jb}$ of the Virasoro
algebra. As these two representations are irreducible, they must
be equivalent, which by our convention on the labelling of
representations, means that $j=\jb$.  Thus the only non-vanishing
components of $A$ in $\calH$ are in diagonal products
$\calV_j\otimes \calV_{j}$ and in each one, $X_A$ is proportional
to the projector $P_j=\sum_{n}|j,n\ket \bra j,n|$. To fix the
normalization we choose $X_A=P_j$ and the corresponding Ishibashi
state is denoted $|j\rrangle$.  This completes the proof
\footnote{Many thanks to G. Watts (private communication) to whom
we owe this elegant derivation. Some elements had appeared
already in M. Bauer's PhD thesis (1989). }
that there is an independent boundary state  $|j\rrangle$ for each
element of the set $\calE =\{ j| (j, \jb=j) \in {\rm Spec}\}$.

The argument is a formal extension of the proof, based on the Schur
lemma, of the existence and uniqueness of an invariant in the
tensor product of finite dimensional representations. It
 extends to the odd spin $s_W$ case
(\ref{IIb}). We have to use the fact 
that $W_n^{\dagger}=(-1)^{s_W}\, U^{-1}\, W_{-n}\, U$ with respect to a
bilinear (or hermitian) form where $U $ is a unitary (or
antiunitary) operator. One exploits the same definition of
the homomorphism $X_A: \calV_{j'} \to
\calV_{j}$, now $\calV_{j'}\,, \calV_{j}$ being highest weight
representations of the
chiral algebra $\calA$ generated by $W_n$.  However $X_A$
corresponds to states in $\calV_{j} \otimes U_{\Omega}\,
U\,\calV_{j'}$,  where $ U_{\Omega}$ is a unitary (antiunitary)
operator implementing the automorphism $\Omega(W_{n}) =
U_{\Omega}\,W_{n}\, U_{\Omega}^{-1}$. The equation (\ref{IIb})
leads to $ W_n X_A=X_A W_n$  again with the result $j'=j$ and
$X_A=P_j$ while the Ishibashi states are given by
$|j\rrangle_\GO=\sum_{n} |j,n \ket \otimes U_{\Omega}\,U\,|j, n\ket$.

The operator $ U$ is in general non-trivial, e.g., for
the $\widehat{sl}(N)_k$ WZW theories $W_n^{\dagger} =
\bar{w}(W_{-n})$ where  $\bar{w}$ is the horizontal
projection of the Chevalley involution of the affine algebra
\cite{Kac}, i.e., it is determined by
$\bar{w}(e^{\alpha_i})= -f^{\alpha_i}\,,$ 
$\bar{w}(f^{\alpha_i})= -e^{\alpha_i}\,,$  
for the simple roots $\alpha_i$ of $sl(N)$, 
where $e^{\alpha}/ f^{\alpha}$ are  raising/lowering operators
respectively, and for the Cartan generators 
$\bar{w}(h^i)=-h^i\,, i=1,2, \dots, N-1.$
The pair $(j, \omega(j))$ characterising $\calE_\GO$ refers to
the eigenvalues of $h^i$ on the first term in $|j\rrangle_\GO$.
If $\Omega$ is the identity, then  $\omega(j)=-j$. If $\Omega=w_0$,
where $w_0$ represents the longest element of the Weyl group,
then $w_0(j)=-j^*$ and  hence 
$\,\omega(j)=w_0 \bar{w}(j)=j^*$. On the other hand 
$\omega(j)=j$ for $\Omega$ coinciding with the
Chevalley automorphism $\Omega=\bar{w}$.  In the last two of
these examples we can identify $\overline{\calV}_{\jb}=
U_{\Omega}\,U\, \calV_{j}$ with a highest weight module
$\calV_{\omega(j)}\,,$ $\omega(j)\in \calI$. It should be
stressed that all these automorphisms $\Omega$ keep invariant the
Sugawara Virasoro generators so the condition (\ref{IIbba}) is also 
 satisfied on the corresponding Ishibashi states.

We shall hereafter drop the explicit dependence on $\GO$.

\medskip

In fact we still have to define a norm (or a scalar product) on
boundary states, in particular on the Ishibashi states.
We have to face two difficulties.
First because of the infinite dimension of the representation
$\calV_j$, the most naive norm,  proportional to $\tr P_j$,
would be infinite. The second problem concerns 
non-unitary representations. In such cases, the hermitian
form on $\calV_j$ is not positive definite, and we may
encounter signs in the norm of states.

The first problem requires some regularization of the naive norm.
Let $\tilde q^\oh=e^{-\pi i\over \tau}$ be a real number,
$0< \tilde q <1$. Then   $\llangle j| \tilde q^{\oh(L_0+\bar L_0
-{c\over 12})} |j\rrangle =\tr P_j \tilde
q^{L_0-{c\over 24}}=\chi_j(\tilde q)$.  
We write in general, allowing some multiplicity 
$\alpha=1,\cdots , N_{jj}$ for the representations: 
\be
\llangle j',\alpha'|
\tilde q^{\oh(L_0+\bar L_0 -{c\over 12})} |j, \alpha\rrangle
= \delta_{jj'}\delta_{\alpha\alpha'}\,\chi_j(\tilde q) \ .
\label{IId}
\ee

The norm of $ |j\rrangle$ should then be some renormalized
version of the $\tilde q\to 1$ limit of (\ref{IId}) \cite{Ishi,
Ca89}, i.e., of the limit in which $q=e^{2 \pi i \tau}$,
 the modular transform of $\tilde q$,  tends to  $0$.
In {\it unitary} theories,
a new scalar product on boundary states may be defined according to
\bea
\llangle j \alpha\|j' \alpha' \rrangle
&=& \lim_{\tilde q\to 1} q^{{c\over 24}}
\llangle j',\alpha'|
\tilde q^{\oh(L_0+\bar L_0 -{c\over 12})} |j, \alpha\rrangle
\nonumber \\
&=&\delta_{jj'}\, \delta_{\alpha\alpha'}\,
\lim_{q\to 0} q^{{c\over 24}} \chi_j(\tilde q)
= \delta_{jj'}\,\delta_{\alpha\alpha'} {S_{1j}} \,,
\label{IIda}
\eea
where we have used the fact that in a unitary theory, the leading
character in the $q \to 0$ limit is that of the identity operator
$\chi_1(q) \approx q^{-{c\over 24}}$.  Note that $ S_{j 1}$ is,
up to a factor $1/ S_{11}$, the quantum dimension of the
representation $j$, a positive number.  Thus in unitary theories
the normalization chosen for $X_j$ is such that the states
$|j\rrangle$ are orthogonal for the scalar product (\ref{IIda}),
with a square norm equal to ${S_{1j}}$.  Although in non-unitary
theories the limit $q\to 0$ in~(\ref{IIda}) does not exist  in
general, due to the existence of representations of conformal
weight $h_i<0$ that will dominate that limit, we may still {\it
define} the norm by the same formula as (\ref{IIda}).
Alternatively, if $j_0$ denotes the unique representation of
smallest conformal weight $h_{j_0}<0$ belonging to $\calE$, and
$c_{{\rm eff}}:= c-24 h_{j_0}$, then we may define
\bea
\llangle j \alpha\|j' \alpha' \rrangle
&=&\delta_{jj'}\, \delta_{\alpha\alpha'}\,
{S_{1j}\over S_{j_0 j}}
\lim_{q\to 0} q^{{c_{{\rm eff}}\over 24}} \chi_j(\tilde q)\ .
\label{IIdaba}
\eea
In all cases we thus have
\be
\llangle j\alpha \|j'\alpha'\rrangle= \delta_{jj'}\delta_{\alpha\alpha'}
{S_{1j}}
\ee
which is now of indefinite sign. In the sequel, we use more compact
notations and the multiplicity label $\alpha$ will be implicit
when referring to $j\in \calE$.

The most general boundary state $|\a\rangle$ satisfying
condition~(\ref{IIb}) must be a linear combination of these
Ishibashi states, which, for later convenience,  we write as
\be
|\a\rangle =\sum_{j\in \calE} {\psi_\a^j \over\sqrt{S_{1j}}}
 \,\, |j \rrangle \ . \label{IIc}
\ee
We denote by $\calV=\{ a\}$ the set labelling the boundary states. 
 We assume that an involution $a \to a^*$ in the set $\calV$ 
is defined and that
$\psi_{a^*}^j=\psi_a^{j^*}=(\psi_a^{j})^*$, where $j\to j^*$ is an
involution in $\calE$~ (in general
$(j,\alpha) \to (j^*,\alpha^*)$, see Appendix~B for examples).
We define conjugate states as
\be
\langle \b| :=\sum_{j\in \calE}
\llangle j| {\psi_{\b^*}^{j}\over\sqrt{S_{1j}}} \ .
\ee
As explained in~\cite{RS1}, this conjugate state may be regarded
as resulting from the action of an antilinear
CPT operation. As a  consequence
\bea
\bra \b\|\a\ket
=\sum_{j\in \calE}{\psi_a^j\left(\psi_\b^j\right)^*\over S_{1j}}
\llangle j \| j\rrangle = \sum_{j\in \calE}
{\psi_\a^j\left(\psi_\b^j\right)^*} \
\eea
so that  the orthonormality of the boundary states is equivalent  to
that of the $\psi$'s.

In some cases,  such as in the computation of
partition functions involving the specialised characters in the
next section, it is sufficient to impose only the Virasoro
condition (\ref{IIbba})  on the  boundary states.
 Then the sum in
(\ref{IIc}), when interpreted in terms of Ishibashi states
pertaining to some extended symmetry,  may include
states $|j\rrangle_{\Omega}$ with different $\Omega$.
 For example, in the minimal $\slh(2)$
models when multiplicities occur in $\calE$,
one can build the Ishibashi states using the Coulomb gas
realisation  with  $\calA=\widehat{u}(1)$.  Then there are two
choices of $\Omega$ keeping $L_n$ invariant. This   allows,  in
particular, the construction of two different Ishibashi states with the
same value of the scaling dimension, i.e., the  explicit resolution of 
the degeneracy of states denoted  $|j, \alpha\rrangle$ above.
Such mixtures of Ishibashi states  may be used in determining the
boundary states of the non-diagonal  $(A, D_{\rm even})$ models.


\subsection{The Cardy equation}


We now consider a conformal field theory on a finite cylinder.
Following Cardy, the partition function may be expressed in two
alternative ways. Regarded as resulting from the evolution of the
system between boundary states $a$ and $b$ under the action of
the Hamiltonian on (i.e. the translation operator along) the
cylinder, it is
\be
Z_{\b|\a}= \langle
\b| \tilde q ^{\oh(L_0+\bar L_0 -{c\over 12})}
|\a\rangle \label{IIe}
\ee
where  $\tilde q^\oh= e^{-2\pi{L\over T}}$
describes the aspect-ratio of the cylinder of period $T$ and length $L$
%
\bookfig{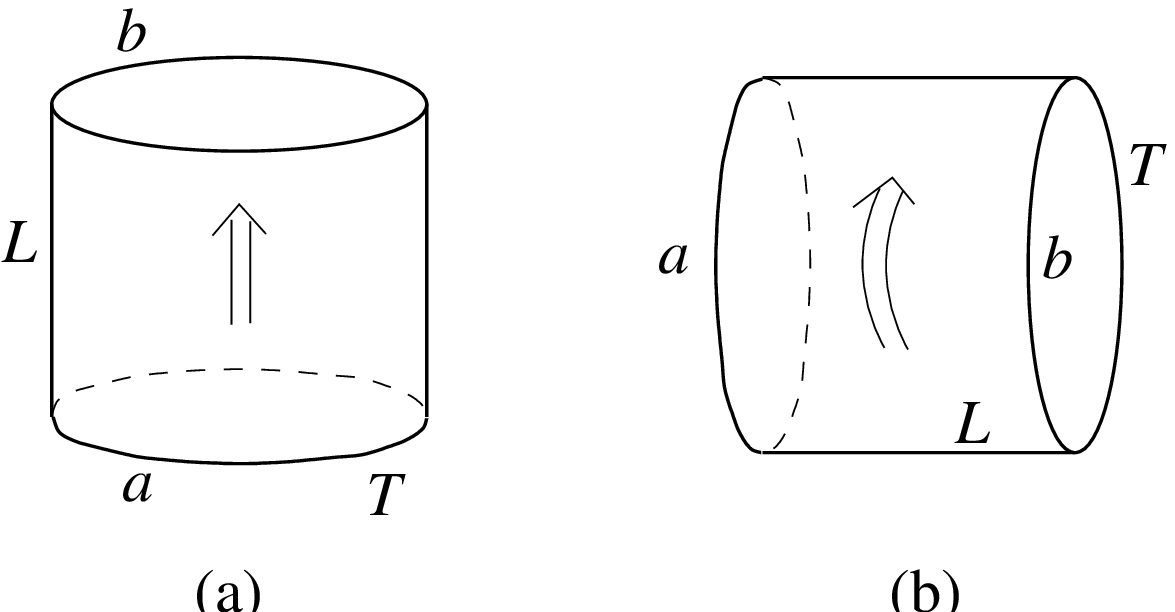}{The Cardy equation}{cardy}
{The two computations of the partition
function $Z_{b|a}$: (a) on the cylinder, between the
boundary states $a$ and $b$,  (b) as a periodic time
evolution on the strip, with boundary conditions $a$ and $b$.}
{4}
as in Figure~\ref{cardy}.
Decomposing the boundary states on the Ishibashi basis and using
(\ref{IId}), one obtains 
\be
Z_{\b|\a}= \sum_{j\in \calE}
\psi_\a^j\, (\psi_\b^j)^*{\chi_j(\tilde q) \over S_{j1}}
\label{IIf}
\ee
where the states $|j\rrangle$ are admissible Ishibashi states of the
system, i.e., the label $j$ runs over the set $\calE$.

On the other hand,  $Z_{\b|\a}$ may be regarded as resulting from
the periodic ``time'' evolution under the action of the
translation operator along the finite width strip in the presence
of boundary conditions $\a$ and $\b$. The latter manifest
themselves only in the nature of the Hilbert space $\calH_{\b\a}$
and its decomposition into representations of a {\it single}
chiral algebra: ${\calH}_{\b\a}=\oplus_i \n_{i \b}{}^{\a}
\calV_i$  with non-negative integer multiplicities
$\n_{i\b}{}^{\a}$.
 If $q=e^{-\pi {T\over L}}$, $Z_{\b|\a}$ is a
{\it linear} form in the characters
\be
Z_{\b|\a}=\sum_{i\in \calI} \chi_i(q) \n_{i\b}{}^{\a} \ . \label{IIg}
\ee

We choose to write
the modular transformation of characters in the form
 $ \chi_i(q)=\sum_j S_{ij} \chi_j(\tilde q)$,  hence
$ \chi_j(\tilde q)=\sum_i S_{ji^*} \chi_i( q)$  . Provided that specialized
characters $\chi_i(q)$ are considered, this complex conjugation
is immaterial, since $\chi_i(q)=\chi_{i^*}(q)$. We shall, however,
make later use of unspecialized characters  (Appendix~A), for which it does
matter.
With this convention,  and assuming  for the time being the independence
of characters,
the two expressions (\ref{IIf}) and (\ref{IIg}) are consistent provided
\be
\n_{i\a}{}^\b
 = \sum_{j\in \calE } {S_{ij}\over S_{1j}} \
\psi_\a^j\, (\psi_\b^j)^*  \ .\label{IIi}
\ee
In the sequel we will refer to this as the Cardy equation.
In the left hand side, we have used the first of the following symmetries
\be
\n_{i\a}{}^\b =\n_{i^*\b}{}^{\a} =\n_{i\b^*}{}^{\a^*}
\ ,\label{IIiac}
\ee
which follow from the properties of the modular matrix and of
the coefficients $\psi_a^{j}$.

The boundary states $|\a\rangle$, $|b\rangle$, are thus such that
$\n_{i\a}{}^{\b}$ is a non-negative integer. Uniqueness of the vacuum implies
$\n_{1\a}{}^\b\le 1 $.  The Cardy equation (\ref{IIi}) is  a {\it
non-linear} constraint on the components of boundary states
$|\a\rangle$ and $|\b\rangle$ on the basis of Ishibashi states.
Note also that it implies that
 $\sum_i  \n_{i\b}{}^{\a} S_{ji}$ vanishes  if $j\notin \calE$ and,
except in cases with multiplicities, must factorize
into a product of contributions of the $\a$ and
$\b$ boundary states, a non-trivial constraint.  Still, these
constraints seem difficult to solve in full generality.

Before we proceed, we have  to pause on  the question of
independence of characters. In general, it is not true that {\it
specialized} characters such as those that we have been using so
far, are linearly independent. For instance, complex conjugate
representations $i$ and $\i^*$ give rise to the same character
$\chi_i(q)$.  Unfortunately,
little is known about unspecialized characters for general
chiral algebras, beside the case of affine algebras.  In Appendix~A, we
show that in
the case of rational conformal field theories with a current algebra, the
previous
discussion may indeed be repeated if the energy momentum tensor of the
theory has been
modified in such a way that unspecialized characters appear. Then
using the known modular transformations of the latter~\cite{Kac},
one derives (\ref{IIi}). We shall therefore assume that
(\ref{IIi}) holds true for general RCFT.

\bigskip
We now return to the Cardy equation (\ref{IIi}), and
assume that we have found an orthonormal set of boundary
states, {\it i.e. } satisfying
\be
\sum_{j\in \calE}\psi_\a^j(\psi_\b^j)^*=\delta_{\a\b}
\label{IIiaa} \ .
\ee
Moreover we make the stronger assumption that we have found a
{\it complete} set of such states, {\it i.e. } satisfying
\be
\sum_{\a}\psi_\a^j(\psi_\a^{j'})^*=\delta_{jj'}
\label{IIiab} \ .
\ee
(Note that this implies that the number of these boundary states must be
equal to the cardinality of $\calE$).

Finally we recall that the ratios $S_{ij}/S_{1j}$, for a fixed
$j\in \calI$, form a one-dimensional representation of the fusion
algebra, as a consequence of the Verlinde formula (\ref{Verlinde}):
\be
{S_{i_1j}\over S_{1j}} {S_{i_2 j}\over S_{1j}}
=\sum_{i_3\in \calI}N_{i_1i_2}{}^{i_3}{S_{i_3 j}\over S_{1j}} \ .
\label{IIiad}
\ee
It follows from (\ref{IIiab}), (\ref{IIiad})
that the matrices $\n_i$, defined by
\be
(\n_i)_\a{}^\b=\n_{i\a}{}^\b\qquad i\in \calI
\ee
also satisfy  the (commuting) fusion algebra
\be
\n_{i_1} \n_{i_2}=\sum_{i_3\in\calI} N_{i_1i_2}{}^{i_3} \n_{i_3}
\label{IIif} \ .
\ee
By (\ref{IIiaa}), $\n_1=I$, the unit matrix, and by
(\ref{IIiac}),  $\n_{i^*}=\n_i^T$.

Conversely, given a set of matrices with non-negative integer
elements, satisfying $\n_{i^*}=\n_i^T$, $n_1=I$ and the fusion
algebra, they form a commuting set, and thus  each $\n_i$
commutes with its transpose. These matrices are thus normal
matrices that may be diagonalized in an orthonormal basis.
Their eigenvalues are of the form $S_{ij}/S_{1j}$  for some $j$,
 and they
may thus be  written in the form (\ref{IIi}).   If one
pretends to determine the spectrum $\calE$ from the $n$'s, one
has to impose also that $j=1$ appears only once in $\calE$, as a
manifestation of the uniqueness of the vacuum.

We thus conclude that the search for orthonormal and complete
solutions to the Cardy equation is equivalent to the search
for $\Bbb N$ valued representations of the fusion algebra
satisfying $\n_i^T=\n_{i^*}$.

This is the first important result of this paper, already presented
succinctly in~\cite{BPPZ}.
The fact that some solutions to the Cardy equation were associated
with representations of the fusion algebra had been noticed
before.  In his seminal paper~\cite{Ca89}, Cardy considered the
case of ``diagonal theories'' (for which $\calE=\calI$) and
showed that the $\n_i$ matrices were nothing other than the fusion
matrices $N_i$, thus obtaining an alternative and more intuitive
derivation  of the Verlinde formula. In an antecedent work by
Saleur and Bauer~\cite{SaBa}, other solutions had been obtained
in non-diagonal theories, starting from their lattice
realization, and the fact that these $\n_i$ coefficients
satisfied the fusion rules had been emphasized in~\cite{DFZ1}.
More recently Pradisi, Sagnotti and  Stanev~\cite{PSS,SS,SS1} proposed a
different argument to the same effect, where a notion of
completeness of boundary conditions is also playing a crucial
role.

\medskip
Solutions such that {\it all} matrices $\n_i$ may be written as
$\n_i=(\n_1)_i \oplus (\n_2)_i$
after the same suitable permutation of rows and columns can be
called reducible. They describe  sets of decoupled boundary
conditions. We thus restrict our attention to irreducible sets of
matrices.


\subsection{WZW $sl(2)$ theories}

For theories with the affine (current) algebra $\slh(2)$
as a chiral algebra, the problem of classifying representations
of the fusion algebra was solved long ago~\cite{DFZ1}.
The integrable highest weight representations of
$\slh(2)_k$ at level $k\in {\Bbb N}$ are labelled by an integer $1\le
j\le k+1$,
$S_{ij}= \sqrt{{2\over k+2}} \sin {ij\pi\over k+2}$
and the Cardy equation says that the generator
$\n_2=\n_{2^*}$ has eigenvalues ${S_{2j}\over S_{1j}}=2 \cos
{\pi j\over k+2}$. The only symmetric irreducible matrices
with non-negative integer entries and eigenvalues less than $2$
are the adjacency matrices of $A$-$D$-$E$-$T$  graphs~\cite{GHJ} 
of Figure~2 (see also Table 1).
The ``tadpole'' graphs are given by $T_n:=A_{2n}/{\Bbb Z}_2$. Here
the level $k$ is related to the Coxeter number by $g=k+2$. Only
the \ade\ solutions are retained as their spectrum matches the
spectrum of $\slh(2)$ theories, known by their modular invariant
partition functions~\cite{CIZ}. For a theory classified by a
Dynkin diagram $G$ of $A$-$D$-$E$ type, the set
$\calE$ is the set of Coxeter exponents of $G$ as in Table 1. The
matrices $\n_i$ are then defined recursively by $\n_1=I$, $n_2=G$ and by
equation (\ref{IIif}) which reduces here to $\n_{i+1}=\n_2
\n_i-\n_{i-1}$, $i=2,3,\ldots,k$. They are the well known ``fused
adjacency matrices''  or ``intertwiners'' $V_i$, studied
in~\cite{DFZ1,PZh} and whose properties are recalled in Appendix~B.
One verifies that all their entries are non-negative integers.
This set of complete orthonormal solutions of the Cardy equation
for $\slh(2)$ theories is unique up to a relabelling of the
states $|\a\rangle$.


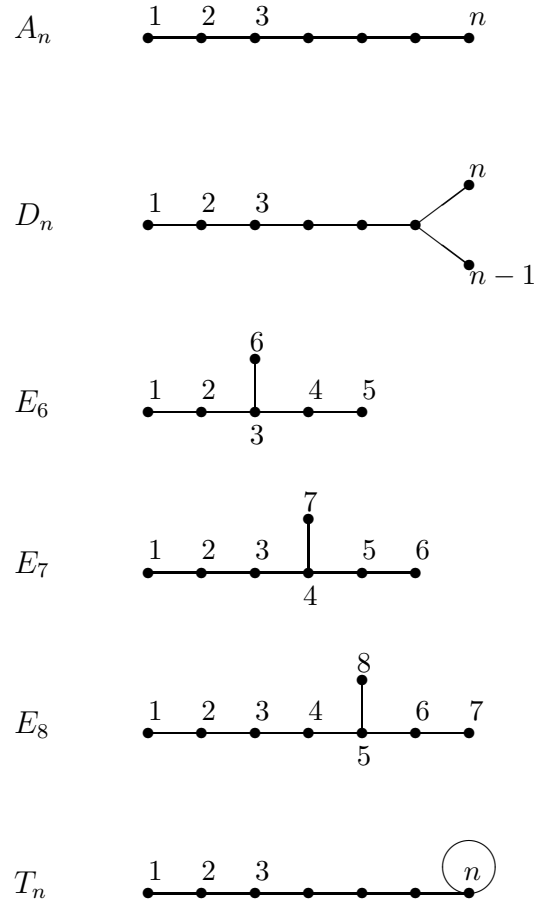
\begin{figure}[p]
\begin{center}
\begin{minipage}[t]{2.5in}

\setlength{\unitlength}{.014in}
\vspace{.2in}
\begin{center}

\begin{picture}(140,280)(-20,-240)

\put(-50,80){$A_n$}
\multiput(0,80)(20,0){6}{\line(1,0){20}}
\multiput(0,80)(20,0){7}{\circle*{4}}
\put(0,85){\small 1} \put(20,85){\small 2} \put(40,85){\small 3}
\put(120,85){\small $n$}

\put(-50,10){$D_n$}
\multiput(0,10)(20,0){5}{\line(1,0){20}}
\put(100,10){\line(4,3){20}} \put(100,10){\line(4,-3){20}}
\multiput(0,10)(20,0){6}{\circle*{4}}
\multiput(120,25)(0,-30){2}{\circle*{4}}
\put(0,15){\small 1} \put(20,15){\small 2} \put(40,15){\small 3}
\put(120,28){\small $n$}\put(120,-12){\small $n-1$}

\put(-50,-60){$E_6$}
\multiput(0,-60)(20,0){4}{\line(1,0){20}}
\multiput(0,-60)(20,0){5}{\circle*{4}}
\put(40,-40){\line(0,-1){20}} \put(40,-40){\circle*{4}}
\put(0,-55){\small 1} \put(20,-55){\small 2} \put(60,-55){\small 4}
\put(80,-55){\small 5}
\put(38,-72){\small 3} \put(38,-37){\small 6}

\put(-50,-120){$E_7$}
\multiput(0,-120)(20,0){5}{\line(1,0){20}}
\multiput(0,-120)(20,0){6}{\circle*{4}}
\put(60,-100){\line(0,-1){20}} \put(60,-100){\circle*{4}}
\put(0,-115){\small 1} \put(20,-115){\small 2} \put(40,-115){\small 3}
\put(80,-115){\small 5}
\put(100,-115){\small 6} \put(58,-132){\small 4} \put(58,-97){\small 7}

\put(-50,-180){$E_8$}
\multiput(0,-180)(20,0){6}{\line(1,0){20}}
\multiput(0,-180)(20,0){7}{\circle*{4}}
\put(80,-160){\line(0,-1){20}} \put(80,-160){\circle*{4}}
\put(0,-175){\small 1} \put(20,-175){\small 2} \put(40,-175){\small 3}
\put(60,-175){\small 4}
\put(100,-175){\small 6} \put(120,-175){\small 7} \put(78,-192){\small 5}
\put(78,-157){\small 8}

\put(-50,-240){$T_n$}
\multiput(0,-240)(20,0){6}{\line(1,0){20}}
\multiput(0,-240)(20,0){7}{\circle*{4}}
\put(0,-235){\small 1} \put(20,-235){\small 2} \put(40,-235){\small 3}
\put(118,-235){\small $n$}
\put(120,-230){\circle{20}}

\end{picture}

\end{center}
\end{minipage}
\end{center}
\caption{The $A$-$D$-$E$-$T$ graphs.}
\setlength{\unitlength}{.01in}
\end{figure}

\begin{table}[p]
\begin{center}
\begin{tabular}{ccc}
\hline \rule[-10pt]{0pt}{25pt}$G$&$g$&$m\in\mbox{Exp}(G)$\\
\hline
\rule[-10pt]{0pt}{25pt}$A_n$&$n+1$&$1,2,3,\ldots,n$\\
\rule[-10pt]{0pt}{20pt}$D_n$&$2n-2$&$1,3,5,\ldots,2n-3, n-1$\\
\rule[-10pt]{0pt}{20pt}$E_6$&$12$&$1,4,5,7,8,11$\\
\rule[-10pt]{0pt}{20pt}$E_7$&$18$&$1,5,7,9,11,13,17$\\
\rule[-10pt]{0pt}{20pt}$E_8$&$30$&$1,7,11,13,17,19,23,29$\\
\rule[-10pt]{0pt}{20pt}$T_n$&$2n+1$&$1,3,5,\ldots,2n-1$\\
\hline
\end{tabular}
\end{center}
\caption{The Coxeter number $g$ and Coxeter exponents $m$ of the
$A$-$D$-$E$-$T$ graphs.}
\end{table}


\subsection{Minimal $sl(2)$ models}

The classification of  $c<1$ minimal models in the bulk is given in
terms of a pair of Dynkin diagrams $(A,G)$ where $G$ is
of \ade\ type~\cite{CIZ}.  Let $h$ be the
Coxeter number of $A_{h-1}$ and $g$ the Coxeter number of $G$ as
given in Table~1. Then the complete \ade\ classification is
\be
\calM(A,G)=\cases{
\calM(A_{h-1},A_{g-1})&\cr
\calM(A_{h-1},D_{(g+2)/2}),\quad g\ \mbox{even}&\cr
\calM(A_{h-1},E_6)&\cr
\calM(A_{h-1},E_7)&\cr
\calM(A_{h-1},E_8)} \label{Minaa}
\ee
with $h,g\ge 2$ and central charges given by
\be
c=1-\disp{6(h-g)^2\over hg} \ . \label{Minab}
\ee
We will use $\calM(A,G)$ to denote these minimal theories.
Since $g$ and $h$ must be coprime and $g$ is
even for all non-$A$ cases, one may always assume, at the price
of a possible interchange in the $(A,A)$ case, that $h$ is odd,
$h=2p+1$.

Some members of these series are identified as follows:
\be
\begin{array}{ll}
\calM(A_2,A_3)=\mbox{critical Ising}&c=1/2\\
\calM(A_4,A_3)=\mbox{tricritical Ising}&c=7/10\\
\calM(A_4,D_4)=\mbox{critical 3-state Potts}&c=4/5\\
\calM(A_6,D_4)=\mbox{tricritical 3-state Potts}\quad&c=6/7\\
\label{Minac}
\end{array}
\ee

We will use $G$ to denote
both the Dynkin diagram and its adjacency matrix. We use $r,
r_1, r_2$ to denote nodes or exponents of $A_{h-1}$; $s, s_1,
s_2$ for the nodes (or exponents) of $A_{g-1}$; $a, a_1, a_2, b$
for the nodes of $G$.  We refer the reader to Appendix~B for more
data on these matrices and their eigenvectors.

If $\Exp(G)$ denotes the set of exponents of $G$ (see Table~1),
the modular invariant partition function of $\calM(A_{h-1},G)$ reads
\be
Z=\oh\sum_{r=1}^{h-1}\sum_{s\in\Exp(G)}|\chi_{rs}(q)|^2 +
\ \mbox{off-diagonal\ terms}\ . \label{Mina}
\ee
The factor $\oh$ removes the double counting due to the
well-known identification of the $(r,s)$ and $(h-r,g-s)$
representations of the Virasoro algebra. The diagonal terms in
$Z$, i.e. the left-right symmetric (highest weight) states in the
spectrum are thus labelled by the set
\be
 \calE=\{j=(r,s)\equiv (h-r,g-s); 1\le r\le  h-1; s\in \Exp(G)\}
\,. \label{Minba}
\ee

Each of the unitary minimal models $\calM(A_{h-1},G)$ with
$g-h=\pm1$ can be realized as the continuum scaling limit of an
integrable two-dimensional lattice model at criticality, with
heights living on the nodes of the graph $G$. In particular, the
critical series with $g-h=1$ is associated with the A-D-E
lattice models~\cite{AndBaxFor87,Pas87a} and the tricritical
series with $g-h=-1$ is associated with the dilute lattice
models~\cite{Roc92,WarNie93}.  In the non-unitary cases the
associated lattice models~\cite{FoBa} possess negative Boltzmann
weights.
In the construction of the corresponding lattice models as well
as in the description of boundary conditions, it turns out that
the two diagrams of the pair $(A,G)$ do not play a symmetric
role.

According to the discussion of the previous section, we have to
study the fusion algebras of minimal models and their
(integer-valued) representations. The Verlinde formula for the
fusion coefficients makes use of the matrix $\calS$ of modular
transformations of characters
\be
\calS_{rs,r's'}=\sqrt{{8\over gh}}(-1)^{(r+s)(r'+s')}
\sin\pi rr'{g-h\over h}\,\sin\pi ss'{g-h\over g} \label{Minb}
\ee
with the restriction $r,r'$ odd (or any equivalent
condition).  The fusion coefficients are then found to be tensor
products of those relative to the $\slh(2)$ algebras of level
$h-2$ and $g-2$, up to a symmetrization which accounts for the
identification $(r,s)\equiv (h-r,g-s)$
\be
\calN_{rs,r's'}{}^{r''s''}=N_{rr'}{}^{r''} N_{ss'}{}^{s''}
+N_{rr'}{}^{h-r''} N_{ss'}{}^{g-s''} \,. \label{Minc}
\ee
This may be regarded as the regular representation of the fusion
matrices $\calN_{rs}$ of the Virasoro algebra of central charge
(\ref{Minab}). Our problem is to find the general non-negative
integer valued representations of this algebra. One observes that
$\calN_{rs}=\calN_{r1}\calN_{1s}$ and that the algebra is thus
generated by $\calN_{21}$ and $\calN_{12}$. Also, the eigenvalues
of $\calN_{12}$ and $\calN_{21}$ are of the form
\def\r{{r'}}\def\s{{s'}}
\bea
{\calS_{12,\r\s}\over \calS_{11,\r\s}}&=& (-1)^{\r+\s}\, 2
\cos\pi \s{g-h\over g} =(-1)^{\r}\,  2 \cos\pi \s{h\over g}
\ , \label{Mind} \\
{\calS_{21,\r\s}\over \calS_{11,\r\s}}&=&
(-1)^\s\, 2 \cos\pi \r{g\over h}
\ . \label{Minds}
\eea
with $1 \le \r \le h-1$, $1 \le \s \le g-1$ and again
$(\r,\s)\equiv (h-\r, g-\s)$.

Turning now to a general (integer valued) representation
$\n_{rs}$ of the fusion algebra, it is still true that it is
generated by $\n_{12}$ and $\n_{21}$. In addition, we want the
spectrum of the $\n_{rs}$ to be specified by the set of
``exponents'' $\calE$ of (\ref{Minaa}), that is $(\r,\s)$ in
(\ref{Mind}-\ref{Minds}), with the eigenvalues labelled by $\s$
appearing with some multiplicity in general.
To remove the redundancy in the labelling
of eigenvalues, we will usually take 
$\r$ odd,  $\r=1,3, \cdots, h-2$,  and
$(\s,\alpha)\in \Exp(G)$. In the sequel, we will drop
this explicit notation for multiplicities.
 We know of course a solution to this problem,
namely
\be
n_{rs}=N_r\otimes V_s + N_{h-r}\otimes V_{g-s}
\ee
in terms of the fusion matrices $N$ of $\slh(2)$ at level $h-2$
and of the intertwiners $V$ of type $G$ introduced in the
previous subsection (see also Appendix~B).
More explicitly, this describes a solution to the Cardy equation
between boundary states $(r_1,a)$ and $(r_2,b)$
\be
n_{rs; (r_1,a)}{}^{(r_2,b)}=N_{rr_1}{}^{r_2}V_{sa}{}^b
+N_{h-r\,r_1}{}^{r_2}V_{g-s\,a}{}^b\ , \label{Minf}
\ee
with $1\le r,r_1,r_2 \le h-1=2p$, $1\le s\le g-1$, and
$a,b$ running over the nodes of the Dynkin diagram $G$.

Because of the properties of the $N$ and $V$ matrices recalled in
Appendix~B, it is readily seen that
\be
n_{rs; (r_1,a)}{}^{(r_2,b)}=n_{rs; (r_1,a)}{}^{(h-r_2,\gamma(b))}=
n_{rs; (h-r_1,\gamma(a))}{}^{(r_2,b)}
\ee
for an automorphism $\gamma$ acting on the nodes
of the graph $G$: this is the identity except   for the $A$, $D_{{\rm odd}}$
and $E_6$ cases, for which it is the natural $\Bbb{Z}_2$ symmetry
of the diagram.  We conclude that
this solution describes boundary states of  $\calM(A_{h-1},G)$
labelled by pairs $(r,a)$ of nodes of the $A_{h-1}$ and of
the $G$ graph, with the identification
\be
(r,a) \equiv (h-r,\gamma(a)) \label{kacsymm} .
\ee
One checks  that the number of independent boundary states
$|(r,a)\rangle$ is
\be
{\rm number \ of \ independent\ boundary \ states\ }=\oh (h-1)n
\label{Mincd}
\ee
with  $n$ the number of nodes of $G$, or the number of its
exponents. This number (\ref{Mincd}) coincides with the number of
independent  left-right symmetric highest weight states $|r,s\rangle
\otimes |r,s\rangle$ in the spectrum of the theory on a cylinder,
i.e. with the cardinality of the set $\calE$, as it should.

With such boundary states, the cylinder partition function reads
\be
Z_{(r_1,a)|(r_2,b)}= Z_{(r_2,b)|(r_1,a)}
= Z_{(r_1,a)|(h-r_2,\gamma(b))} = \sum_{r,s}
\chi_{r s}(q) N_{rr_1}{}^{r_2} V_{s a}{}^b
\,. \label{IIIa}
\ee
Here the sum runs over $1\le r\le h-1$, $1\le s\le g-1$.

Let us look more closely at  (\ref{Minf}).
There exists a basis in which (\ref{Minf})
takes  a factorized form. Indeed
one may use the identifications $(r,s)\equiv (h-r,g-s)$
and (\ref{kacsymm}) to restrict $r, r_1, r_2$ to odd values
(recall that $h=2p+1$ is odd). Then $N_{h-r r_1}{}^{r_2}=0$,
the  r.h.s. of (\ref{Minf}) factorizes and
the following expressions 
\be 
\Psi^{(r''s'')}_{(r,a)}= \sqrt{2} S_{r r''} \psi_a^{s''}, \qquad r,r''~{\rm odd},
\ s'' \in \Exp(G)\ ,  \label{jbsduty}
\ee 
written in terms of the modular matrix of $\slh(2)$ at level $h-2$
and of eigenvectors $\psi$ of $G$, are readily seen to be eigenvectors
of $n_{rs}$. Their eigenvalue is of the form $\calS_{rs,r's'} /
\calS_{11,r's'}$ after some reshuffling $r'',s'' \to r',s'$.

One also shows (see Appendix~C) that there exists a basis
in which
\bea
n_{12}&=& {I}_p \otimes G =\pmatrix{G  & & & \cr
                  & G & & \cr
                  &  & \ddots & \cr
                  & & & G\cr} \ , \\
n_{21} &=& T_p \otimes \Gamma = \pmatrix{0 & \Gamma && \cr
	  		\Gamma & 0 &\Gamma & \cr
			 & \ddots& \ddots & \Gamma \cr
			 & & \Gamma& \Gamma\cr}
\eea
in terms of the tadpole $T_p$ adjacency matrix and of
$\Gamma$, the matrix that realizes the
 automorphism $\gamma$: $\Gamma_a{}^b=\delta_{a\gamma(b)}\,$.

\medskip
Conversely, suppose we only know that the representation $n_{rs}$
has a spectrum specified by the set of exponents $\calE$.
The question is: are these spectral data  sufficient to guarantee
that the only $\n_{rs}$ are of the form (\ref{Minf}) in a certain basis?
 A proof of this fact is relegated to Appendix~C. Notice that
our discussion has assumed the classification of modular
invariants to be known. It should be possible to extend it as in the case
of WZW $\slh(2)$ models  and to classify the representations of the
Verlinde algebra without this information. A few spurious cases
involving tadpoles etc. would then have to be discarded.

To recapitulate, we have proved that the only representations of
the fusion algebra of minimal models are given by (\ref{Minf}).
To our knowledge, this is the first proof of the
uniqueness of these (complete orthonormal) boundary states
of minimal models.

\vskip5mm
Some physical intuition about the meaning of these boundary
conditions may be helpful. For this we appeal to the lattice
realization of the minimal model as a generalized height model on
the graph $G$ (see~\cite{BPZ}).  A
boundary condition of the type $(1,a)$ describes a {\it fixed} boundary
condition, where  the height of the model is fixed to value $\a$
on the graph. The interpretation of the other label $r$ is less
intuitive. The boundary condition $(r,a)$ is realized by
attaching an $r$-times fused weight to height $\a$.

The expression (\ref{IIIa})
for the cylinder partition function
encompasses and generalizes cases that were already
known:
\begin{itemize}
\item From the work of Saleur and Bauer~\cite{SaBa}
who discussed boundary conditions in lattice height models of
\ade\ type on a cylinder in which the heights on the boundaries
are fixed to the values $\a$ respectively $\b$. They showed that in the
continuum limit, the partition function reads $$ Z_{\b|\a}=
\sum_s V_{sa}{}^b \chi_{1s}\ . $$
\item From the work of
Cardy~\cite{Ca89} who showed how to construct new boundary
conditions by  fusion.
\item From the work of Pasquier and
Saleur~\cite{PaSa}, who interpreted the pair of relations
\bea
Z^{(A_{h-1},G)}_{(1b)|(1a)}&=&\sum \chi_{1s} V_{sa}^{\ \ b}
\label{IIIb}\\
Z^{(A_{h-1},A_{g-1})}_{(1,s)|(1,1)}&=&\chi_{1s}\label{IIIc}
\eea
as expressing the decomposition of the representation of the
Temperley-Lieb algebra on the space of paths from $\a$ to $\b$ on
graph $G$ onto the irreducible ones on the paths from 1 to $s$ on
graph $A_{g-1}$, see point (ii) at the end of Appendix~B.
\end{itemize}

{\bf Examples}\\
Let us illustrate these expressions of boundary states by a few
simple cases. In the Ising model (the $(A_2,A_3)$ minimal model),
$h=3$, $G=A_3$, thus $n=3$ and  there are $\oh (3-1)\times 3 =3$
boundary states, generally denoted~\cite{Ca89} $+$,$-$ and $f$.
On the lattice, the first two describe fixed boundary conditions
 on the spin $\sigma=1$ or $-1$
respectively, while $f$ corresponds to free boundary conditions.


It is then
instructive to consider two related examples, see also~\cite{AfflOS98,
FuchsS98}. The first is
the $c=2$ $D_4$ solution of  $\slh(2)_4$ at level 4, and the
other is its cousin, the $c=4/5$ minimal (3-state Potts) model,
already mentioned in Section 2.4 and labelled by the pair $(A_4,D_4)$. In the
former case, we find four boundary states, labelled by 1 to 4,
that we attach to the nodes of the $D_4$ diagram. All these
states satisfy the required boundary conditions. The set
of exponents is $\calE=\{1,3,3,5\}$.  But this $D_4$
$\slh(2)$ model is also known to result from the conformal
embedding of $\slh(2)_4$ into $\slh(3)_1$.
Regarded as an $\slh(3)$ theory, the model admits {\it three}
boundary states satisfying the more restrictive $\slh(3)$
conditions $(L_n-\overline{L}_{-n})|\a\ket=
(J_n+\Omega \overline{J}_{-n})|\a\ket=0$,
where the choice of $\Omega $ corresponds to the diagonal set
$(j,j)$.
These three boundary states  may be regarded as the three nodes
of a triangular graph $\calA^{(4)}$ (see Appendix~D and Figure~10),
or as the three
extremal nodes of the $D_4$ diagram that have survived the
additional $\slh(3)$ constraint.

The discussion of the Potts model is quite parallel. From the minimal
model standpoint, it is the $(A_4,D_4)$ model, $h=5$,
$n=4$ and there are 8 boundary states~\cite{AfflOS98, FuchsS98}:
\bea
&&
\phantom{A} A=(1,1)=(4,1),\quad
\phantom{A} B=(1,3)=(4,3),\quad
\phantom{A} C=(1,4)=(4,4)\nonumber\\
&&BC=(2,1)=(3,1),\quad
AC=(2,3)=(3,3),\quad
AB=(2,4)=(3,4)\\
&&\qquad\qquad ABC=(1,2)=(4,2),\quad
N=(2,2)=(3,2) \nonumber
\eea
On the lattice, the first three $A,B,C$ describe fixed boundary conditions
where the ``spin'' takes at each site of the boundary
one of the three possible values. The mixed boundary conditions
$AB$, $BC$, $AC$ describe boundary conditions where the
spin on the boundary can take on two values independently.
The boundary conditions $ABC$ and $N$ are free  boundary conditions
but for $N$ the weights depend on whether adjacent spins are
equal or not.

The model may also be regarded as the simplest $\calW_3$ model.
In that picture, one may impose more stringent boundary
conditions.
Only the six states denoted above $A$, $B$, $C$,
$AB$, $BC$, $AC$ satisfy the additional condition
$(W^{(3)}_n+ \Omega \overline{W^{(3)}}_{-n})|a\ket =0$.
 They correspond
to the extremal nodes of the pair $(T_2,D_4)$ or,
alternatively, to the nodes of the pair $(T_2, \calA^{(4)})$.
As will be discussed in more detail in Section 3,
the subset of these nodes, to be denoted $T$, can be
identified in both examples with
the representation labels of the corresponding extended chiral algebra.
The matrix elements $\psi_a^j$ for $a\in T$ satisfy~\cite{PZ2}
the 
relation
${\psi_a^j \over \sqrt{S_{1 j}}}
={S_{a \{j\}}^{ext}  \over \sqrt{S_{\{1\} \{j\}}^{ext}
 } }\,,$ where $\{j\}$
denotes the orbit of the exponent $j$ with respect to the  ${\Bbb Z}_2$
automorphism and $S_{a \{j\}}^{ext}
$  is the modular matrix of the extended
theory. This relation
 implies that $|a\rangle = \sum_{ \{j\}}\, {S_{a \{j\}}^{ext}
 \over
\sqrt{S_{\{1\} \{j\}}^{ext}
} }
 \sum_{j\in \{j\}}\, |j\rrangle$, i.e., we can identify
 $\sum_{j\in \{j\} } |j\rrangle =
| \{j\} \rrangle$ with an extended Ishibashi state.
The missing boundary condition corresponds to a twisted boundary
condition  from the point of  view of the extended algebra.

We conclude that, as expected, the number and nature of the boundary
states reflect the precise conditions that  they are supposed to
satisfy.



\sect{Graph Fusion Algebras}

According to the discussion of Section 2, given a certain chiral algebra
$\calA$, the sets of complete orthonormal boundary states of
RCFTs consistent with this algebra are classified by
representations of the Verlinde algebra of $\calA$, on matrices $\n_i$
with non-negative integral entries.
Or stated differently: given a certain RCFT with a chiral algebra
$\calA$, the sets of complete orthonormal boundary states of
this theory consistent with this algebra are classified by
representations of the Verlinde algebra of $\calA$, on matrices $\n_i$
with non-negative integral entries, with eigenvalues
specified by the diagonal part $\calE$ of the spectrum.
These matrices may thus
be regarded as the adjacency matrices of a collection of $|\calI|$
graphs.  In practice,
it is sufficient to look
at the smaller number of matrices representing the generators
of the fusion ring. For example,
one matrix in the case of $sl(2)$ considered above, or
the $N-1$ matrices associated with the fundamental representations,
in the case of $sl(N)$.

The simplest case is given by the regular
representation of the fusion algebra, when the matrices
$\n_i$ are the Verlinde matrices themselves, $\n_i=N_i$. This is
the case of so-called diagonal theories,
when all representations of the set $\calI$
appear once in the spectrum ${\rm Spec}=\{(i,i)| i\in \calI \}$.
This may be regarded as the case of reference from several
points of view: it was the first case analysed in detail~\cite{Ca89};
the corresponding graphs are playing a central role; and
finally in that case, Cardy was able to provide a
physical argument explaining why the fusion matrices arise
naturally. It is the purpose of this section
to extend these considerations to more general solutions.
We shall find that the role of the fusion matrices in the
arguments of Cardy is now played by two sets of matrices. The
first is the set of matrices $\n_i$ that describe the coefficients
of the cylinder partition function; the second is a new set
of matrices $\hN_\a$, forming  what is called the graph fusion algebra.

On the other hand, since we know that the cylinder partition functions,
or equivalently the matrices $\n_i$, contain some information
about the bulk theory, through the knowledge of the
diagonal spectrum $\calE$, it is expected that this classification
of boundary conditions should have some bearing on the
classification of bulk theories, namely on the classification
of torus partition functions and on bulk structure constants.
Remarkably, this programme
 works even better than expected and the two classification
problems seem to be essentially equivalent, at least for type I
theories (see end of Section 1). This will be
explained in Section 3.3  below.

\subsection{More on graphs and intertwiners}

Suppose we have found a solution to the Cardy equation, namely a
set of $n\times n$ matrices (\ref{IIi}), $(\n_i)_a{}^b$, $i=1,
\cdots, |\calI|$, $a,b=1,\cdots, n$. What was said in detail in
Section 2 and in Appendix~B in the case of $\slh(2)$ can be repeated
here.  As their entries are non-negative integers, these matrices
may be regarded as adjacency matrices of a set of $|\calI|$
graphs $G_i$, with $n=|G_i|\equiv |G|$ nodes.  We shall refer
collectively to these $|\calI|$  graphs as ``the graph $G$'',
whereas the basic solution provided by the $N$'s themselves will
be called ``the $A$ graph'', (borrowing the notation from the
$sl(2)$ case). The eigenvalues of the matrices $\n_i$ are
specified by a set $\Exp(G)$ in the sense that they are of the
form $S_{ij}/S_{1j}$, $(j,\alpha)\in \Exp(G)$.  Moreover $\Exp(G)=\calE$
if the RCFT is given and the diagonal spectrum $\calE$ is known.
But in general, the determination of the set $\Exp(G)$ is part of the
problem. The fundamental relation $\sum_b (\n_i)_a{}^b
\, (\n_j)_b{}^c=\sum_k N_{ij}{}^k (\n_k)_a{}^c$ may be
interpreted in two ways:
\begin{itemize}
\item Regarded as $|G|\times |G|$ matrices,
the matrices $\n_i$ form a representation of the fusion algebra
(\ref{IIif}).
\item Regarded as a $|\calI|\times |G|$ rectangular
matrix, each matrix $\tilde\n_a$ for $a$ fixed,
$(\tilde\n_a)_j{}^b:= \n_{ja}{}^b$ intertwines the representatives
$N_{i}$ and $\n_i$ in the two representations $N_i \tilde\n_a=
\tilde\n_a \n_i$, or more explicitly
\be
\sum_k N_{ij}{}^k \n_{ka}{}^c = \sum_b \n_{ja}{}^b \n_{ib}{}^c \ .
\ee
 We shall thus occasionally refer to the matrices $\n_i$ as
``intertwiners''.
\end{itemize}

The case of graphs and intertwiners pertaining to $\slh(2)$
theories has been discussed at length in Section 2 and in Appendix~B.
In Appendix~D, we outline the
discussion of $\slh(3)$. In that case, the fusion algebra
is generated by two matrices $\n_{(2,1)}$ and $\n_{(1,2)}$
(labelled by the two fundamental (shifted)
weights of $sl(3)$), and as these two representations are
complex conjugate to one another, the matrices $\n_i$ are
related by transposition  $\n_{(2,1)}=n^T_{(1,2)}\, $. Then according
to~(\ref{IIif}),
$n_i$ is given by the same polynomial of $n_{(1,2)}$ and $n_{(2,1)}$
with integral coefficients as that representing
$N_i$ in terms of $N_{(1,2)}$ and $N_{(2,1)}$. It is thus
sufficient to list all possible graphs representing the  matrix $\n_{(2,1)}$,
provided one checks that all $n_i$ have non-negative integral
entries.
In contrast with the case of $\slh(2)$, no complete solution
is known for $\slh(3)$. The current state of the art is presented in
Appendix~D with  tables, figures and relevant comments.


\subsection{Graph fusion algebras}

To see what is playing the role of the fusion algebra
in the argument of Cardy, we have to introduce the graph
fusion algebra.  The graph fusion algebra, as first discussed by
Pasquier~\cite{Pa}, is a fusion-like algebra
attached to a connected graph $G$. Let $\psi_\a^j$ be the
common orthonormal eigenvectors of the adjacency matrices
$G$ labelled by $j\in\Exp(G)$. In general, these eigenvectors can
be complex.  In the case of degenerate eigenvalues the associated
eigenvectors need to be suitably chosen.
We assume that the graph has a distinguished
node labelled $1=1^*$ such that $\psi_1^j>0$, for all $j\in
\Exp(G)$.

One then defines the numbers
\be
\hN_{ab}{}^c=\sum_{j\in\Exp(G)}{\psi_{a}^j
\psi_{b}^j(\psi_{c}^j)^*\over \psi_{1}^j}
\label{graphfusmats}
\ee
and the matrices $\hN_\a$ with elements
$(\hN_\a)_\b{}^c=\hN_{ab}{}^c$ satisfy
$\hN_{ab}{}^c=\hN_{ac^*}{}^{b^*}$ and $N_a^T=N_{a^*}$.
Because of orthonormality, $\hN_1=I$.  Since each matrix $\hN_a$
has a single non-vanishing entry in the row labelled $1$,
namely $(\hN_a)_1{}^b=(\hN_1)_a{}^b=\delta_{ab}$,
the matrices $\hN_a$ are linearly independent. The $\hN_{ab}{}^c$ are
the structure constants of the graph fusion algebra satisfied by
the $\hN$ matrices
\be
\hN_\a \hN_\b =\sum_c \hN_{ab}{}^c \hN_c \label{IIza}
\ee
which is an associative and commutative algebra.  Of course, if
the graph $G$ is of type $A$, this boils down to the ordinary
Verlinde fusion algebra since the matrix $\psi$ of eigenvectors
is nothing but the modular matrix $S$.

Since the $\hN_\a$ and $\n_i$ matrices have the same
eigenvectors, it is easy to derive the matrix relation
\be
\n_i \hN_\a= \sum_\b \n_{i\a}{}^\b \hN_\b\ . \label{IIzc}
\ee
In particular for $a=1$, $\hN_1=I$, and all $\n_i$
appear as linear combinations with non-negative integer
coefficients of the $\hN$'s
\be
\n_{ia}{}^b=\sum_c \n_{i1}{}^c\hN_{ca}{}^b \label{IIzd}
\ee
Alternatively (\ref{IIzc}) may be used as a starting point
to reconstruct the graph algebra, as explained in  Section 3.5.

It should be stressed that the definition of a graph fusion
algebra is not unique. In general, it depends on the choice of
the distinguished node 1 and, when there are degenerate
eigenvalues, also on the choice of the eigenvectors
$\psi_\a^j$. To view the graph fusion algebra as a proper
fusion algebra we would like the structure constants
$\hN_{ab}{}^c$ to be non-negative integers. But even the
rationality of these numbers is not obvious and it is therefore
surprising that, for  appropriate choices of the $\psi$'s and of
node 1 and for {\it most } cases,
they turn out to be integers of either sign.
Among all the examples known to us in $\slh(N)$ theories, $2\le N\le 5$,
  it fails in only two cases: the graph called $\calE^{(12)}_5$ in Figure~11,
for which there is no node 1 satisfying $1=1^*$, and whose
$\hN$ algebra involves fractions of denominator 4; and a graph 
in the $\slh(4)_4$ theory,~\cite{PZ2},
in which half-integer $\hN_{ab}{}^c$ of either sign occur.
Adopting  (\ref{IIif}), (\ref{IIzc}) in the framework of
subfactors theory the latter example has been  reinterpreted by
Xu ~\cite{Xu} by trading commutativity of the graph fusion
algebra for integrality.

 Finally the non-negativity
of the $\hN$ is only possible for certain graphs which we call proper fusion  
graphs.  For example, for the $sl(2)$ theories,
the $A$-$D$-$E$ graphs that admit a proper graph fusion algebra
are
\be
\mbox{proper  $A$-$D$-$E$  graphs}=
A_n,D_{2q},E_6,E_8 \,. \label{properADE}
\ee
The choice of distinguished node for the $sl(2)$
$A$-$D$-$E$ graphs is explained in Appendix~B.

We note that the set of proper $sl(2)$ fusion graphs
matches the modular invariant partition functions
 listed as ``type I'' at the end of Section 1. 
The situation is somewhat different for $sl(3)$ graphs, 
for which we have to introduce a further distinction. 
In this case, some graphs with non-negative $\hN$'s 
are not associated with type I theories. We reserve
the terminology ``type I graph'' for those graphs
associated with type I theories (see Appendix~D and Tables).  
Moreover, as is clear from  the $\slh(4)_4$ example above, 
some type I
modular invariant partition functions are associated with
graphs with non-integer and/or non-positive $\hN_{ab}{}^c$. In the following,
we discard these exceptional cases and restrict ourselves
to type I graphs that are associated with type I RCFTs.
In general, the question of precisely which
graphs admit type I  fusion algebras should be related to
the classification of type I RCFTs, and thus is a very
interesting open question.


\subsection{Fusion rules and block characters}


Given a solution to the Cardy equation, that is a set of partition
functions
\be
Z_{\a|\b}(q)=\sum_{i\in\calI} \n_{i\a}{}^{\b}\, \chi_i(q)\label{cylPFs}
\ee
and the corresponding graphs, we assume as in Section 3.2 above
that there exists a special node called 1. We then introduce
the combinations of characters (or ``block characters'')
\be
\chit_\c(q)=\sum_{i\in\calI}\hV_{i\c}\,\chi_i(q)
\ee
where
\be
\hV_{i\c}=\n_{i1}{}^\c
\ee
is referred to as  the basic  intertwiner.
Thanks to (\ref{IIzd}), (\ref{cylPFs}) may be rewritten as
\be
Z_{\a|\b}(q)=\sum_\c \hN_{\c\a}{}^\b \chit_\c(q)\label{FuseRules}
\ee
where the coefficients are now given by
the structure constants of the graph fusion algebra of $G$.

Equation (\ref{FuseRules}) is a mathematical identity and as such
is valid and consistent independent of the choice of the distinguished
node and eigenvectors of $G$. Physically, however, the case where
the $\hN_{\a\b}{}^\c$ are non-negative integers is the most
interesting.  In that case, following Cardy's
discussion~\cite{Ca89}, it is suggested that $\hN_{\a\b}{}^\c$
gives the number of times that the propagating mode or
representation $\c$ appears in the strip or cylinder with
boundary conditions $\a$ and $\b$. Thus if $G$ is a
type I graph, {\it i.e.} if the structure constants $\hN_{\a\b}{}^\c$
are non-negative integers, 
we have a possible interpretation: the nodes $\a$ of the
graph(s) $G$ under consideration label a class of representations
of some extended chiral algebra. The blocks $\chit_a$ are their
characters, and the integer coefficients $\hN_{\a\b}{}^\c$ are
their fusion coefficients.

To probe this interpretation, let us see how it confronts the
results ``in the bulk'', in particular how it is consistent with
the form of the torus partition function.  There, it has been
observed already long ago that (for type I theories)
the torus partition function (cf (\ref{IIabc})) may be recast in the form
\be
Z_{{\rm torus}}= \sum_{\a\in T} |\chit_\a|^2\ , \label{subset}
\ee
i.e. as a diagonal sum over {\it a subset} $T$ of block
characters.  The subset $T$ corresponds to a subalgebra of the
$\hN$ algebra, in the sense that if $\a,\b\in T$,
$\hN_{\a\b}{}^\c\ne 0$ only if $\c\in T$.  This  
interpretation of $n_{i1}{}^c$ as a multiplicity of
representation  $i$ in the block $c$,
that was
first observed empirically~\cite{DFZ2}, was subsequently derived
in a variety of cases of type I $sl(N)$ theories based either
on conformal embeddings or on orbifolding~\cite{PZ2,PZ3}.  More
recently, it 
appeared as an important ingredient in the 
investigation
of the algebraic structure underlying these theories~\cite{Xu,BE,BEK}.

The following interpretation is thus suggested.  The nodes
$\a\in T$ label representations of the maximally
extended algebra $\calA'$ of the RCFT (of type I) under
consideration. The subalgebra of the $\hN$ algebra is the
conventional fusion algebra of this RCFT. The other nodes $\a
\not\in T$ might label other ``twisted'' representations.
The entire $\hN$ algebra would
describe the fusion of all, twisted and untwisted,
representations of $\calA'$. This interpretation in terms of
twisted representations seems corroborated by the fact that some
$\chit_\a$ are known also to occur in partition functions on a
torus in the presence of twisted boundary conditions.
The fact that general boundary conditions on a cylinder also
appeal to these representations was first observed in the Potts
model in~\cite{AfflOS98}. See for some work in this direction~\cite{H},
and the more systematic developments~\cite{FS98} 
along the lines of~\cite{DVVV}.  The concept of twisted representations
in other cases, like conformal embeddings, remains to be understood.
\vskip1cm
Having discussed the situation for type I
theories and graphs, we return to RCFTs
of type II.  There the situation is more elusive.  On the one
hand, as discussed above, the boundary conditions on a cylinder
are labelled by nodes of an improper graph $G$, and although
we can still write an expression of the form (\ref{FuseRules}),
its physical interpretation is unclear.  On the other hand, from
 (\ref{IIabd}), we know~\cite{Ocn} that the torus partition function may
be expressed in terms of block characters pertaining to a
``parent'' type I theory with  graph $G'$
\be
Z_{{\rm torus}}= \sum_{\a\in T} \chit_\a(q)
\left(\chit_{\zeta(\a)}(q)\right)^* ,
\label{parent}
\ee
where $T$ is once again a subset of the nodes of $G'$
corresponding to a subalgebra of the $\hN$ algebra, and $\zeta$
is an automorphism of that subalgebra
$\hN_{\zeta(\a)\zeta(\b)}{}^{\zeta(\c)}= \hN_{\a\b}{}^{\c}$.

For example, the $\slh(2)_{16}$ theory labelled by the Dynkin
diagram $E_7$ is known to be related in that way to the $D_{10}$
theory. Their respective  torus partition function read
\bea
 Z^{(D_{10})}_{{\rm torus}} &=&
|\chi_{1} +\chi_{17}|^2+ |\chi_3+\chi_{15}|^2
+|\chi_{5} +\chi_{13}|^2+|\chi_{7} +\chi_{11}|^2
+2|\chi_{9}|^2 \\
&=& \sum_{\a=1,3,5,7,9,10\atop \a\in D_{10} } |\sum_i
\hV^{(D_{10})}_{i\a }
\chi_i|^2 \label{IIIdsix}\\
 Z^{(E_7)}_{{\rm torus}} &=&
|\chi_{1} +\chi_{17}|^2+|\chi_{5} +\chi_{13}|^2+|\chi_{7}
+\chi_{11}|^2
+|\chi_{9}|^2 +\left((\chi_{3} +\chi_{15}) \chi_{9}^* +
c.c.\right)\\
&=& \sum_{\a=1,3,5,7,9,10\atop \a\in D_{10} }
\left(\sum_i \hV^{(D_{10})}_{i\a} \chi_i \right)
\left(\sum_i \hV^{(D_{10})}_{i\zeta(\a)} \chi_i \right)^* \ ,
\label{IIIesev}
\eea
with $\zeta$ exchanging the two nodes 3 and $10$ of the $D_{10}$
diagram.

It seems that the parent graph $G'$ also plays a role for
cylinder partition functions of type II theories.  Indeed, to obtain cylinder
partition functions expanded with non-negative coefficients in terms of block
characters, we just have to  expand in the block characters of $G'$.
Specifically, we find
$$
Z^{(G)}_{b|a}(q)=\sum_{c\in G'} n^{(GG')}_{ca}{}^b \hat{\chi}^{(G')}_c(q)
$$
where the $G$-$G'$ intertwiners are given by
$$
n^{(GG')}_{ca}{}^b=\sum_{m\in \Exp(G)}
{\psi^{(G')}_c{}^m\over \psi^{(G')}_1{}^m}
\psi^{(G)}_a{}^m {\psi^{(G)}_b{}^m}^*
$$
These turn out to be non-negative integers $n^{(GG')}_{ca}{}^b\ge 0$ and
satisfy the
$G'$ graph algebra. Here it is
assumed that the distinct exponents of $\Exp(G)$ are in $\Exp(G')$ and that
the sum is
over exponents of $G$ counting multiplicities. Moreover, if there is more
than one
eigenvector of $G'$ corresponding to $m\in\Exp(G)$, then any of these
eigenvectors can
be matched with the given $m\in\Exp(G)$. The formula can be derived in the
same way as
our previous formulas. In particular, this formula applies for
$G=E_7$ and
$G'=D_{10}$. In this case there is an ambiguity as to which $D_{10}$
eigenvector
is taken for $m=9$ but in fact one can take either. The matrices
are changed by the $Z_2$ symmetry but the cylinder partition
functions agree.  The formula also holds for the type II
$sl(3)$ theories $G={\calE}_2^{(12)}$, ${\calE}_4^{(12)}$ and 
${\calE}_5^{(12)}$ and $G'={\calD}^{(12)}$.

\bigskip
%
\bookfig{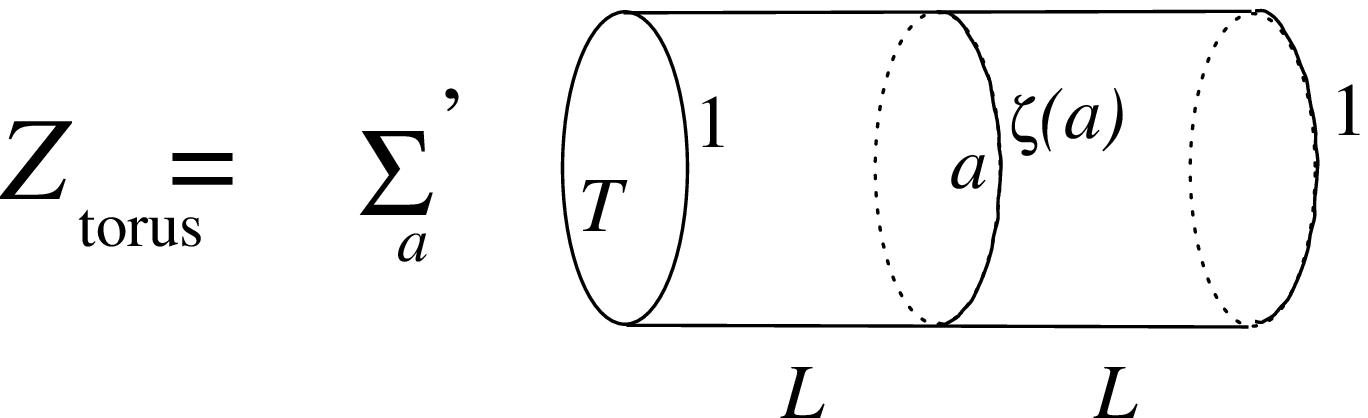}{from cylinder to torus}
{cylinder}
{The torus partition function reconstructed from two cylinder partition
functions.}
{3}
Putting everything together,
we finally observe that in general for a ``rectangular'' torus
with two periods $2L $ and $iT$, made by pasting
together two cylinders,    (see Figure~\ref{cylinder}),
\be
Z_{{\rm torus}}^{(G)}
=\sum_{\a\in T} Z_{\a|1}^{(G')}
Z_{\zeta(\a)|1}^{(G')}
\ ,
\ee
i.e. the partition function may be obtained as the sum over a
special set of boundary conditions of cylinder partition functions. This
expression is of course deeply rooted in all the connections
between bulk and boundary theories, open and closed strings, etc,
but still we find its simplicity intriguing.


\subsection{Examples}

More examples can be given to the previous general scheme.
\begin{itemize}
\item $\slh(N)$ : the classification of the representations
of the fusion algebra of  $\slh(N)_k$  is a well posed
problem on which we have only partial results. 
In particular, classes of graphs pertaining to $\slh(3)$
as well as some cases for higher $N$ have been
expounded from various standpoints in~\cite{DFZ1,DFZ2,PZ1,PZ2,PZ3,Ztani},
(see Appendix~D). In all known cases, the previous
discussion may be repeated: intertwiners, type I graphs,
and other concepts introduced above,  still apply.
  We refer the reader to the above references.
\item The case of $\slh(N)_1$ may be described in detail.
The representations of $\slh(N)_1$
are labelled by an integer $0\le i\le N-1$ (we depart here from our
previous convention, with $i=0$ denoting the identity). The
fusion rules are isomorphic to the addition of integers modulo
$N$, and the algebra defined by
$N_i N_j= N_{i+j\, \mod N}$ is thus generated by $N_1$, $N_j=(N_1)^j$.
The eigenvalues of $N_j$ are $\exp 2i\pi jl/N$.
The regular representation is provided by $N\times N$ matrices,
generated by $(N_1)_i{}^j= \delta_{j, i+1\,\mod N}$.
All the previous eigenvalues are reached once and we may thus say that
the system has ``exponents'' $l=0, \cdots , N-1$ ($\mod N$).
In general, a representation $\{\n_i\}$ of the fusion algebra
is associated with each divisor $q$ of $N=p.q$,
including $q=1$ and $q=N$: $q$ denotes the order of the matrix $\n_1$
which is $q\times q$ dimensional and such that $(\n_1)_\a{}^\b
=\delta_{\b-\a, 1\,\mod q}$,
for a labelling of the nodes $a,b=1, \cdots ,q$.
(This exhausts all integer-valued representations of the algebra.
Indeed the conditions that $\n_1^T= \n_{N-1}$ and $\n_1 \n_{N-1}={I}$
imply that the only entries of $\n_1$ are $0$ and $1$, and that $\n_1$
is a permutation matrix. Being of order $q$ and indecomposable,
$\n_1$ is a matrix of a cycle of length $q$. Q.E.D.)
Obviously the matrices $\n_i=(\n_1)^i$ are all integer-valued,
$\n_0={I}$, and $\n_i^T=\n_{i^*} =\n_{N-i}$.
The graph of adjacency matrix $\n_1$ is an oriented $q$-gon.
In that case, we may say that the $q$ exponents are
$0,p, \cdots, p(q-1)$.

This census of representations of the $\slh(N)_1$ fusion
algebra matches almost perfectly that of modular invariant partition
functions carried out by Itzykson~\cite{Itz} and Degiovanni~\cite{Degio}.
We recall that according to these authors,
a different modular invariant is associated with each divisor of $n$,
where $n= N$ if $N$ is odd and  $n=N/2$ if it is even. Thus, only the case
$N$ even, $q=1$ has to be discarded in our list of representations
of the fusion algebra, as it does not correspond to a modular invariant.
\end{itemize}


\subsection{More on graph algebras}
%

In Section~3.2, we have introduced the matrices $\hN$ by
(\ref{graphfusmats}) and
derived (\ref{IIzc}).
Instead of looking at the graph as a collection of points we can look at it
as a
collection of matrices $\hN$, providing a basis of a
commutative, associative algebra with identity,
and an action of the intertwiners $\n_i$ given by (\ref{IIzc}), that is, we
take
(\ref{IIzc}) as a starting point.
Given the graph $G$, in particular the coefficients $\n_{i 1}{}^c$,
it is possible in many cases to invert (\ref{IIzd}) and solve for $\hN_a$
as linear combinations with integral coefficients of the intertwiners $\n_i$,
or equivalently, as polynomials of the fundamental adjacency matrices.
Similarly the relation (\ref{IIzc}) written in terms of the eigenvalues
$\gamma_j(i)={S_{ji}\over S_{1i}}$,
$\hat{\gamma}_a(i)={\psi_a^i\over \psi_1^i}$
\be
\gamma_j(i)\ \hat{\gamma}_a(i)=\sum_b\ \n_{j a}{}^b\ \hat{\gamma}_b(i)\,,
 \quad i\in
\calE\,, \, j\in \calI\,, \label{V-N}
\ee
is a recursive relation determining (the rows of) the eigenvector matrix
$\psi_a^i$. In general, typically in the presence of degenerate
eigenvalues, the matrix $\psi_a^i$ is not determined uniquely, or
alternatively, (\ref{IIzd}) cannot be inverted for all $a\in \calV$.
For the type I cases, however, as explained above,
  there exists an extended fusion algebra  isomorphic to
a subalgebra of the graph algebra,  so that the extended fusion
matrices $N_{B_i}^{ext}$  can be identified  with a subset $\hN_a$
with the nodes $a\in T\subset \calV$~\cite{DFZ2}.
In most of these type I cases one can solve for all $\hN_a$ in terms of
the $\n_i$'s and  $N_{B_i}^{ext}$, or alternatively express
$\psi_a^j$ in terms
of the modular matrices $S_{ij}\,, S_{B_s B_l}^{ext}$. A particularly simple
subclass of Type I for which one can go quite far in the programme of
reconstructing the graph G and all the related structures is presented by the
orbifold theories, in particular the ones associated with groups generated by
simple currents.  In our approach they can be described by graphs obtained
by `orbifolding' the fundamental graphs of the initial diagonal theory,
the simplest example being provided by the WZW $sl(2)$ $D_{2l}$ series
obtained
by ``orbifolding" the Dynkin diagram $A_{4l-3}$ over the ${\Bbb Z}_2$
group generated by the automorphism $\gamma$. In these
cases as well as in 
their  $sl(N)$ generalisations defined in \cite{Kos} involving the group
${\Bbb Z}_N$, one can algorithmically  construct the eigenvector matrix, see
Appendix~B for an illustration in the simplest $N=2$ case. In a
different approach, 
using tools similar to 
the original orbifold treatment of~\cite{DVVV},
an elegant general formula for the eigenvector matrix was derived recently in
\cite{FS98}. It should be noted that the same graph (orbi)folding procedure
leads also to type II graphs, e.g., the $sl(2)$ $D_{odd}$ series, or
their $sl(3)$ generalisations for $k\not = 0$ mod $3$, see Appendix~D.

We have assumed up to now in this discussion that the graphs are already
known.
On the other hand the relations (\ref{IIzc}), (\ref{IIif})
 can be taken as the
starting point for finding new graphs, typically ``exceptional" graphs not
covered by the previous  orbifold constructions.
Since any graph in the vicinity of the identity resembles
the original ``diagonal" ($A$) graph, one can first
 try to identify $\n_i$'s for which the r.h.s. of (\ref{IIzd}) reduces to
one term, i.e., $\n_{i 1}{}^a =\delta_{a a_i}$ and hence one can identify
 $\n_i=\hN_{a_i}$. According to (\ref{V-N}) this also determines
$\psi_{a_i}^j$ by $\gamma_i(j)$ once $\psi_{1}^j$ is known. This is a problem
which is reduced to the computation of some Verlinde fusion multiplicities.
Indeed let us take the first matrix element $a=b=1$ of the matrix relation
(\ref{IIif}) we have
\be
\sum_c \, \n_{i1}{}^c  \n_{j1}{}^c = \sum_{l}\,
N_{l i}{}^j \, \n_{l 1}{}^1=  \sum_{l\in \rho}\,
N_{l i}{}^j\,\label{irred}
\ee
where in the last sum $\rho=\{l\in \calI | n_{l 1}{}^1\ne 0
 \}$ and $l$ is counted $n_{l 1}{}^1$ times.
Let us assume first that in (\ref{irred}) $i=j$.
Whenever the r.h.s. of (\ref{irred})
is equal to $1$, since
by definition $\n_{i1}{}^a$ are integers, the l.h.s. summation
reduces to one 
term, i.e., we recover  $\n_{i 1}{}^a =\delta_{a a_i}$. Furthermore
plugging this into the l.h.s. of (\ref{irred})
taken for $j\not = i$ we recover $n_{j 1}{}^{a_i}$ as being given
by the sum of Verlinde fusion multiplicities in the  r.h.s.
 of (\ref{irred}), i.e., we determine the multiplicity with which
$N_{a_i}$ appears in $n_j$, see (\ref{IIzd}).  Similarly, a value $2$ or
$3$ for the
r.h.s. of (\ref{irred}) with $i=j$ would lead to $2$, respectively $3$
terms in
(\ref{IIzd}), while $4$ could be
interpreted either as leading to $4$ terms with multiplicity one, or $1$ term
with multiplicity two, i.e., $\n_{i 1}{}^a =2 \delta_{a a_i}$.
What we only need
in order to check all these possibilities is to know the content of
the set $\rho,$
 i.e., $\n_{i 1}{}^1$. This data is provided in type I theories for all of
which
$\rho$ encodes the content of the identity representation of the extended
algebra. More generally, $\n_{i 1}{}^a =\mbox{mult}_{B_a}(i)$, identifying
$a$ with a representation
$B_a$ of the extended algebra. The relation (\ref{irred}) and
its consequences just described are the first steps
in a  consistent algorithmic procedure proposed by Xu~\cite{Xu} 
in the   abstract framework of subfactors theory
(see also \cite{BE,BEK}  for further developments).
In particular, the subset
of $n_i$ which can be identified with some $\hN_{a_i}$ are related to
``irreducible" sectors 
 with the sum in the l.h.s. of (\ref{irred}) interpreted as a
scalar product $(a_i, a_j)$.  The algorithm reduces
systematically the determination of $n_i\,, \hN_a$ in type I
cases to data provided by the Verlinde fusion matrices $N$ and
$N^{ext}$. All graphs previously found in \cite{DFZ1,DFZ2,PZ2}
were recovered in ~\cite{Xu} by this method and
a new  example corresponding to the $\widehat{sl}(4)_6$ modular
invariant was found in \cite{PZ3}.

Finally let us point out that to some extent this algorithm for finding
new solutions  of the equations (\ref{IIzc}) and (\ref{IIif}), i.e., new
graphs,
 can be applied to type II cases where
we do not know a priori $\n_{i 1}{}^1$, i.e., the set $\rho$.
One can start with some trial set and
compute $\sum_\alpha (\psi_1^{(j,\alpha)})^2=S_{1j} \sum_{l\in\rho}
 S_{l j}$. 
A first consistency check is that  $\sum_{j\in \calE} (\psi_1^{j})^2=1$.
 Then one can proceed as in type I. For example
the $E_7$ Dynkin diagram may be reconstructed using $\rho=\{1,9,17\}$
\footnote{Note added in proof: this was independently discussed
in the recent paper~\cite{BEK2}, 
see also ~\cite{Ocn}. The paper ~\cite{BEK2}
 provides a systematic approach in the framework
of the subfactors theory to both types of modular invariants.}.
Different (consistent) choices of the set $\rho$ might lead to
the same graph, reflecting the possibility of different choices
of the identity node.

In some simple cases it is possible to recover a complete set
of boundary conditions by applying formula (\ref{IIzc}) 
to a known subset in such a way that only one term appears
in the sum in the r.h.s. In terms of the equivalent formula 
(\ref{V-N}) for
the eigenvalues we obtain a new solution $\psi_b^i\,, b=b(a,j)\,, $
by ``fusing'' a given one $\psi_a^i\,$
 with the Verlinde eigenvalue $\gamma_j(i)$. This
seems to be the idea of the  so called ``Affleck fusion
conjecture'' \cite{Affl}, 
which clearly has a restricted application, with the general formulae
(\ref{IIzc}),  (\ref{V-N}) being the correct substitute for it.

Another approach to constructing Type I graphs was discussed in \cite{PZ2}
and used to find new solutions for higher rank cases. It is based
on the use of a  relation for the structure constants of the
Pasquier algebra, the  dual of the graph algebra, with structure constants
labelled by elements
in the set Exp(G)
and given by a  formula analogous to   (\ref{graphfusmats}),
however with the summation 
running over the nodes of the graph; this algebra will be discussed
further in section 4.4.1 below.


\sect{Bulk and Boundary Operator Algebras}

In this section we investigate the algebras of fields in the
presence of boundaries and the equations for their structure constants
resulting from duality constraints.
Our discussion parallels that of Cardy and Lewellen~\cite{CL},
but generalises it in two respects: to  higher rank and to
non-diagonal theories.  This results in additional multiplicities
associated with the more general representations of the Verlinde
algebra (\ref{IIif}).  Our presentation makes use of concepts
used by Moore and Seiberg for bulk RCFTs   and extends them
appropriately for this new setting. This leads to a richer
structure in the equations and the appearance   of a triplet of
algebras $(n_i, \hN_a, M_j)$. Separately these algebras have
appeared before but the inter-relation between these algebras has
not been shown in this context.

\subsection{Ground state degeneracies }

As stressed by Affleck and Ludwig~\cite{AL}, the logarithm
of the partition function, in the limit $L/T\to \infty$,
contains not only the universal term proportional to $L$ and to the central
charge (in unitary theories), but also an $L$ independent term,
interpreted as a
boundary condition dependent ``ground state degeneracy'' $\ln g_\a g_\b$.
Indeed in that limit
\be
\ln Z_{\b|\a} \sim {c\over 24}\, 4\pi\, {L\over T} + \ln \psi_\a^1
+\ln \psi_\b^1 -\ln S_{11} \label{IVia}
\ee
where as before we denote by 1 the representation of conformal weight 0,
corresponding to the identity operator.
We therefore identify
\be
g_\a ={\psi_\a^1 \over\sqrt{S_{11}}}\ . \label{IVib}
\ee

Thus in unitary minimal models, using (\ref{jbsduty}) and (\ref{Minb}),
we have the following expression for the boundary states
\be
|(r,a)\ket=\sum_{r',s'\atop r' {\rm odd}, s'\in \Exp(G)}
 2^{{1\over 4}} { S^{(h)}_{r r'}\,\psii{a}{s'}\over \sqrt{
S^{(h)}_{1 r'} S^{(g)}_{1 s'}}}
\, | r',s' \rrangle
\ee
and their $g$ factor
\be
g_{(r,a)}={\llangle 1,1| (r,a)\rangle\over \sqrt{
S^{(h)}_{1 1} S^{(g)}_{1 1}}}
= 2^{{1\over 4}}
{ S^{(h)}_{r 1}\,\psii{a}{1}\over \sqrt{
S^{(h)}_{1 1} S^{(g)}_{1 1}}}\ ,\label{IVga}
\ee
in terms of the modular matrices $S^{(h)}$ and $S^{(g)}$ of
$\slh(2)$ at levels $h-2$ and $g-2$, $|g-h|=1$.
For example, for the critical 3-state Potts model, we obtain 
$$ g_A=({5-\sqrt{5}\over 30})^{1\over 4}=0.550936
\qquad (g_A:g_{AB}:g_{ABC}:g_N)=(1:{1+\sqrt{5}\over 2}:
\sqrt{3}:{1+\sqrt{5}\over 2}\sqrt{3}) $$
in agreement with~\cite{AfflOS98}.
As a particular case of (\ref{IVga}), the ratio
$g_{(r,a)}/g_{(1,a)}$ equals $S^{(h)}_{r1}/S^{(h)}_{11}$, in agreement with
(A.3) of \cite{AfflOS98} and the fact that one obtains the boundary state
$|(r,a)\rangle $ by fusion (in the sense of Cardy) of boundary states
$(1,a)$ and $(r,1)$.

In non-unitary cases, these expressions have to be slightly amended.
If $j_0$ denotes the representation of smallest conformal weight $h_{j_0}<0$
and assumed to belong to $\calE$, then
\be
\ln Z_{\b|\a} \sim {c_{{\rm eff}}\over 24}\, 4\pi\, {L\over T}
+ \ln \psi_\a^{j_0}
+\ln \psi_\b^{j_0} -\ln S_{1 j_0} \label{IViaa}
\ee
with $c_{{\rm eff}}:= c-24 h_{j_0}$. Also
\be
g_\a ={\psi_\a^1 \over\sqrt{S_{ 1 j_0}}}\ . \label{IViba}
\ee

For simplicity of notation
in most of what follows we shall restrict to unitary theories.
Denoting $g_a=\bra\un \ket_a$ from now on,
\be
\lim_{L/T\to \infty}\, Z_{\b|\a}\,e^{-{\pi c\over 6}\,
\, {L\over T}  }/g_b= \bra\un \ket_a\,.
\ee
One can consider furthermore the partition function with some
field insertions at the same limit \cite{CL}, \cite{RS1},
\cite{RS2};  we shall normalise them similarly so that only a
dependence on $g_a$ is retained, i.e., $\bra\un \ket_a\,$ will
coincide with the  1-point function of the identity operator in
this limit.

\bigskip

\subsection{Bulk and boundary fields, OPE}

\subsubsection{Boundary fields}

According to Cardy \cite{Ca89},  boundary conditions
can be interpreted as created
by the insertion of fields ${}^b\Psi^a_{j,\zb}(x)\,$  living on
the boundary, Im $z=0$, $x=$Re $z$ of the upper half-plane $z\in  H_+$.
Here $j\in \calI\,,$ $a,b\in \calV\,,$
  and $\beta $ accounts for the  multiplicity
$\n_{j a}{}^b$ of such fields, to be called ``of type''
 $ \Big({b\,\phantom{c}\atop j\,a}\Big) \,.$
Thus $\beta$ can be interpreted as a ``coupling'' index
$\zb=1,2\dots, \n_{j a}{}^b\,$ and the boundary fields as
 kind of   ``chiral vertex operators'' (CVO) associated
with a second type of couplings $ \Big({b\,\phantom{c}\atop
j\,a}\Big)_{\zb}\,,$ $a,b\in \calV\,, j\in \calI$.
This is  a formal analogy since
 the boundary  states $|a\ket, |b\ket$ labelled by  $a,b$
are  superpositions of
Ishibashi states.  The multiplicity  index $\zb$ is traditionally omitted,
but it should be stressed that even
in the $sl(2)$ case, in all but the diagonal cases ($\calE =
\calI$), there are always some non-trivial multiplicities
$\n_{ja}{}^b > 1$, so most of the time we shall retain this index. Since
$\n_{1 a}{}^b=\delta_{ab}$ the index  associated with the coupling $
\Big({b\,\phantom{c}\atop 1\,b}\Big) \,$ takes only one
value and  will be denoted $\zb=\un_b$, or just $\un$,
or,  altogether omitted. On the other hand
in all non-diagonal cases there is a non-trivial subset $\{
{}^{1}\Psi^{1}_{j}\,, j\in \rho\}$ of boundary fields, with
$1\in \rho\subset \calI\,$, 
where the set $\rho$ has been introduced in Section 3.5.

To make contact with Section 2.2,
consider the finite strip $w={L\over\pi}$ log $z $
equipped with the  Hamiltonian $ H_{ab}= {\pi \over L} \, (L^{(H)}_0
-{c\over 24})$.
The space of states is generated  by all the descendent  states
 created  for
fixed $a,b$ from the (properly normalised) vacuum state
${}^b\Psi^a_{j,\beta}(0)|0\rangle \  $
by the modes of the  Virasoro algebra
generating the  real analytic  conformal transformations.
This includes besides the sum over $j\in \calI$
a summation over the multiplicity $\n_{j a}{}^b$
of these states for fixed $a,b,j$, i.e., we can think of
the Vir representation spaces $\calV_{j\,,\beta}$
as being labelled by pairs $(j,\beta)$.
The ``dual vacuum state'' is defined by a boundary field
placed at infinity
$\sum_{\zb'}\, \lim_{x\to\infty}\,c_{a,b,j;\zb,\zb'}\,
x^{2 \triangle_j } \, \bra 0| {}^a\Psi^b_{j^*,\zb'}(x)$ where
$c_{a,b,j;\zb,\zb'}$ is a normalisation constant
and $\zb'=1,\cdots, \n_{j^* b}{}^{a}$ is an index of type
$ \Big({a\,\phantom{c}\atop j^*\,b}\Big)\,$.
Accordingly the trace of the operator $e^{-T \,H_{ab}}$
computed  imposing the periodicity
$w\sim w+T$ in time direction can be written as  a sum over
$(j,\beta)$ of characters $\chi_{j,\beta}\,,$ with the summation
over $\zb$ leading to (\ref{IIg}), with $a$ and $b$ exchanged.

Boundary fields ${}^{b}\Psi^{a}_{j, \zb}(x)$ appear as  ordinary fields
$\Psi_{j}(x)$ decorated by a pair of indices
$a,b$ according to some rules.  In the limit $L/T\to \infty$
their $1$-, $2$- and $3$-point functions  are given by
the  corresponding invariants of
$\Psi_{j}$ with respect to $sl(2,\Bbb{R})$
with  normalisation coefficients depending on $a,b$.
As for the ordinary fields the $1$-point function is non-zero
only for the identity operator
\be
\bra 0| {}^{b}\Psi_{j, \za}^a(x)|0\ket=\delta_{j 1}\,
\delta_{ba} \, \delta_{\za \un_a}\, \bra \un \ket_a
\ee
with the restriction on $a,b$ coming from $\n_{1 a}{}^b=\zd_{ab}$.
As for the ordinary CVO the product $
{}^{b}\Psi_{i, \za_1}^c\, {}^{d}\Psi_{j, \za_2}^a\, $
of two boundary fields is defined only for coinciding $c=d$. Similarly
the initial and the final indices in a vacuum expectation value
of a product of boundary fields, are restricted to coincide
(due to the periodicity in the strip time direction, the boundary
half-line being effectively closed)
but in distinction with the ordinary CVO they can be arbitrary
and not just equal to the identity $1$. The $2$- and
$3$-point functions read
\be
\bra 0|{}^{a}\Psi_{j, \za_1}^b(x_1)\, {}^{b}\Psi_{i,
\za_2}^c(x_2)|0\ket
=\delta_{j i^*}\, \zd_{a c}\,{C_{ i^*i\,; \za_1\, \za_2}^{ab}
\over |x_{12}|^{2\triangle_j}}\ \,,  \quad
x_1\not =x_2\,, \label{twoinv}
\ee

\be
\bra 0|{}^{a}\Psi_{i, \za_1}^b(x_1)\, {}^{b}\Psi_{j,
\za_2}^c(x_2)\, {}^{c}\Psi_{k, \za_3}^d(x_3)|0\ket_t = \delta_{ad}
{C_{ijk\,;\za_1\, \za_2\, \za_3\,;t}^{abc}\over
|x_{12}|^{\triangle_{ij}^k}\,
|x_{23}|^{\triangle_{jk}^i}\, |x_{31}|^{\triangle_{ki}^j} }\,,
\quad x_1\not =x_2\not =x_3\not =x_1\,, \label{threeinv}
\ee
where $\Delta_{ij}^k= \Delta_i+\Delta_j-\Delta_k$ and $x_{ij}=x_i-x_j$.

The functions (\ref{twoinv}), (\ref{threeinv}) are invariant with
respect to $SL(2,\Bbb{R})$, with  representations denoted by a pair
$(\delta, \ze=\pm)$, see  \cite{Gelf}; here we choose $\ze=1$
corresponding to taking the modulus of the multiplier of the
$SL(2,\Bbb{R})$  transformations
and the expressions in (\ref{twoinv}), (\ref{threeinv})
imply trivial monodromy of the boundary field correlators.
In the $\widehat{sl}(n)_k$ WZW models the fields carry an
additional tensor index, or, in a functional realisation, depend on
an additional (multi)variable $X$ accounting for the
representations of the ``isospin'' $sl(n,\Bbb{R})$ algebra and the
$n$-point functions  involve also $n$--point invariants with
respect to this algebra.
For example,  in the $\widehat{sl}(2)_k$ WZW case the fields
$\Psi_j(x,X)$ can be described in terms of a pair of real
variables~\cite{FZ},  the coefficients in the 
polynomial expansion with respect to $X$ representing
the horizontal algebra descendants. 
 In this case the isospin  labels are
$2j=0,1,\dots , k\,,$ and the $2$- and $3$-point invariant
correlators contain additional factors $X_{il}^{2j}$ along
with any $x_{il}^{-2\triangle_j}$.
For simplicity  we adapt the notation
for the minimal $W_n$ models rather than
their WZW counterparts, omitting the explicit indication
of the ``isospin''  variables and the corresponding invariants.
To keep track of the various possible three-point invariants, we
shall retain the multiplicity index $t$ as in (\ref{threeinv}).

Up to the normalisation constant and up to phases,
(\ref{threeinv}) is the 3-point invariant function 
$\bra 0|\phi_{i\,,\un_i'}\,\phi_{j\,, t}
\,\phi_{k\,,\un_k} |0\ket $   of the ordinary CVO
$\phi_{j\,, t}(x)$.
Here $t$ is a coupling index of type
$ \Big({i^*\,\phantom{c}\atop j\,k}\Big)\,,$ $t=1,2,\dots,  N_{jk}{}^{i^*}$.
Two kinds of permutations act on these couplings, see~\cite{MS1}:
$\sigma_{23}\, $  $ \Big({p\,\phantom{c}\atop i\,j}\Big)\,
\to \Big({p\,\phantom{c}\atop j\,i}\Big)\,,$ and
$\sigma_{13}: \Big({p\,\phantom{c}\atop i\,j}\Big)\,
\to \Big({j^*\,\phantom{c}\atop i\,p^*}\Big)\,$.
For simplicity in the sequel we denote
the one value  indices indicating couplings with
one label of type $\calI$ set to $1$, like $\sigma_{23}(\un_i)\,,
\un_i'\,, $ or, $\un_a$
(corresponding to couplings of type
 $ \Big({i\,\phantom{c}\atop i\,1}\Big)\,,$
  $ \Big({1\,\phantom{c}\atop i\,i^*}\Big)\,,$
 or,  $ \Big({a\,\phantom{a}\atop 1\,a}\Big)\,$ resp.),
simply by $\un$.

Motivated by the form of the 3-point function, the
operator product expansion (OPE) of (primary) boundary fields ${}^{b}\Psi_{i,
\za}^c(x)$ is defined according to
\bea
{}^{b}\Psi_{i, \za_1}^c(x_1)\, {}^{c}\Psi_{j, \za_2}^a(x_2)\,
= \sum_{p\,,\zb\,,t}\
{}^{(1)}F_{c p}\left[\matrix{i&j\cr b&a} \right]_{\za_1\,
\za_2}^{\zb\   \ t }\  \sum_{P}\,
\bra  p,P| \phi_{i, t}(x_{12})|j,0\ket\ {}^{b}\Psi_{p,
\zb; P}^a(x_2) \label{IVa}\\
=
\sum_{p,\zb, t }\
{}^{(1)}F_{c p}\left[\matrix{
i&j\cr b&a} \right]_{\za_1\, \za_2}^{\zb\  \ t}\
{1\over |x_{12}|^{\triangle_i+\triangle_j-\triangle_p}}\
{}^{b}\Psi_{p, \zb}^a(x_2) + \  \dots \,, \nonumber
\eea
where  $P$ is an index for the descendent states of the
representation $\calV_p$ with $p\in \calI$.
The indices $\za_1\,,\, \za_2\,,\, \zb$ account for the
multiplicity of vertices of type
$ \Big({b\,\phantom{c}\atop i\,c}\Big)\,,$
$ \Big({c\,\phantom{a}\atop j\,a}\Big)\,,$
$ \Big({b\,\phantom{a}\atop p\,a}\Big)\,,$
 respectively, $a,b,c \in \calV\,,$
 i.e., $\alpha_1=1,2\dots, \n_{i c}{}^b\,,$  etc,
while $t$ is that  of a standard vertex
$ \Big({p\,\phantom{c}\atop i\,j}\Big)\,,$
$t=1,2, \dots N_{ij}{}^p$. We will often restrict for simplicity
to the $sl(2)$ case, so that
the index $t$ can be omitted. From the 1-point function
\bea
&&{}^{(1)}F_{c p}\left[\matrix{
i&1\cr b&a} \right]_{\za_1\, \za_2}^{\zb\ \ t}=\zd_{pi}\,
\zd_{ac}\ \zd_{\za_2 \,\un}\, \zd_{t \,\un}\, \zd_{\za_1 \zb} \,,
\nonumber \\
&&{}^{(1)}F_{c p}\left[\matrix{
1 &j\cr b&a} \right]_{\za_1\, \za_2}^{\zb\ \ t}=\zd_{p\,j}\,
\zd_{b\,c}\ \zd_{\za_1\, \un}\, \zd_{t\, \un}\, \zd_{\za_2\, \zb}
 \,. \label{ident}
\eea

 With the normalisation of the CVO $ \Big({p\,\phantom{c}\atop
i\,j}\Big)_t\,$ implied by the second equality in (\ref{IVa}) the
numerical coefficients ${}^{(1)}F_{c p}\left[\matrix{ i&j\cr b&a}
\right]_{\za_1\, \za_2}^{\zb\  \ t}$ with $i,j,p\in \calI$,
$a,b,c \in \calV$,  represent the OPE coefficients of the
boundary fields and their determination is part of the problem.
They are reminiscent of the matrix elements of the fusing (or
crossing) matrices $F$ (whence the notation here for this
``second'' fusing matrix), which serve as OPE  coefficients of
the usual CVO \cite{MS1}.  The definition (\ref{IVa}) extends to
descendent fields in the l.h.s. as for the usual CVO.
Symbolically (\ref{IVa}) can be written as
\bea
 \Big({b\,\phantom{c}\atop i\, c}\Big)_{\za_1\,, \,x_1}\
\Big({c\,\phantom{c}\atop j\, a}\Big)_{\za_2\,, \,x_2}
= \sum_{p\,, \zb\,,  t}\
{}^{(1)}F_{c p}\left[\matrix{
i&j\cr b&a} \right]_{\za_1\, \za_2}^{\zb\  \ t}\
 \Big({p\,\phantom{c}\atop i\,j}\Big)_{t\,, \,x_{12}}\
\Big({b\,\phantom{c}\atop p\, a}\Big)_{\zb\,, \,x_2} \label{moorese}
\eea
and can be depicted similarly as  the standard Moore-Seiberg diagrams,
see Figure~\ref{moores}. Denoting by $\calU_{p a}^{b}$ the space of
boundary fields of type
$\Big({b\,\phantom{a}\atop p\, b}\Big)_{\za_1 }$
we have dim $\calU_{p a}^{b}=\n_{p a}{}^{b}$ while the space of standard
CVO, $U_{i j}^{p}$, has dimension given by the Verlinde fusion
multiplicity $N_{i j}{}^{p}$, dim $U_{i j }^{p}=N_{i j}{}^{p}$.
Thus we can interpret  ${}^{(1)}F$ as a linear operator
\be
\oplus_{c}\, \calU_{i c}^{b}\, \otimes\calU_{j a}^{c} \rightarrow
\oplus_{p}\, U_{i j}^{p} \, \otimes\calU_{p a}^{b}\,,
\ee
 the dimension of the two sides being identical,
according to  (\ref{IIif}).
Given the 1- and 2-point correlators above the computation
of the general boundary field $n$-point functions is reduced to the
computation of the conformal blocks of the standard CVO $
\Big({p\,\phantom{c}\atop i\,j}\Big)_{t\,, \, x}\,.$
%
%
\bookfigh{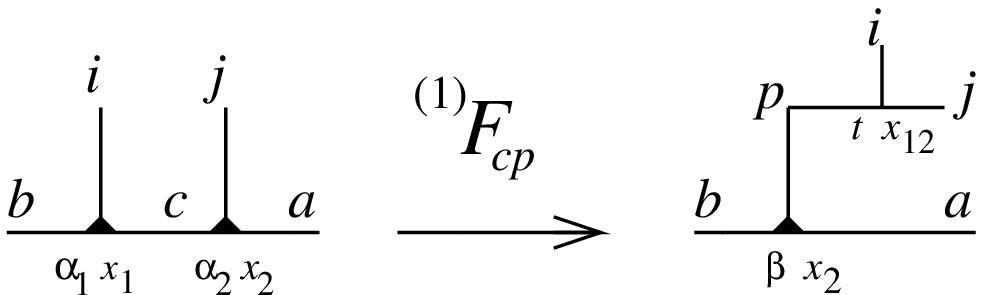}{Graphical representation of \ref{moorese}}
{moores}{ Graphical representation of (\ref{moorese}).  To stress the
presence of two types of vertices, we distinguish them
explicitly {\it on this particular figure only}
 }{2}

Comparing with the $2$-- and $3$--point functions, see
Figure~\ref{twthrpt}(a) and
(b), we have
\be
C_{ i i^*\,; \za_1\, \za_2}^{ab}=
{}^{(1)}F_{b 1}\left[\matrix{
i&i^*\cr a&a} \right]_{\za_1\, \za_2}^{\un_a \un_i' }\
\bra \un \ket_a \,,
\ee

\bea
C_{ijk\,;\za_1\, \za_2\, \za_3\,;t}^{abc}=\sum_{\zb}\,
{}^{(1)}F_{bk^*}\left[\matrix{ i&j\cr a&c} \right]_{\za_1\,
\za_2}^{\zb\, \sigma_{23}\sigma_{13}(t)} \,
{}^{(1)}F_{c 1}\left[\matrix{ k^*&k\cr a&a} \right]_{\zb\,
\za_3}^{\un_a \un_{k^*}' } \, \bra \un \ket_a
\nonumber \\ =
\sum_{\zg}\,
{}^{(1)}F_{ c i^*}\left[\matrix{ j&k\cr b&a } \right]_{\za_2\
\za_3}^{\zg \,  t}\ \,
{}^{(1)}F_{b 1}\left[\matrix{i& i^*\cr a&a} \right]_{\za_1\,
\zg}^{\un_a \un_{i}' }\,  \bra \un \ket_a \,.
\label{tpc}\,
\eea
%
\bookfigh{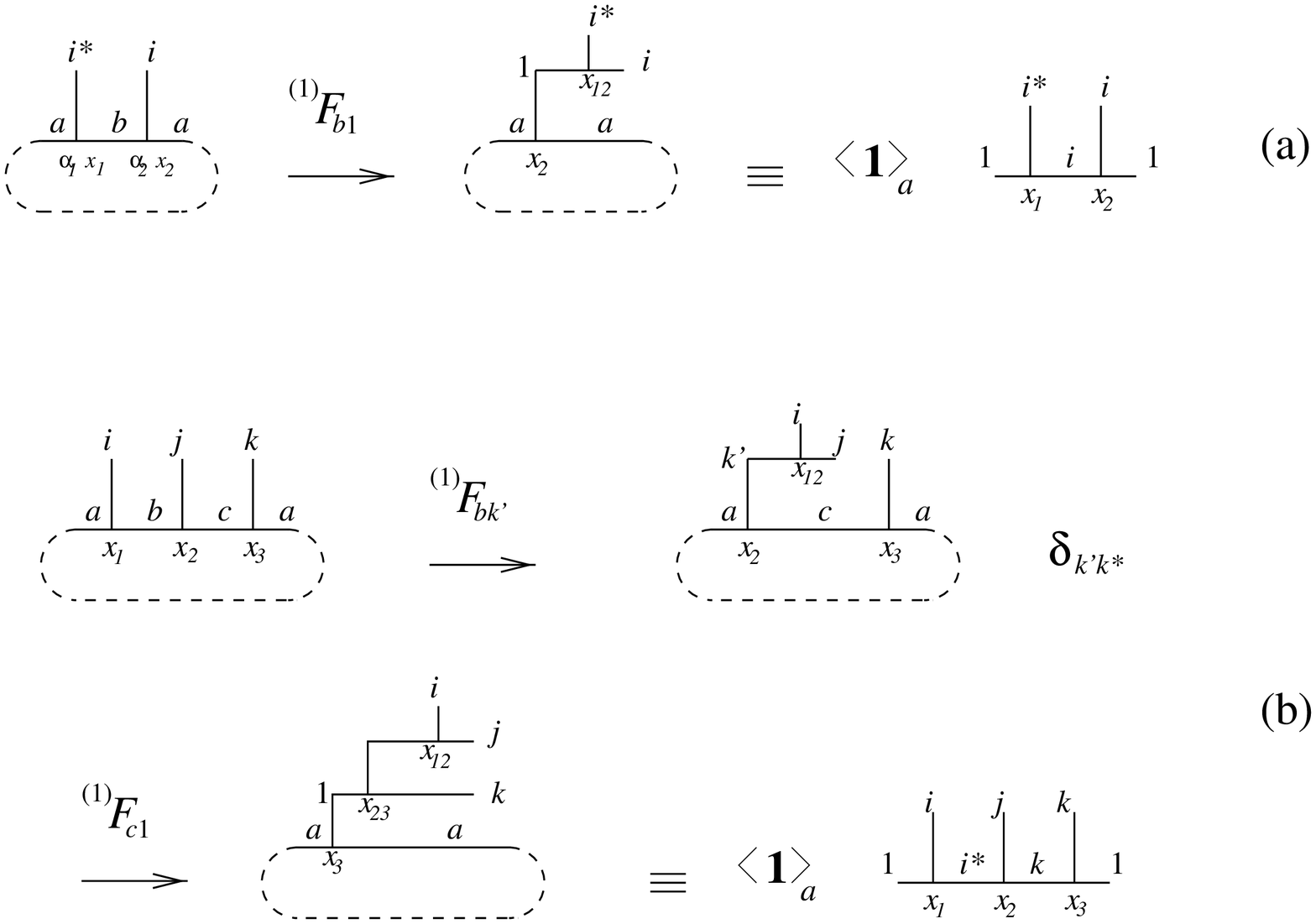}{two and three point function}{twthrpt}{ (a)
and (b):  boundary field 2- and 3-point  functions }{10}


The $2$-- and $3$--point normalisation coefficients are
assumed to satisfy the symmetry conditions
\bea
C_{i^*i \,; \za\, \zb}^{ab}=C_{i i^* \,; \zb\, \za}^{ba}
&&= (C_{i^* i\,; \sigma_{13}(\zb^*)\, \sigma_{13}(\za^*)}^{a b })^*
\,, \nonumber \\
\label{trpc} \\
%
C_{ijk\,;\za_1\, \za_2\, \za_3;t}^{abc}=
C_{jki\,; \za_2\, \za_3\,\za_1; 
\sigma_{13}\sigma_{23}(t)}^{bca}
&&= \Big( C_{k^*j^* i^*\,;\sigma_{13}(\za_3^*)\, \sigma_{13}(\za_2^*)\,
\sigma_{13}(\za_1^*); \sigma_{13}(t^*)}^{ c b a} \Big)^*
\nonumber
\eea
%
The first equalities in (\ref{trpc}) are 
cyclic symmetry relations,  
see Figure~\ref{cyclic}, while the 
 second equalities come from an antilinear  (``CPT'') transformation,
 which in particular sends the field ${}^{b}\Psi_{j, \zb}^c(x)\,$
to its conjugate 
$ {}^{c}\Psi_{j, \zs_{13}(\zb^*)}^b(-x)$
with  multiplicity indices 
consistent with
$\n_{j a}{}^c=\n_{j^* c}{}^{a}=\n_{j^* a^*}{}^{c^*}$ .

%
%
\bookfigh{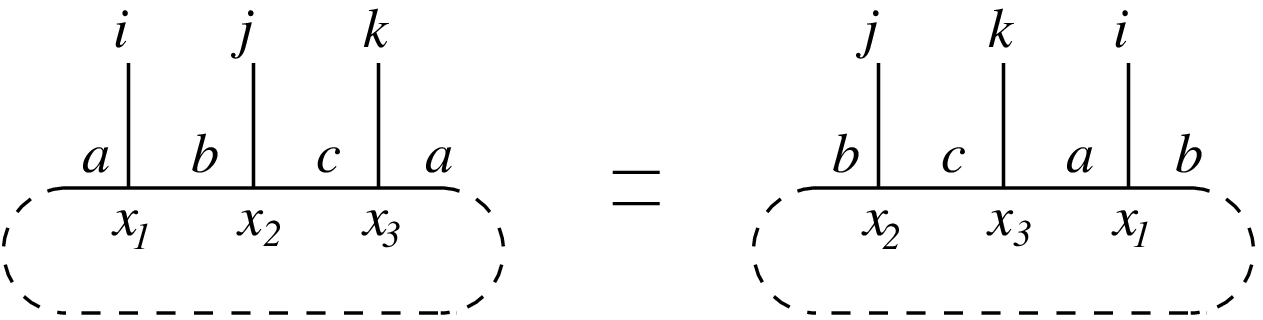}{cyclic symm}{cyclic}{ The cyclic 
symmetry of 3-point functions }{2}
%
%
%
The  cyclic symmetry relations  imply
\be
{}^{(1)}F_{c 1}\left[\matrix{ j&j^*\cr a&a} \right]_{\za_1\,
\za_2}^{\un_a \un_j' }\ \bra \un \ket_a \, =
{}^{(1)}F_{a 1}\left[\matrix{ j^*&j\cr c&c} \right]_{\za_2\,
\za_1}^{\un_c \un_{j^*}' }\ \bra \un \ket_c \,, \label{cycI}
\ee
\bea
\sum_{\zb}\,
{}^{(1)}F_{ak^*}\left[\matrix{ j&s\cr b&c} \right]_{\zd\,
\za}^{\zb \,t}\ \,
{}^{(1)}F_{c 1}\left[\matrix{ k^*&k\cr b&b} \right]_{\zb\,
\zb_2}^{\un_b \un_{k^*}' } \, \bra \un \ket_b
\nonumber \\ =
\sum_{\zg}\,
{}^{(1)}F_{b s^*}\left[\matrix{ k&j\cr c&a } \right]_{\zb_2\
\zd}^{\zg \,\sigma_{23} \sigma_{13}(t)} \,
{}^{(1)}F_{c 1}\left[\matrix{s& s^*\cr a&a} \right]_{\za\,
\zg}^{\un_a \un_{s}' }\, \bra \un \ket_a  \label{cyc}\,,
\eea
while the the second equalities in (\ref{trpc})  lead to
\be
{}^{(1)}F_{ak}\left[\matrix{ j&s\cr b&c} \right]_{\zd\,
\za}^{\zb t}\ \,
= \Bigg({}^{(1)}F_{a k^*}\left[\matrix{ s^*&j^*\cr c&b}
\right]_{\sigma_{13}(\za^*)\, \sigma_{13}(\zd^*)}^{\sigma_{13}(\zb^*)
\sigma_{23}(t^*)}\Bigg)^*
\ .
\ee
 Combining (\ref{cycI}) and  (\ref{cyc}), one recovers (\ref{tpc}).

\medskip


\subsubsection{Bulk fields and bulk-boundary coefficients}

We turn to the second ingredient of the Cardy-Lewellen boundary CFT, the bulk
fields.
The half-plane bulk fields  $\Phi_I
(z,\bar{z})\,,$  $ z=x+i
y\in H_+\,, \bar{z}=x-i y$ transform under a representation of
$L^{(H)}$ \cite{Ca84} realised by differential operators
\be
L_n^H= L_n(\triangle_i\,,z) +L_n(\triangle_{\bar{i}}\,,\bar{z})
\ee
and characterised by a pair $
I=(i,\bar{i})$ of weights. In cases
when there is more than one field with the same labels
$(i,\bar{i})$ a more involved notation like $I=
(i,\bar{i};
\alpha)$ is needed, but usually  omitted for simplicity. For type I
theories as well as for arbitrary scalar fields both $i,\bar{i}
\in \calE$, while in general $i,\bar{i} \in \calI$.

The invariance with respect to the  subalgebra spanned by
$L_{\pm1, 0}^{(H)}$ determines the $1$--point function
of $\Phi_{(i, \bar{i})}(z\,,\bar{z})$ as well as the  $2$--point function
$\bra  {}^{a}\Psi_{p, \za}^a \, \Phi_{(i, \bar{i})}\ket\,,$ e.g.,
\bea
&&\bra 0|{}^{a}\Psi_{p, \za}^a(x_1)\, \Phi_{(i, \bar{i})}
 \,(z, \bar{z})|0 \ket
\label{mixtp}\\\nonumber
&&={C_{p, (i, \bar{i}), \za\,, t}^a \over (z-\bar{z})^{\triangle_i+
\triangle_{\bar{i}}-\triangle_p}\,
(x_1-z)^{\triangle_i+\triangle_p-\triangle_{\bar{i}} }\,
 (x_1-\bar{z})^{\triangle_{\bar{i}}+\triangle_p-\triangle_i }
}\,, \quad x_1> {\rm Re} \, z \,,
\eea
while $\bra  \Phi_{(i, \bar{i})}\,  {}^{a}\Psi_{p, \za}^a \ket\,$
is defined for ${\rm Re} \, z > x_1$ by the analogous expression
with  $x_1-z\,,$  $ x_1-\bar{z}$ replaced by
$z -x_1\,,  \bar{z}-x_1$. 
Requiring the symmetry of this function
 under the exchange of the two fields, i.e., the independence of
the ordering,  leads to
the constraint $\triangle_i-\triangle_{\bar{i}}\in
\Bbb{Z}$. 

The r.h.s.  of (\ref{mixtp}) is the 3-point block of the standard CVO
$\bra 0|\phi_{p\,,\un_p'}(x_1)\,\phi_{i\,, t}(z)
\,\phi_{\bar{i}\,,\un_{\bar{i}}}(\bar{z}) |0\ket \,, $ with
$t$ a coupling index of type
$ \Big({p^*\, \atop i\,\bar{i}}\Big)\,,$ $t=1,2,\dots,
 N_{i \bar{i}}{}^{p^*}$.
Consistently with this the (primary) bulk field can be
represented  for small $z-\bar{z}$ via the decomposition
\bea
\Phi_{(i, \bar{i})}(z\,,\bar{z})
&&
=\sum_{a,\za\,, p\in \calI\,, t}\
{}^{a, \za}B_{(i, \bar{i})}^{p, t}\, \sum_{P}\,\bra  p, P|
\phi_{i,t}
({z-\bar{z}})|\bar{i},0 \ket\
{}^{a}\Psi_{p,\za;P}^a(\bar{z})\label{IVb}\\
&&=
\sum_{a,\za\,, p\in \calI\,, t}\
{}^{a, \za}B_{(i, \bar{i})}^{p, t}\,
{1\over
(z-\bar{z})^{\triangle_i+\triangle_{\bar{i}}-\triangle_p}}\,
{}^{a}\Psi_{p, \za}^a(\bar{z}) + .... \nonumber
\\&&=
\sum_{a,\za\,, p\in \calI\,, t}\
{}^{a, \za}B_{(i, \bar{i})}^{p, t}\,
{1\over
(2i y)^{\triangle_i+\triangle_{\bar{i}}-\triangle_p}}\,
{}^{a}\Psi_{p, \za}^a(x) + .... \nonumber
\eea
which extends to descendents.
Here ${}^{a}\Psi_{p, \za}^a(z)$
are ``unphysical'' generalised CVO obtained extending to the
(full) plane the boundary fields of the previous section.
Their OPE is determined by the same  fusing matrix ${}^{(1)}F$,
i.e., as in (\ref{IVa}), the latter extended to complex arguments
$z_i\,,$ ${\rm Re} \, z_{12} > 0$, with $|x_{12}|$ replaced by
$z_{12}$;  we shall need only this fusing property.

The constants ${}^{a, \za}B_{(i, \bar{i})}^{p, t}\, $ (``{\it bulk-boundary
reflection coefficients}'') in this decomposition depend on two
couplings of different types, $\Big({p\,\phantom{c}\atop
i\,\bar{i}}\Big)_t$ and
$\Big({a\,\phantom{c}\atop p\,a}\Big)_{\alpha}$.  Note that
the coefficients used here  differ by a phase from the traditionally
 normalised
coefficients \cite{CL}, \cite{L}, which will be denoted  ${}^{a,
\za}B_{(i, \bar{i})}^{p, t}(CL)$, i.e.,
\be
{}^{a, \za}B_{(i, \bar{i})}^{p, t}=e^{i {\pi\over 2}
 (\triangle_i+\triangle_{\bar{i}}-\triangle_p)}
\ {}^{a, \za}B_{(i, \bar{i})}^{p, t}(CL) \label{CLchoice}\,.
\ee

%
%
\bookfigb{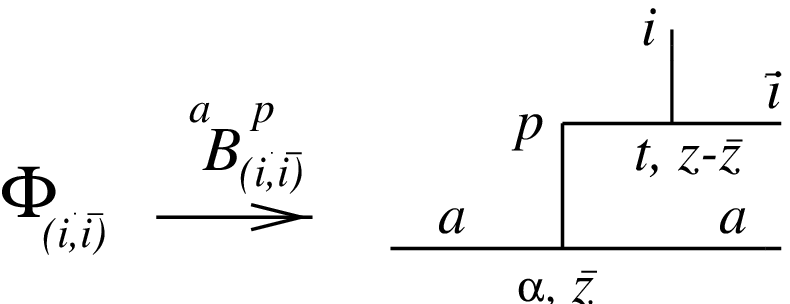}{bulk-bdy}{bbc}{ Graphical representation
of the decomposition (\ref{decompo}) of  bulk fields }{2}
%
%
%
The decomposition  (\ref{IVb}),  symbolically written as
\bea
\Phi_{(i, \bar{i})}(z,\bar{z})=
\sum_{a,\za\,, p\,, t}\
{}^{a, \za}B_{(i, \bar{i})}^{p, t}\,
 \Big({p\,\phantom{c}\atop i\,\bar{i}}\Big)_{t\,, \,z-\bar{z}}\
\Big({a\,\phantom{c}\atop p\, a}\Big)_{\alpha\,, \,\bar{z}}\,,
\label{decompo}
\eea
see Figure~\ref{bbc},
reduces the computation of the $n$--point functions
of $\Phi_{(i, \bar{i})}$   to the computation of the blocks of the
generalised  CVO  ${}^a\Psi_{p,\za}^b(z)$, which combined with
their OPE (the extension from $|x_{12}|$ to $z_{12}$ of
(\ref{IVa}))  allows to recover all correlators in terms
of standard conformal blocks.
The invariant 1-point function projected onto the boundary state
$a$ reads
\be
\bra \Phi_{(i, \bar{i})}(z,\bar{z})\ket_a=  \zd_{i^*\,\bar{i}}\,
{ {}^{a, \un}B_{(i, \bar{i})}^{1, \un}\over
(z-\bar{z})^{2\triangle_i}}\, \bra \un\ket_a \,. \label{opf}
\ee
Omitting the trivial  indices 
 and simplifying the label $(i, i^*)$ to $i$,
one has in particular ${}^{a}B_{1}^{1}=1$ for any $a$.

\vskip5mm

The OPE of the half-plane  bulk fields $\Phi_{(k,\bar{k})}(z,\bar{z})$
is defined according to
\bea
&&\Phi_{(k,\bar{k})}(z_1,\bar z_1)
\, \Phi_{(l,\bar{l})}(z_2, \bar z_2) \nonumber \\
&&=\sum_{j\,, \bar{j}\,,
t\,, \bar{t}}\, D_{(k,\bar{k}) (l,\bar{l})}^{(j,\bar{j});
t,\bar{t}}\,
\sum_{J\,, \bar{J}}\, \bra  j,J| \phi_{k,t}
(z_{12})|l,0 \ket\  \bra  \bar{j},\bar{J}|
\phi_{\bar{k},\bar{t}}
(\bar{z}_{12})|\bar{l},0 \ket\
 \Phi_{(j,\bar{j});(J,\bar{J})}(z_2,\bar z_2)  \nonumber \\
 &&=
 \sum_{j\,, \bar{j}\,, t\,, \bar{t}}\,
 {D_{(k,\bar{k}) (l,\bar{l})}^{(j,\bar{j}); t,\bar{t}}\,
 \over
z_{12}^{\triangle_{kl}^j}\, \bar{z}_{12}^{\triangle_{\bar{k}
\bar{l}}^{\bar j}} }\,
 \Phi_{(j,\bar{j})}(z_2,\bar z_2)   + .... \label{ope}
\eea
The coefficients
 $D_{(k,\bar{k}) (l,\bar{l})}^{(j,\bar{j}); t,\bar{t}}$
are related to the full-plane bulk OPE coefficients, see below.


\subsection{Boundary CFT duality relations}
\subsubsection{Cardy-Lewellen  equations rederived}

We collect in this section the set of equations  
resulting from the sewing constraints on the various OPE expansions
 \cite{L}; some of these equations can be interpreted
as expressing 
locality (symmetry) of the boundary CFT correlators. For simplicity
of notation we shall sometimes omit the explicit indication of
the coupling indices  of type
$\Big({i\,\phantom{c}\atop j\,k}\Big)\,$ and the corresponding
summations, i.e., the equations will be written
 essentially for the simplest $sl(2)$ case.
However we shall keep the charge conjugation in the indices of
$\calI$ so that the general formulae can be easily recovered.
In the  $sl(2)$ WZW case the braiding phases are given  by
the shifted scaling dimensions $\triangle_j^{Sug} - j$, instead
of  $\triangle_j^{Sug}$ (since the pair of coordinate and isospin
variables is moved as a whole). Then formulae work equally well with
the same fusing and braiding matrices,  without additional signs,
 as for the corresponding 
subfamily $(1,2j+1)$ of fields in Virasoro minimal models.

Applying (\ref{IVa}) in different ways to the $4$--point function
of boundary fields $\Psi$, that is demanding associativity,
 leads to a relation connecting the two
types of fusing matrices ${}^{(1)}F$ and $F$, the fusing
matrix for the ordinary CVO, which reads symbolically
\bea
F\ {}^{(1)}F\ {}^{(1)}F\  ={}^{(1)}F\ {}^{(1)}F\ \label{pentag}
\eea
or, more explicitly,
\bea
\sum_{m, \, \zb_2, t_3, t_2}\,
F_{m p}\left[\matrix{
i&j\cr l&k} \right]_{t_2\, t_3}^{u_2\, u_3 }\
{}^{(1)}F_{b l}\left[\matrix{
i&m\cr a&d} \right]_{\za_1\, \zb_2}^{\zg_1\, t_2 }\
{}^{(1)}F_{c m}\left[\matrix{
j&k\cr b&d} \right]_{\za_2\, \za_3}^{\zb_2\, t_3 }\
\nn \\
=\sum_{\zb_1}\,
{}^{(1)}F_{c l}\left[\matrix{
p&k\cr a&d} \right]_{\zb_1\, \za_3}^{\zg_1\, u_2 }\
{}^{(1)}F_{b p}\left[\matrix{
i&j\cr a&c} \right]_{\za_1\,\za_2}^{\zb_1\, u_3 }\,. \label{mpt}
\eea
The identity (\ref{mpt}), when restricted to the $sl(2)$ case,
 is a  slightly simplified version of the
equation (L 3.29) in~\cite{L} and can be also obtained from the latter
using the relation (\ref{tpc}) and dropping a (non-zero) factor of
type ${}^{(1)}F_{a1}$. The direct derivation of this pentagon-like
identity depicted in Figure~\ref{pentagon}  is analogous to the
derivation of the standard pentagon equation for the fusing
matrices $F$ since the boundary field $n$-point blocks are
analogs of the ordinary $(n+2)$-point conformal blocks with
an additional constraint due to the delta function in the 2-point
boundary block.  \footnote{`Mixed' pentagon identities analogous
to (\ref{pentag})  appear in
 the framework of `weak Hopf algebras' as part of a `Big
Pentagon identity' \cite{BS}.  The counterparts of  ${}^{(1)}{F}$ 
are interpreted as kind of ``3j - symbols'' along with the standard
interpretation of the fusing matrices $F$ as ``6j - symbols''. 
} The  relation  (\ref{tpc}) is reproduced from the pentagon
identity  (\ref{mpt}) for particular values of the indices.
%
%
\bookfigh{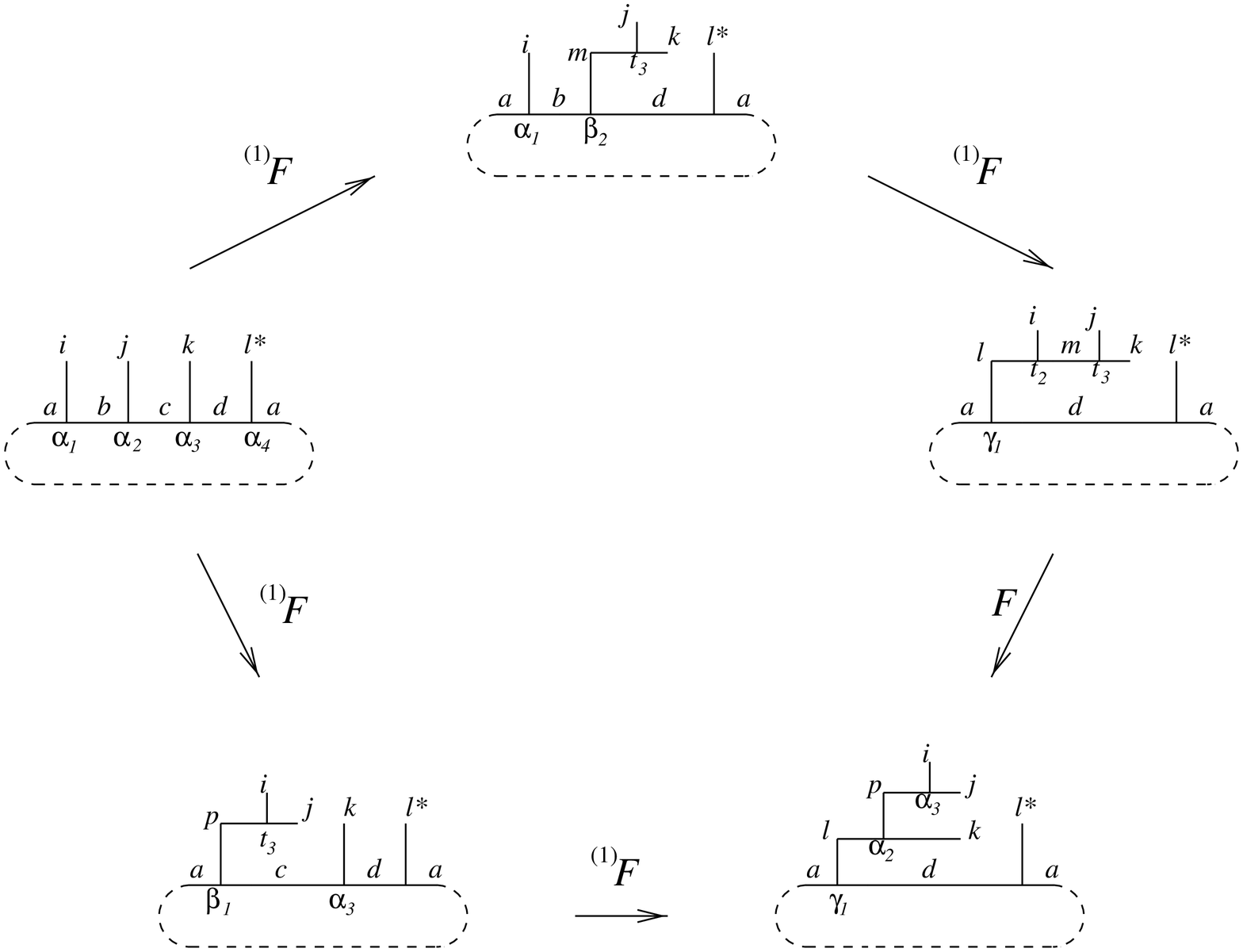}{mixed pentagon}{pentagon}{The ``mixed''
pentagon identity }{10}

Imposing the symmetry of
the 3-point function
  $\bra  {}^{b}\Psi_{j,\zd}^a(x_1)\, \Phi_{(i, \bar{i})}(z, \bar z)\,
  {}^{a}\Psi_{k, \zg}^b(x_2) \ket =$
  
  $\, \bra {}^{b}\Psi_{j,\zd}^a(x_1)\, {}^{a}\Psi_{k, \zg}^b(x_2)  \Phi_{(i,
\bar{i})}(z, \bar z)\, \ket$
  one derives following Figure~9,
\bea
&&\sum _{\zb, \zb'} \, 
{}^{b, \zb}B_{(i, \bar{i})}^{p}\ \bra \un \ket_b\
{}^{(1)}F_{b 1}\left[\matrix{
p^*&p\cr b&b} \right]_{\zb'\, \zb}^{\un\  }\ 
{}^{(1)}F_{a p^*}\left[\matrix{
j&k\cr b&b} \right]_{\zd\, \zg}^{\zb'\  \ }
\nonumber
=
\sum_{s\,,  \,\za, \za'}\,  
{}^{a, \za}B_{(i, \bar{i})}^{s}\ \bra \un \ket_a \
{}^{(1)}F_{a 1}\left[\matrix{
s^*&s\cr a&a} \right]_{\za'\, \za}^{\un\  }\\
&& 
{}^{(1)}F_{b s^*}\left[\matrix{
k&j\cr a&a} \right]_{\zg\, \zd}^{\za'\  \ }\
\sum_{m}\,
e^{i \pi(2 \triangle_i -2 \triangle_m + \triangle_k +\triangle_j
-\triangle_p)}\,
F_{s m}\left[\matrix{j &i \cr k^*&\bar{i}\cr } \right]\,
F_{m p^*}\left[\matrix{ k&j \cr \bar{i}^*&i } \right]\,.\label{eqI}
\eea
%
%
\bookfigh{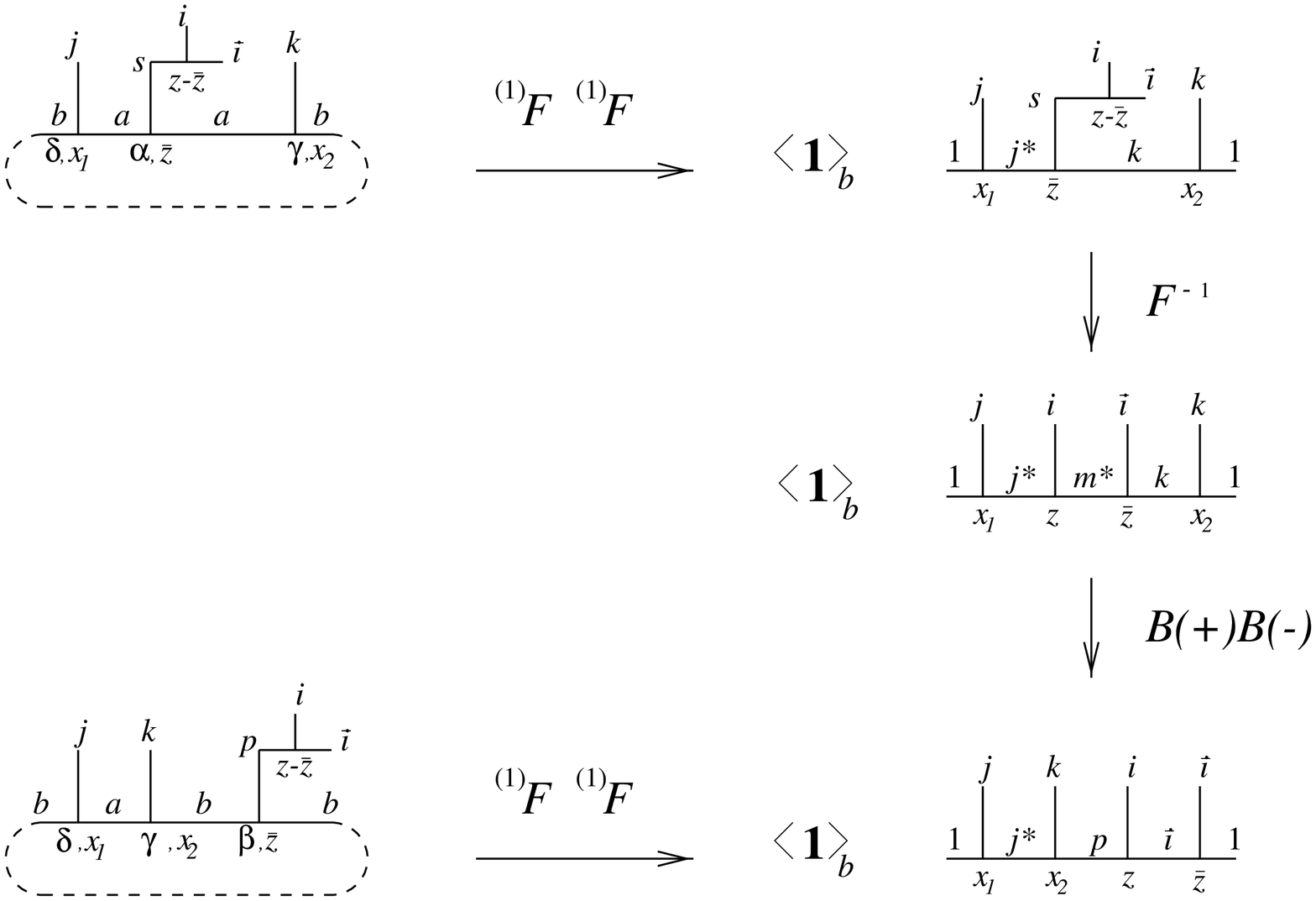}{derivation of eqI}{derivation}{Derivation of
(\ref{eqI}) }{10}
\noindent
On Figure~\ref{derivation} the braiding matrices $B(\pm)$ appear, see
Appendix~E. In the r.h.s. of (\ref{eqI}) we have also used 
the cyclic symmetry relations (\ref{cycI}),(\ref{cyc}). Using
furthermore  these relations 
a factor of type ${}^{(1)}F_{a 1}$ (and the related
summation)  can be  dropped in both sides
of (\ref{eqI}) 
which leads to a slighly simplified version as compared with the
original equation (L 3.32) in \cite{L}.  
\omit{
Exchanging $\Phi_{(i,
\bar{i})}$ with the first boundary field instead with the second
leads to a similar relation, combining which with (\ref{eqI})
recovers the first of the  standard symmetries of the fusing
matrices in  (\ref{symf}).
}
\medskip

{}From the  2-point function $\bra \Phi_{(k, \bar{k})}(z_1, \bar z_1)
\Phi_{(l, \bar{l})}(z_2, \bar z_2)\ket_a $ using either the OPE
formula (\ref{ope}), or (\ref{IVb}) (follow Figure 10 with $i=1$),
 we obtain
\bea
&&\sum_{\za\,, \zb}\,
{}^{a, \za}B_{(k, \bar{k})}^{r, s_1}\ \,  {}^{a, \zb}B_{(l,
\bar{l})}^{r^*, s_2}\ \,
{}^{(1)}F_{a 1}\left[\matrix{
r&r^*\cr a&a} \right]_{\za\, \zb}^{\un_a \,  \un_r'}
\label{eqII}\\
&&=
\sum_{j}\,{}^{a}B_j^{1}\,
e^{i \pi( \triangle_k + \triangle_{\bar{l}} - \triangle_r
-\triangle_j)}\,  \sum_{t, \bar{t}}\,
D_{(k,\bar{k}) (l,\bar{l})}^{(j,j^*); t,\bar{t}}\ \,
F_{j r}\left[\matrix{
\bar{k}&k\cr \bar{l}^*&l} \right]_{\sigma_{13}(\bar{t})\ \
t}^{\sigma_{23} \sigma_{13}(s_2) \ \sigma_{23}(s_1)}\,,
 \nonumber
\eea
or, equivalently,
\bea
&&D_{(k,\bar{k}) (l,\bar{l})}^{(j,j^*); t,\bar{t}}\ \, 
{}^{a}B_j^{1}\,
e^{i \pi(\triangle_k + \triangle_{\bar{l}} - \triangle_j)}
\label {eqIIa} \\
&&=
\sum_{\za\,, \zb\,, r\,, s_1\,, s_2}\,
e^{i \pi \,\triangle_r}\,
{}^{a, \za}B_{(k, \bar{k})}^{r, s_1}\ \,  
{}^{a, \zb}B_{(l, \bar{l})}^{r^*, s_2}\ 
{}^{(1)}F_{a 1}\left[\matrix{
r&r^*\cr a&a} \right]_{\za\, \zb}^{\un_a \,  \un_r'}\,
F_{r^* j}\left[\matrix{k&l\cr\bar{k}^*& \bar{l}}
 \right]_{\sigma_{13}(s_1)\  \ s_2}^{\sigma_{12}(\bar{t})  \ t}\,.
\nonumber
\eea

\medskip
{}Lastly from the 3-point function
$\bra  \Phi_{(k, \bar{k})}(z_1)\,  \Phi_{(l, \bar{l})}(z_2)
\, {}^{a}\Psi_{i, \zg}^a(x)\ket$, we obtain, 
see Figure~\ref{derivatio}
\bea
&&\sum_{\za\,, \zb}\,
{}^{a, \za}B_{(k, \bar{k})}^{r}\ \,  
{}^{a, \zb}B_{(l, \bar{l})}^{t}\ \,
{}^{(1)}F_{a i^*}\left[\matrix{r&t\cr a&a} \right]_{\za\, \zb}^{\zg }
\label{eqIII}\\
&&=
\sum_{j\,,\bar{j}\,}\, D_{(k,\bar{k}) (l,\bar{l})}^{(j,\bar{j})}
{}^{a, \zg}B_{(j, \bar{j})}^{i}\, 
e^{i \pi(\triangle_{k} - \triangle_r-\triangle_j)}\,
\sum_{s}\, e^{i \pi\, \triangle_s}\, 
F_{\bar{j} s^*}\left[\matrix{j&\bar{k} \cr i^*&\bar{l} } \right]
F_{j^* r^* }\left[\matrix{ l&s\cr k^*&\bar{k}} \right]\,
F_{s t}\left[\matrix{l&\bar{l}\cr r^*&i} \right]
 \nonumber
\eea
For $i=1$ (\ref{eqIII})
reduces to  (\ref{eqII}).
The sum over $s$ in the r.h.s. of (\ref{eqIII})
  represents up to phases one of the
sides in (an auxiliary) hexagon  identity, resulting
in permuting $\{l\,,i\,,\bar{k}\}$ to $\{\bar{k}\,,i\,,l\}$ i.e.,
can be written as $B_{23}(-) B_{12}(-) B_{23}(-)$ and thus can be
replaced by $B_{12}(-) B_{23}(-) B_{12}(-)$. This gives an
alternative representation of the r.h.s. of  (\ref{eqIII})
obtained from the above by replacing everywhere $(k\,, \bar{k}\,,
j\,, r)\to (\bar{l}\,, l\,,\bar{j}\,, t^*)$. This is the original
form  of the equation (L 3.35) in \cite{L}, when (\ref{CLchoice})
is inserted,  with furthermore inverse 
operator ordering convention and opposite overall sign of the phase.

%
\bookfigh{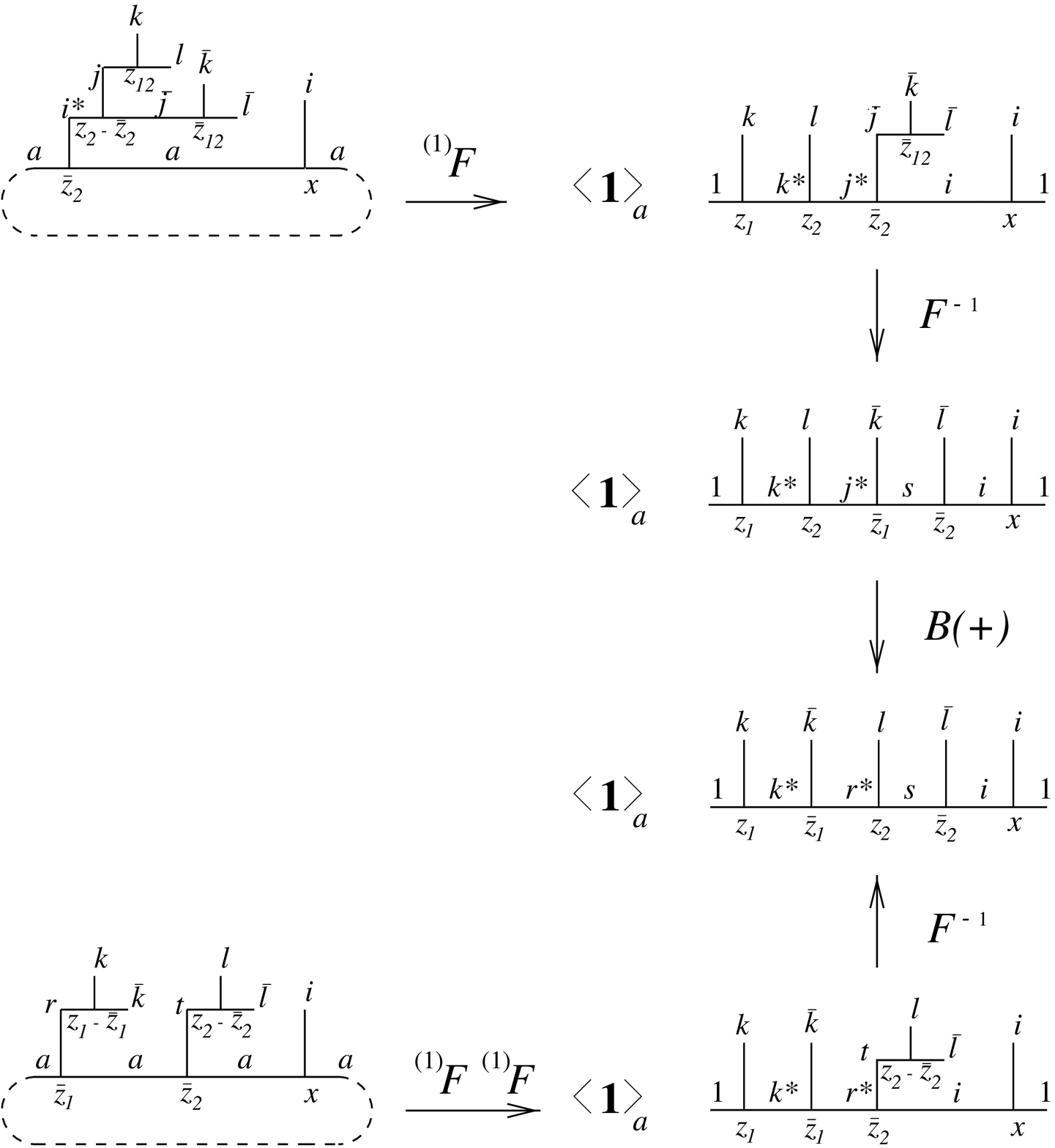}{derivation of eqIII}{derivatio}
{Derivation of (\ref{eqIII}) }{13}

The locality of this 3-point function (the symmetry under the exchange
of the fields $ \Phi$)  implies also
\bea
D_{(k,\bar{k}) (l,\bar{l})}^{(j,\bar{j}); t,\bar{t}}\ \,
=(-1)^{s_k+s_l-s_j}\ D_{ (l,\bar{l})
(k,\bar{k})}^{(j,\bar{j});
\sigma_{23}(t),\sigma_{23}(\bar{t})}\,, \qquad
s_k:=\triangle_k-\triangle_{\bar{k}}\,,
\eea
while from the associativity of the OPE (\ref{ope}) one obtains
in particular the relation
\bea
D_{(k,\bar{k}) (l,\bar{l})}^{(j,\bar{j}); t,\bar{t}}\
\,D_{(j,\bar{j}) (j^*,\bar{j}^*)}^{(1,1)}=
D_{ (l,\bar{l}) (j^*,\bar{j}^*)}^{(k^*,\bar{k}^*); \sigma_{13}\sigma_{23}(t),
 \sigma_{13}\sigma_{23}(\bar{t})}\
\,D_{(k^*,\bar{k}^*) (k,\bar{k})}^{(1,1)}\,. \label{asoc}
\eea

All the above equations hold true as well with a sign
 $\epsilon=\pm 1$, inserted in the exponents of all the phases
in these equations including (\ref{CLchoice}),
and replacing the bulk-boundary coefficients $B$ with $B_{\epsilon}$.
Thus when rewritten in terms of
the Cardy-Lewellen normalised coefficients $B(CL)$ all equalities
are true for both choices of sign.

\medskip

{\small 
{\it Remark:}
The Lewellen equations in the  diagonal $sl(2)$ case  were
recently confirmed in \cite{R}. The seemingly different
version of equation (\ref{eqI})
in  \cite{PSS}
is in fact equivalent to the  original Lewellen equation, after
taking into account one of the duality relations (a hexagon identity,
see (\ref{Racah}) in  Appendix~E) for the braiding matrices.
On the other hand  the
derivation of the versions of  (\ref{eqII}), (\ref{eqIII}) in
\cite{PSS} appears to  be  affected
by a  missing phase
in the intermediate (and needless) formula (17) in \cite{PSS}.
This phase   is
compensated in the final formulae following from (\ref{eqII})
(like (\ref{pa}) below) by another phase due to the
presumably neglected difference in the normalisation of the
bulk-boundary coefficients (like in (\ref{CLchoice})) as compared
with that in \cite{CL,L}. }


\subsubsection{More pentagon relations}

Before we turn to a discussion on the implications of the
Cardy-Lewellen equations
we shall introduce one more ingredient to the scheme.
 It is natural to assume that there exists a
``third fusing matrix'', a matrix inverting ${}^{(1)}F$,
$\  {}^{(3)}{F}\, {}^{(1)}F= I ={}^{(1)}F \, {}^{(3)}{F}$,
or more explicitly, 
\be
\sum_{b\,, \zb_2\,, \zb_3} {}^{(3)}{F}_{p b^* }\left[\matrix{
 c^*&k\cr a^*&j} \right]_{\zs_{23}(\za_2) \
t'}^{\zs_{12}(\zb_3)\, \zs_{12}(\zb_2) }\ 
{}^{(1)}F_{b s}\left[\matrix{
k&j\cr c&a} \right]_{\zb_2\, \zb_3 }^{\zs_{13}(\zg_2)\, t}=\zd_{ps}\,
\zd_{\za_2\, \zg_2}\, \zd_{t'\, t} \,, \label{inver}
\ee
\bea
{}^{(3)}{F}_{p b }\left[\matrix{
a&1\cr c&k} \right]_{\za \, t}^{\zb\, \zg}&&= \zd_{a\, b}\,
 \zd_{k\,p}\, \zd_{\za\, \zb}\, \zd_{t\, \un}\,
  \zd_{\zg\,\un }\,, \nonumber\\
{}^{(3)}{F}_{p b }\left[\matrix{
a&j\cr c&1} \right]_{\za \, t}^{\zb\, \zg}&&=
\zd_{c\, b}\,  \zd_{j\,p}\, \zd_{\za\, \zg}\,
\zd_{t\, \un }
\,          \zd_{\zb\, \un } \,.
\nonumber
\eea
Along with the standard CVOs $\Big({p\,\phantom{c}\atop
j\,k}\Big)\,$,   this matrix involves  new ``couplings''of type
$\Big({c\,\phantom{c}\atop a\,p}\Big)_{\zb'}\,$ which can be
thought of as obtained by a  permutation $\sigma_{23}$ from the
boundary fields $\Big({c\,\phantom{c}\atop p\,a}\Big)_{\zb}\,,$
whence the notation $\zb'=\sigma_{23}(\beta)$. The matrix ${}^{(3)}{F}$ 
satisfies a ``mixed'' pentagon identity 
analogous to (\ref{pentag})
\bea
 {}^{(3)}{F}\ {}^{(3)}{F}\  F\   ={}^{(3)}{F}\ {}^{(3)}{F}\,.
\label{inv}
\eea
Furthermore multiplying both sides of (\ref{mpt}) with
 ${}^{(3)}{F}_{m'c^*}\left[\matrix{
 b^*&j \cr d^*&k } \right]_{\zd \, t}^{\zs_{12}(\za_3)\,
\zs_{12}(\za_2) } $ and summing over $c, \alpha_2\,, \alpha_3\,,$ 
 using (\ref{inver}) in the r.h.s.,
we obtain another equation of  similar form
\bea
{}^{(3)}{F}\ {}^{(1)}F\ {}^{(1)}F\ ={}^{(1)}F\ F\ 
\,, \nonumber
\eea
which implies various useful
relations obtained for particular values of the indices. One of
them reproduces the inverse property of ${}^{(3)}{F}$,
another one reads
\bea
 \sum_{\zb }\
{}^{(3)}{F}_{1 b^* }\left[\matrix{a^*&j^*\cr a^*&j} 
\right]_{\un \,\un}^{\zs_{23}(\za)\, \zs_{23}(\zb) }\
{}^{(1)}F_{a 1}\left[\matrix{
j&j^*\cr b&b} \right]^{\un\,\un}_{\zs_{13}(\zg)\, \zs_{13}(\zb) }\
={1 \over d_j }\ \delta_{\za, \zg}\label{Va}
\eea
where $d_j=S_{j1}/S_{11}=F_{1 1 }\left[\matrix{
j&j^*\cr j&j} \right]^{-1}$ is the quantum dimension. It
 furthermore implies
\bea
d_j \ \sum_{\za, \zb}\
{}^{(3)}{F}_{1 b^* }\left[\matrix{
a^*&j^*\cr a^*&j} \right]_{\un \,\un}^{\zs_{23}(\za)\, \zs_{23}(\zb) }\
{}^{(1)}F_{a 1}\left[\matrix{
j&j^*\cr b&b} \right]^{\un\,\un}_{\zs_{13}(\za)\, \zs_{13}(\zb) }\
=n_{j a}^b \,.  \label{Vb}
\eea
%
Using the inverse matrix ${}^{(3)}{F}$, as
introduced here,  the half-plane bulk field can
be also written as a product of  generalised boundary fields (a
``bilocal'' operator)
\bea
\Phi_{(i\,, \bar{i}) }^H(z,\bar{z})
= \sum_{a\,,b\,, \za\,,
\zb}\, \Big(
\sum _{p\,, \zg\,,t}\,
{}^{a, \zg}B_{(i, \bar{i})}^{p, t}\,
{}^{(3)}{F}_{p b^* }\left[\matrix{
a^*& i \cr a^*& \bar{i} } \right]_{\zs_{12}\zs_{23}(\zg)\ 
t}^{\zs_{12}\zs_{23}(\zb)\, \zs_{12}\zs_{23}(\za) }\, \Big)\,
  {}^a\Psi_{i,\za}^b(z)\, {}^b\Psi_{\bar{i},\zb}^a(\bar{z})\,,
\label{bbdy} 
\eea
which reproduces the small  $z-\bar{z}$  expansion in (\ref{IVb});
compare (\ref{bbdy}) with the chiral decomposition of the (full) plane
physical fields 
\bea
\Phi_{(i, \bar{i})}^P(\zeta,\bar{\zeta})=\sum_{k,\bar{k}, l, \bar{l},t,t'}\,
{}^{(pl)}D_{(j, \bar{j}) (k,\bar{k})}^{(l,\bar{l}); (t,t')}\
{}^l\phi_{i\,, t}^k(\zeta) \otimes {}^{\bar l}\phi_{\bar{i}\,, t'}^{\bar
k}(\bar{\zeta})\,. \nonumber
\eea

Finally we shall exploit the inverse ${}^{(3)}{F}$ of the
matrix ${}^{(1)}F$ to rewrite (\ref{eqI}) in another equivalent form
to be used in the next section. Namely
we apply the inverse  to 
${}^{(1)}F_{b s^* }$ in the r.h.s. and obtain
\bea
&&\sum_{\za}\,   {}^{a, \za}B_{(i, \bar{i})}^{s}\,
\bra \un \ket_a\,
{}^{(1)}F_{a 1}\left[\matrix{
s^*&s\cr a&a} \right]_{\zs_{13}(\za')\, \za}^{\un\  \ }\,
\sum_{ m\,}\,
e^{i \pi(2 \triangle_i -2 \triangle_m + \triangle_k + \triangle_j
- \triangle_p)} \,
F_{s m}\left[\matrix{j &i \cr k^*&\bar{i}\cr } \right]\,
F_{m p^*}\left[\matrix{ k&j \cr \bar{i}^*&i } \right]
\nonumber \\
 \label{eqIa}\\
&&=\sum _{b\,, \zb\,,\zb'} 
{}^{b, \zb}B_{(i, \bar{i})}^{p}\, \bra \un \ket_b\,
{}^{(1)}F_{b 1}\left[\matrix{
p^*&p\cr b&b} \right]_{\zb'\, \zb}^{\un\  \ }\,
\sum _{\zg\,, \zd} \, 
{}^{(3)}{F}_{s^* b^* }\left[\matrix{a^*&k\cr a^*&j} 
\right]_{\zs_{23}(\za') \,}^{\zs_{23}(\zd)\,\zs_{23}(\zg) }\,
{}^{(1)}F_{a p^*}\left[\matrix{
j&k\cr b&b} \right]_{\zs_{13}(\zd)\, \zs_{13}(\zg)}^{\zb' \   }\,
\nonumber \\ 
\nonumber \\ 
&&=\sum _{b\,, \zb\,,  \zg'\,, \zd}  {}^{b, \zb}B_{(i,
\bar{i})}^{p}\, \bra \un \ket_b \,
{}^{(1)}F_{b k^*}\left[\matrix{
p&j\cr b&a} \right]_{\zb\, \zs_{13}(\zd)}^{\zg' \  \ }\,
\sum_{\zg}
{}^{(3)}{F}_{s^* b^* }\left[\matrix{a^*&k\cr a^*&j} 
\right]_{\zs_{23}(\za') \,}^{\zs_{23}(\zd)\, \zs_{23}(\zg) }\,
{}^{(1)}F_{a 1}\left[\matrix{ k^*&k\cr b&b} 
\right]_{\zg'\, \zs_{13}(\zg)}^{\un\  \ } \,.
\nonumber
\eea
In the third line we have used the  symmetry
relation (\ref{tpc}).

%
%

\subsection{Consequences of the bulk-boundary equations}
\subsubsection{The Pasquier algebra and its dual}

In this section we  analyse  some important consequences of the
set of equations derived.  We start with  equation (\ref{eqIa}),
an inverted version of the first Lewellen bulk-boundary  equation
(\ref{eqI}), in which we take $s=1=p$. This implies $k=j^*\,,
\bar{i}=i^*$, $\za=\un=\za'=\zb=\zb'$.  The sum over
$m$ in the l.h.s. is proportional to the modular matrix $S_{ji}$,
see (\ref{torb}), while the sums over the coupling indices
$\zd\,, \zg$ are worked out using (\ref{Vb}), the final result
being 
\bea
{S_{ji}\over S_{1i}}\ {}^{a}B_{i}^{1} \,
\bra \un \ket_a = \sum_{b} \,
\n_{j a}{}^b\  {}^{b}B_{i}^{1} \, \bra \un \ket_b\,.
\label{V-NI}
\eea
For simplicity we have done this computation in the $sl(2)$
case but it extends straightforwardly to arbitrary rank leading
to the same formula.

Comparing   (\ref{V-NI}) with (\ref{V-N}), we see that it can be
identified with the realisation (\ref{V-N}) of the relation
(\ref{IIzc}) in terms of the eigenvalues $\hat{\gamma}
_a(i)={\psi_a^i\over \psi_{\un}^i}$ of the graph algebra matrices
$\hN_{a}$.  Namely we can identify the ratio $ {}^{b}B_{i}^{1} \,
\bra \un \ket_b/ {}^{a}B_{i}^{1}\, \bra \un \ket_a$  with the
ratio $\hat{\gamma}_b(i)/ \hat{\gamma}_a(i)$.  Recalling  the
expression for $\bra \un \ket_a$ in (\ref{IViba})  we find a
relation between the boundary state coefficients  $\psi_a^i$  and
the bulk-boundary coefficients ${}^{a}B_i^{1}$
\be
{}^{a}B_{i}^{1}=e^{i \pi \triangle_i} \ {}^{a}B_{i}^{1}(CL)=
{\psi_a^{i}\over \psi_a^{\un}}\ e^{i \pi \triangle_i} \,
\sqrt{C_{ii^*}\over d_i}\,,  \label{idr}
\ee
where  for the time being $C_{ii^*}$ is an arbitrary constant,
$C_{11}=1$.  Conversely, if we assume the identification
(\ref{idr}) (as, e.g., derived in the $sl(2)$ case by other means
in \cite{CL}, with $C_{ii^*}=1$, see also \cite{RS1,RS2}) we
recover the relation (\ref{V-N}), or, (\ref{IIzc}) directly from
one of the bulk-boundary equations. As  discussed in Section 3.5,
from this relation we reconstruct the graph algebra.

In the diagonal case  $\calE=\calI$, where $\psi_a^j=S_{aj}$, the
relation (\ref{V-N}) coincides with the Verlinde formula, i.e.,
the standard fusion algebra realised by its characters.  On the
other hand the Verlinde formula is known
\cite{PSS} to be recovered from the diagonal version of the other
bulk-boundary equation, the  Cardy-Lewellen equation (\ref{eqII})
to which we now turn.  This equation simplifies for $r=1$,
leading to  $\bar{k}=k^*\,, \, \bar{l}=l^*$. Using (\ref{idr})
and denoting $p_i(a)={\psi_a^{i}/ \psi_a^{1}}$ the equation
(\ref{eqII}) turns into
\be
p_k(a)\, p_l(a)=\sum_{j}\,
M_{kl}{}^j\, p_j(a)\,,
\label{pa}
\ee
where
\bea
M_{kl}{}^j &&=   \sum_{t,  \bar{t}}\,
d_{(k,k^*) (l,l^*)}^{(j,j^*); t,\bar{t}} \nonumber \\
&&: =
\sqrt{d_k\, d_l \over d_j} \, \sqrt{ C_{j j^*}\over C_{kk^*}\, C_{l
l^*} }\,   \sum_{t,  \bar{t}}\,
D_{(k,k^*) (l,l^*)}^{(j,j^*); t,\bar{t}}\ \,
F_{j 1}\left[\matrix{
k^*&k\cr l&l} \right]_{\sigma_{13}(\bar{t})\   t}^{\un\, \un}\,,
 \label{mstr}
\eea
and $M_{kl}{}^j=0$ if the corresponding Verlinde
multiplicity $N_{kl}{}^j$ vanishes.
Alternatively, inverting (\ref{pa}),
\be
M_{kl}{}^j=\sum_{a\in \calV}\, {\psi_a^{k}\, \psi_a^{l}\, \psi_a^{j
*}\over
\psi_a^{1}}\quad k\,,l\,,j\in \calE\,. \label{pac}
\ee

Let us first look at the diagonal case in which according to
(\ref{pac}) the constants $M_{kl}{}^j$ coincide with the Verlinde
fusion rule multiplicities $N_{kl}{}^j$. This is confirmed also
directly by the alternative expression (\ref{mstr}) provided by
equation (\ref{eqII}) as we shall now show.  In   the diagonal
case, denoting $D_{(k,k^*) (l,l^*)}^{(j,j^*);
t,\bar{t}}=C_{(k,k^*) (l,l^*)}^{(j,j^*); t,\bar{t}}$,  we can use
the inverted equation (\ref{eqIIa}) taken  for $a=1$, a choice
which trivialises all summations, since $\n_{r 1}{}^1=N_{r
1}{}^{1}=\zd_{1 r}$, with the result (pointed out in the $sl(2)$
case in \cite{R})
\bea
{{}^{1}B_j^{1}\over {}^{1}B_k^{1}\,{}^{1}B_l^{1} }\,
e^{i \pi(\triangle_k + \triangle_{l} - \triangle_j)}\,
C_{(k,k^*) (l,l^*)}^{(j,j^*); t,\bar{t}}\ \,
=
F_{1 j}\left[\matrix{
k&l\cr k&l^*} \right]_{\un \ \un}^{\sigma_{12}(\bar{t}) \ t}\,.
\label{eqIIc}
\eea
Taking in particular $j=1$ (\ref{eqIIc}) gives
\bea
{}^{1}B_k^{1}\,{}^{1}B_{k^*}^{1} =
e^{2  \pi i \triangle_k }\ d_k\
C_{(k,k^*) (k^*,k)}^{(1,1)}\,. \label{IVaa}
\eea
Comparing with (\ref{idr}) taken in the diagonal case we see that
we can identify the undetermined constant $C_{ii^*}$ with the
normalisation constant of the bulk $2$--point function.  We shall
retain this identification of  $C_{ii^*}$ in the non-diagonal
cases (at the same level as the given diagonal case) which
amounts to setting the relative $2$-point normalisation to $1$.
Combined with (\ref{asoc}) and (\ref{idr}) the relation
(\ref{eqIIc}) leads to a symmetry of the fusing matrices
analogous to the cyclic symmetry (\ref{cycI})
\bea
F_{1 j}\left[\matrix{
l&k\cr l&k^*} \right]_{\un \ \un}^{t_1\ t_2}\
d_l=
F_{1 l}\left[\matrix{
j&k^*\cr j&k} \right]_{\un \ \un}^{t_2\ t_1}\
d_j \,. \label{cycII}
\eea
Inserting (\ref{eqIIc}) back into (\ref{mstr}) and using
(\ref{cycII}) reduces the sum over $\bar{t}$ to the standard
pentagon identity specialised for some choice of the indices
(cf. the analogous relation (\ref{Va})).  Finally we are left
with the sum  over the coupling index
$t=t\Big({j\,\phantom{c}\atop k\, l}\Big)\,,$ which reproduces
the Verlinde multiplicity $N_{kl}{}^j$ and completes the
argument; alternatively 
the same conclusion is achieved using the simple choice of gauge
(\ref{cgau}). 

Note that the relation (\ref{eqIIc}), with (\ref{IVaa}) accounted
for, is a linear version of the standard (quadratic) relation for
the full plane diagonal OPE coefficients which results from
locality of the (full) plane bulk fields $4$--point functions,
see Appendix~E.  In the $sl(2)$ case this identifies the  OPE
coefficients of the half- and full-plane diagonal bulk fields.
The identification extends to the nondiagonal $sl(2)$ scalar OPE
coefficients as can be seen generalising to $2$-point bulk
correlators the computation  of the limit $L/T \to \infty$ of
the $1$-point correlators in  \cite{CL},  leading  to
(\ref{idr}).

\medskip
In the general (non-diagonal) $sl(2)$ cases characterised by a
fixed level (central charge)  we can express the fusing matrix in
the r.h.s. of (\ref{mstr}) in terms of that in the r.h.s. of
(\ref{eqIIc}). Using once again the $sl(2)$ versions of the
identities just described, we express it in terms of the diagonal
OPE coefficients at the same level, obtaining for $N_{kl}{}^j=1$
\be
M_{kl}{}^j=d_{(k,k) (l,l)}^{(j,j)}=
D_{(k,k) (l,l)}^{(j,j)}\, /C_{(k,k) (l,l)}^{(j,j)}\,.
\label{relat}
\ee
The relative scalar OPE coefficients $d_{(k,k) (l,l)}^{(j,j)}$
have been computed for the $sl(2)$ WZW and the Virasoro
(unitary) minimal models, see, e.g., \cite{PZ1} for an exhaustive
list of references. Now using the expression for the eigenvectors
(\ref{jbsduty}), they  can be computed as in (\ref{pac}) for all
minimal models.

The matrices $(M_k)_l{}^j=M_{kl}{}^j$ can be seen as a matrix
realisation of an associative commutative algebra with identity,
distinguished basis and an involution $*$. In (\ref{pa}) the
algebra is realised by its 1-dimensional representations
(characters) given by ratios of elements of the eigenvector
matrix defining $\n_i$.  This  algebra, traditionally called the
``Pasquier algebra'' (``$M$''--algebra), is dual in the sense of
ref.~\cite{BI} to the graph $\hN$-- algebra considered in Section
3 but unlike its dual, its structure constants are not in general
integral, but rather algebraic numbers.  In the simplest $sl(2)$
case the squares $(M_{kl}{}^j)^2$ of these constants are rational
numbers for all \ade\  cases; this rule persists for most of the
$sl(3)$ cases but is broken by two of the graphs $\calE^{(12)}_1$
and $\calE^{(12)}_2$ corresponding to the exceptional  modular
invariant at level $k+3=12$, see Appendix~D.  The type II $sl(2)$
cases, $D_{\rm odd}$  and $E_7$ are again distinguished by the
fact that the sign of some of the multiplicities $M_{kl}{}^j$ is
negative and this is a basis independent statement in the sense
that there is no choice of basis to make all $M_{kl}^j$
non-negative, contrary to the Type I cases $D_{\rm even}\,,$
$E_6\,, E_8$, and this is a general feature of type II theories.
\medskip

The formula (\ref{relat}) extends beyond the $sl(2)$ case for
$(k,l,j)$ such that $N_{kl}{}^j=1$, i.e., in cases with trivial
Verlinde multiplicity the matrix elements $M_{kl}{}^j$ provide
the (relative) OPE coefficients $d_{(k,k^*) (l,l^*)}^{(j,j^*)} $.
For non-trivial Verlinde multiplicities $N_{kl}{}^j>1$ the
relation between the constants $M_{kl}{}^j$ (\ref{pac}) and the
OPE coefficients is not so direct.  Let us give a $sl(3)$ WZW
example which illustrates the relation (\ref{mstr}).  There are
three graphs found in \cite{DFZ1} which correspond to the
exceptional  block-diagonal modular invariant at level $k+3=12$,
see Appendix~D, where these graphs are denoted by
$\calE_i^{(12)}\,, i=1,2,3$. One can pick up triplets of weights
$(i,j,l)$ such that the Verlinde multiplicity of the diagonal
$sl(3)$ model at level $k+3=12$ is trivial, $N_{ij}{}^l=1$ and
check the values of the corresponding Pasquier algebra structure
constants $M_{ij}{}^l$ for each of the three graphs. The result
is that, comparing in particular $\calE_1^{(12)}\,$ and
$\calE_3^{(12)}\,,$ there exist such triplets leading to
different values of $M_{ij}{}^l$ for the two graphs.  Since for
trivial Verlinde multiplicities the formula (\ref{mstr}) gives a
direct relation between the two types of constants, $M_{ij}{}^l=
d_{(i,i^*) (j,j^*)}^{(l,l^*); \un,\un}$, this result suggests
that there are two different solutions for the bulk OPE
coefficients in this case.  Only one of these two non-diagonal
solutions, namely the one which can be associated with the type I
graph $\calE_1^{(12)}$ was recovered in \cite{PZ2}, exploiting a
set of equations for the $M$ -algebra structure constants. This
set was derived from the bulk CFT locality equations assuming an
additional (quadratic) constraint on the OPE coefficients in
theories with an extended symmetry; some of its consequences
were also reproduced in the abstract framework of \cite{Xu}, in
particular the relation $\n_{i 1}{}^a=$ mult$_a(i)$ discussed in
Section 3. Precisely this relation fails (and hence the
assumptions on the OPE coefficients in \cite{PZ2}) for the graph
$\calE_3^{(12)}$, which otherwise satisfies all the requirements
of type I.

\medskip

We conclude with a comment on  the OPE coefficients. As discussed
in \cite{RS1,RS2} one can relate in the limit $L/T \to \infty$
the correlators of the half- and full-plane bulk fields
$\Phi_I^H(z,\bar{z}) $ and $\Phi_{I'}^P(\zeta,\bar{\zeta}) $,
looking at the two dual representations of the  partition
function with field insertions; in particular (\ref{idr}) was
recovered in this way. Though this transformation needs to be
elaborated for higher rank cases it seems reasonable to expect
(and in agreement with (\ref{eqIIc})) that using the two choices
of the automorphism $\Omega$, discussed in Section 2.1, we can
identify in this way the OPE coefficients of the two bulk fields
with either $I'=I=(j,\bar{j})$  or $I'=(j,\bar{j}^*)$.

\bigskip

{\small {\it A bit of history}: The algebra (\ref{pa})  defined
through the eigenvectors of the \ade\ Cartan matrices  first
appeared in the context of the $sl(2)$ \ade\ lattice models
proposed by Pasquier ~\cite{Pa}  a short time before the Verlinde
fusion rule formula (the ``A'' algebra in the $sl(2)$ case) was
found.  The interpretation in terms of a pair of dual C
(`Character')--algebras was proposed in \cite{DFZ2} in the
discussion of the set of graphs found in \cite{DFZ1} as a
generalisation of the Dynkin diagrams associated with the modular
invariants of $sl(3)$ WZW and minimal models.  The fact that the
relative scalar OPE coefficients $d_{(k,k) (l,l)}^{(j,j)} $ of
all \ade\ series of the $sl(2)$ WZW (or the subfamily of fields
$(1,s)$ in unitary minimal models) coincide in a suitable basis
with the Pasquier algebra (\ref{pa}) structure constants
$M_{kl}{}^j$  was the main result of \cite{PZ1}. It was
established through a case by case check, supported by a lattice
model derivation in which the same coefficients appear
considering representations of the Temperley-Lieb algebra.  CFT
locality constraints resulting in formulae quite similar in
spirit to (\ref{mstr}) were furthermore exploited in \cite{PZ2},
\cite{PZ3} as an ingredient in the construction of generalised
Pasquier algebras and thus of new examples of graphs related to
$sl(n)$ modular invariants, extending the results in \cite{DFZ1},
\cite{DFZ2}.  The authors of \cite{PZ1},
\cite{PZ2}, \cite{PZ3} were however not aware of the parallel
development of boundary CFT, and in particular of \cite{CL},
\cite{L}, where the equation (\ref{eqII}) first appeared. The
importance of the algebra obtained from this equation at $r=1$
was recognised and stressed in  \cite{PSS}, where a
representative example of the $sl(2)$ WZW  $D_{\rm odd}$ series
was considered, for which the set $\calV$ and the characters
$\hat{\chi}_{_{i}}(a)$ of the algebra were explicitly described.
Presumably the authors of \cite{PSS} were not aware of the
general \ade\ result in \cite{PZ1}.  In the same framework of
boundary CFT the Pasquier algebra reappeared recently in a
systematic study of orbifold theories, see \cite{FS98} and
references therein, under the name  ``(total) classifying
algebra''.}


\subsubsection{Relation to the Moore-Seiberg  set of duality
equations}

We have seen that the two Lewellen bulk-boundary equations
(\ref{eqI}) and (\ref{eqII}) when restricted to some particular
values of the indices $p$ in ${}^{b}B_{(i,
\bar{i})}^{p}$ become in some sense ``dual'' to each other, recovering
the two dual C-algebras, the graph and Pasquier algebras.
These algebras are identical in the
diagonal case, reproducing the Verlinde fusion algebra, which
suggests that in this case the above two equations might be
related.

On the other hand let us recall that the original derivation \cite{Verl,
MS1} of the Verlinde formula relies on the use of one of the
basic Moore-Seiberg duality relations, namely the equation resulting from
the modular property of the two-point functions on the torus,
 see (\ref{tor}).
It involves the fusing/braiding matrices $F$ or $B$ and the
modular matrix $S_{ij}(p)$ (in general $S_{ij}^{t'\, t}(p)$)
for the $1$-point functions $\chi_j^{(p)}(\tau,z)$
 on the torus, the index $p$ standing for the
(representation) label of the inserted CVOs
$\Big({i\,\phantom{c}\atop p\, i}\Big)_{t\,, \,z'}\,$ and
$\Big({j\,\phantom{c}\atop p\, j}\Big)_{t\,, \,z}\,,$
$\log z'=(\log z)/\tau\,,$ see \cite{MS1}.

The alert reader may have already noticed the full analogy between
the Moore-Seiberg torus identity (\ref{tor}) and the second version
(\ref{eqIa}) of equation (\ref{eqI}).
It suggests that the quantity taking over the role of the modular matrix
$S(p)$ is the  bulk-boundary reflection
coefficient $B^p$. In the diagonal case this correspondence
is precise, i.e., the two are identical up to a constant.

Indeed first note that equation (\ref{eqIa}), still considered in
the general (non-diagonal) case, simplifies for  $s=1$, that is $k=j^*\,,
\bar{i}=i^*$, $\za'=\un$.  Inserting in the first line  the
expression for the modular matrix $S_{ij}(p)$, see (\ref{mtor}), 
 and  using in the third line (\ref{Va}),  we obtain an
expression for the modular matrix $S(p)$,
\bea
&&
{}^{a}B_{i}^{1}\,\bra \un \ket_a\,
{1\over  F_{ j^*1 }\left[\matrix{ j&j^* \cr p&p} \right]}\,
{S_{ji}(p^*)\over S_{i1}}
=\sum _{b\,,\zb} \, {}^{b, \zb}B_{(i, i)}^{p}\
\bra \un \ket_b\, \sum_{\zd}\, {}^{(1)}F_{b j}\left[\matrix{
p&j\cr b&a} \right]_{\zb\, \zd}^{\zd \  \ }\,. \label{Vc}
\eea

Let us concentrate now on the diagonal case
$\calI=\calE$. The sums in the r.h.s. of (\ref{Vc}) can be
reduced to one term choosing $a=1$ and using (\ref{ident}), since
$\n_{j 1}{}^b=N_{j 1}{}^b=\delta_{j b}$, $\zd=\un$. Alternatively,
one can take $a=1$ directly in the original equation (\ref{eqI})
--  the resulting (linear) formula for $B^p$ in terms of the
fusing matrices $F$ (instead of the formula for its square
derived in \cite{CL}) was  first explicitly written down by
Runkel \cite{R}.
More explicitly we have in the $sl(2)$ case,
\bea
 \, {}^{j}B_{(i, i)}^{p}\ \ =
{{}^{1}B_{i}^{1}\,\over d_i\, d_j}
{1\over F_{ i1 }\left[\matrix{i&i \cr p&p} \right]}\,
{S_{ij}(p)\over S_{11}}
= {{}^{1}B_{i}^{1}\,\over d_i\, d_j}
{1\over F_{ j1 }\left[\matrix{j&j \cr p&p} \right]}\,
{S_{ji}(p)\over S_{11}}\,.
\label{refid}
\eea
Both (\ref{Vc}) and (\ref{refid}) easily extend  beyond the $sl(2)$
case. In particular
restoring all coupling indices the latter formula reads in general
\bea
{}^{j,\zb}B_{(i, i^*)}^{p^*, t}\ = {}^{1}B_{i}^{1}\, \sum_u\,
F_{ 1 p}\left[\matrix{j&j^*\cr j&j} \right]_{\un
\ \ \un}^{u\ 
\sigma_{13}\sigma_{23}(\zb)}\
{S_{ji}^{u \ \sigma_{23}\sigma_{13}(t)}(p)\over S_{1i}}\,.\label{refidg}
\eea
With the help of one of the consequences of the pentagon identity,
(\ref{refidg})  can be inverted and
brought into a form analogous to that of (\ref{refid}).

The coincidence of the two seemingly very different quantities,
the coefficients $B^p$ in the expansion (\ref{IVb}) of the
half-plane bulk field and the
modular matrix $S(p)$ of the torus $1$--point blocks, is quite
surprising and needs a better understanding.  We were led to this
observation trying to find a connection between the two duality
schemes, the one of Moore-Seiberg involving the
torus, the other, of Cardy-Lewellen, involving
the cylinder.  Thus to bring equation
(\ref{eqIa}), derived from the first of the Lewellen
bulk-boundary equations (\ref{eqI}), into a form identical to the
original Moore-Seiberg torus duality relation we  furthermore need to
identify the three fusing matrices,
${}^{(1)}F\,, F\,, {}^{(3)}{F} \,,$
i.e.,
\be
{}^{(1)}F_{b p}\left[\matrix{
k & j \cr a& c} \right]_{\alpha\ \beta}^{\gamma\ t }=
F_{b p}\left[\matrix{
k & j \cr a& c} \right]_{\alpha\ \beta}^{\gamma\ t }\,, \quad
{}^{(3)}F_{p b }\left[\matrix{
c & j \cr a& k} \right]^{\alpha'\ \beta'}_{\gamma'\ t' }
=F_{p b}\left[\matrix{
c & j \cr a& k} \right]^{\alpha'\ \beta'}_{\gamma'\ t' }\,.
\label{fusid}
\ee
This identification is consistent
since in the diagonal case $\calI=\calE$ both  mixed pentagon
identities (\ref{mpt}) and (\ref{inv}) then become  the ordinary
Moore-Seiberg pentagon identity; see also \cite{R},
where the identification of the boundary field OPE coefficients
with the fusing matrices in the $sl(2)$ case was first established by a
more elaborate argument. 

It is now straightforward to show that the first and second lines
in (\ref{eqIa}) reproduce  the two sides of the Moore-Seiberg identity
(\ref{tor}).  Taking into account  (\ref{refid}) and
(\ref{fusid}) the second Lewellen bulk-boundary equation
(\ref{eqII}) is seen also to be a consequence of the first, i.e.,
of the same Moore-Seiberg torus duality relation. To show this, one has to
insert in (\ref{eqII}) the expression (\ref{eqIIc}) for the OPE
coefficients and to compare the equation with (\ref{eqIa}) with
$p=1$, see also Appendix~E.

We thus see that in the diagonal case the two basic bulk-boundary
equations (\ref{eqI}), (\ref{eqII}) are not independent and are
equivalent to one of the basic Moore-Seiberg duality relations.  The third
bulk-boundary equation (\ref{eqIII}), a more general version of
(\ref{eqII}), is an identity which
 involves again only the Moore-Seiberg duality matrices $F\,,
B\,, S$,
and thus can be expected, following the completeness argument of
\cite{MS1}, to be derivable using the basic Moore-Seiberg duality relations.
This in particular implies that any solution of the
set of Moore-Seiberg (chiral) duality relations provides a solution of the
diagonal  boundary CFT equations.

{\small{ Remark: Rewritten in terms of $S(p)$ the diagonal case
Lewellen equation (\ref{eqII})
can be also interpreted as a generalised Verlinde fusion formula
with (non-integral) ``multiplicities'' $\calF_{qk}^l$ given by
some particular $F$ matrix elements.
The matrices $\calF_q$ are ``diagonalised'' by $S(p)$ with the
usual eigenvalues $S_{qi}(1)/S_{1i}(1)$. Because of this they
realise another representation of the usual Verlinde algebra.
This formula, which derives from the Moore-Seiberg torus duality
identity, appears to have been already considered, following a
different motivation, in \cite{BV}.}}

 \medskip

We conclude this section with a few comments on the general
non-diagonal cases. The Cardy-Lewellen boundary CFT can be looked
at as a purely 
``chiral'' alternative  of the usual CFT approach in which we
combine left and right chiral blocks imposing consistency
conditions.  It has its  `price' in that everything effectively
``splits''-- the set $\calI$ is replaced by two ``dual'' sets
$\calV$ and $\calE$ (for type I, while for type II we have to
retain the whole $\calI$ to describe non-scalar fields), there
are   two representations of the Verlinde fusion algebra and a
related new fusion algebra (at least in type I cases); there are
two types of ``chiral vertex operators'', new duality matrices,
in particular a second fusing matrix ${}^{(1)}F\,$ and its
inverse, $  {}^{(3)}{F}, $ along with the standard $F$,
satisfying new duality relations, the mixed pentagon relations,
generalising one  of the basic genus zero polynomial identities;
instead of one relation involving the modular matrix $S(p)$,
there are two independent relations -- the two bulk-boundary
equations  in which the role of  $S(p)$ is taken over by the 
reflection coefficients $B^p$.
It remains to find a consistent solution of the equations at
least in the $sl(2)$ case.  Some of the ingredients are already
known and have been recalled above. In particular the solution
for the $D$-series has just been obtained by Runkel~\cite{Runk}.

\sect{Conclusions and Outlook}

In this paper we have reexamined various aspects of boundary
effects in RCFTs. We have in particular analyzed the consistency
conditions that determine the allowed boundary states and fields
 and their characteristic data, OPE coefficients, etc.
We have seen that boundary conditions are naturally associated
with a graph, or a collection of graphs, whose spectral properties
(eigenvalues)
encode the diagonal spectrum of the bulk theory. This legitimates empirical
observations made previously on the role of graphs in the classification
of RCFTs. We have seen that the torus partition function may be
fully reconstructed from the information contained in these graphs.
We have seen also that in several cases ($\slh(2)$,
$\slh(N)_1$ theories), this approach provides a substantially simpler
route to the classification of RCFTs than the study of bulk properties
(modular invariants\dots).  We have finally seen
that further important information
about some boundary effects ($g$-factors, boundary structure constants)
is also  encoded in the spectral properties (eigenvectors) of these
graphs. The bottom line of this analysis is that a triplet of
matrix algebras ($\n_j$, $\hN_a$, $M_i$) plays a central  role
in the whole discussion.  These algebraic structures have been
also confirmed by the detailed analysis of the 
basic equations of the boundary field theory. 
In the diagonal case the triplet of
algebras reduces to one, the Verlinde fusion algebra.
Accordingly, we have seen that in this case
the basic boundary CFT Lewellen equations can be 
identified with a  set of genus $0$ (the pentagon) and genus 
$1$ duality identities of Moore-Seiberg. This leads to an
identification of some of the basic notions in the two
approaches, namely, the boundary fields OPE coefficients
$^{(1)}F$ and the bulk-boundary reflection coefficients $B^p$,
with the chiral CFT fusing matrix $F$ and the modular matrix
$S(p)$, respectively (see the  text for precise formulae).

The more general representations $\n_i$ of the Verlinde fusion
algebra and the dual pair $\{ \hN\,, M\}$ of associative,
commutative (semisimple) algebras have been introduced in earlier
work on  bulk  (and later on boundary) 
conformal field theories, but it seems
to us that the consistency of the whole scheme now appears in its
full generality and that boundary RCFTs reveal these features in
a simpler and more compelling way than in the bulk. In a loose
sense, the boundary effects expose better the underlying chiral
structure of the theory and its algebraic pattern. This should
certainly not come as a surprise, as this is in the same spirit
as the old connection between open and closed strings. 

The study of a RCFT through its boundary conditions, its algebra triplet, etc,
still requires a lot of work. The derivation of the Cardy equation
relies on a technical assumption that has been only
partially justified, namely the proper definition of unspecialized
characters with linear independence and good modular properties for
general chiral algebras.
Also it would be good to have a better understanding of the
completeness assumption: given a certain number of boundary conditions
satisfying the Cardy equation, is it obvious that we may always
supplement them into a complete set in the  sense discussed in Section 2?
Then many questions have been only partially treated:
Justify in full generality the validity
of expressions~(\ref{subset}, \ref{parent}) which have been
established so far only for particular cases; understand better the
nature and fusion rules of ``twisted'' block representations
that appear in this discussion;
set up a general scheme for the systematic classification of integer
valued representations of fusion algebras;
set up with more rigour the formalism of generalized
chiral vertex operators, their fusing matrices and the ensuing
duality equations as a consistent chiral approach,  alternative to the
Moore-Seiberg scheme,
 etc, such are some of the
outstanding problems that are awaiting a proper treatment.

Also it remains to see how our discussion of boundary
conditions must be generalized in theories where there is no choice
of a common diagonalising matrix $\psi_a^j$ leading to
$\hN$ algebra with integer structure constants. 
In the approach of~\cite{Xu, BE}, in which the 
numbers $\hN$ are integers, one has 
to drop the axiom of commutativity, replacing this algebra
by a non-commutative structure. In that respect,
a better understanding of the relation of  our work with other more
abstract approaches --- Ocneanu theory of subfactors, weak Hopf
algebras --- would be most profitable.

Directions for future work also include the discussion of
other cases: rational or irrational theories at $c=1$, $c=2$,
or $\calN=2$ superconformal  theories are particularly
important cases in view of their physical applications to
condensed matter or to string theory. The generalization to
other types of twisted boundary conditions along the
cycle of the cylinder, as examined recently in~\cite{Ruelle},
might constitute another useful approach. Finally the
parallel discussion of these boundary conditions and
algebraic structures in lattice models should be extremely
instructive and will be the object of a forthcoming publication.



\section*{Appendix~A: The Cardy Equation}
\rnc{\theequation}{A.\arabic{equation}}\setcounter{equation}{0}

In this Appendix, we rederive the Cardy equation (Section 2.2) in the presence
of sources, which have the effect of introducing unspecialized
characters in the partition function.
We restrict to a conformal field theory with a current
algebra. Let $\{J^\alpha\}$ denote the generators in the Cartan
subalgebra, and $\nu_\za$ be ``charges'' coupled to them.
We consider the theory on the cylinder $L\times T$
of Section 2.2, call $w=u+i v$ the local variable, $0\le v\le L$,
$u$ periodic of period $T$,   and modify the energy-momentum tensor $T(w)$
into
\bea
T'(w)&=& T(w) -{2i\pi \over T} \sum_a \nu_\za J^\za(w) -{k\over 2}\sum_\za
\Big({2\pi \nu_\za\over T}\Big)^2 \label{Aaa}\\
\bar T'(\bar w)&=& \bar T(\bar w) -{2i\pi \over T} \sum_\za
\nu_\za\bar J^\za(\bar w)
-{k\over 2}\sum_\za \Big({2\pi \nu_\za\over T}\Big)^2\ . \label{Aab}
\eea
As an elementary calculation shows,
the last term is dictated by the requirement that $T'$ satisfies
the conventional OPE of an energy momentum tensor. The central charge
is not affected by the additional terms.

One then computes the evolution operators in the two channels of
Section 2.2, see Figure~\ref{cardy}.
For the cylinder, mapped to the plane by $\zeta =e^{-2\pi i w/T}$,
the Hamiltonian reads
\bea
 H^{{\rm cyl}}
&=&{1\over 2\pi }\int_0^{-T} du (T'(w)+{\bar T} '(\bar w))=
 {1\over 2\pi }\oint dw (T'(w)+{\bar T} '(\bar w))\\
&=& \big({2\pi\over T}\big) \Big(L_0^{{\rm(P)}}+\bar L_0^{{\rm(P)}}
-{c\over 12} +\sum_\za \nu_\za (J_0^{{\rm (P)}\,\za}
-\bar J_0^{{\rm (P)}\,\za})\Big)
\eea
Note that the additional term in (\ref{Aaa}) and (\ref{Aab})
has not contributed to the integral over a closed cycle.
 Taking into account the fact that on boundary states $L_0^{{\rm(P)}}=
\bar L_0^{{\rm(P)}}$ and $J_0^{{\rm (P)}\,\za}
=-\bar J_0^{{\rm (P)}\,\za}$, we find that
 the first expression of the partition function reads
\be
Z_{b|a}=\bra b|e^{-{4\pi L\over T}(L_0-{c\over 24}+\sum_\za
\nu_\za J_0^\za)} |a\ket
= \sum_{j\in \calE}
\psi_\a^j\, (\psi_\b^j)^*{\chi_j(\tilde q, \nu \tilde \tau) \over S_{j1}}
\ee
where we have defined
\be
\chi_j( q, z):= \tr_{\calV_j} q^{L_0-{c\over 24}} e^{2\pi i\sum_\za
z_\za J_0^\za }\ ,
\ee
and as above $\tilde q=e^{-4\pi{L\over T}}$, hence $\tilde \tau=
2i{L\over T}$.

In the other channel, the time evolution on the strip is
described by the Hamiltonian
\be
H_{ba}= \Big( \int_0^{iL}{dw\over 2\pi i} T'(w)+
\int_0^{-iL} {d\bar w\over 2\pi i}{\bar T}'(\bar w)\Big)
\ee
and upon mapping on the upper half plane $H$ by
$z=e^{\pi{w\over L}}$, we find
\bea
H_{ba}&=&
{\pi\over L}\big(L_0^{{(H)}} -{c\over 24}\big) -{2i\pi \over T}\sum_\za
\nu_\za J_0^{{H}\za} -{L\over 2\pi}k \sum_\za
\Big({2\pi \nu_\za\over T}\Big)^2
\eea
where now the additional piece in (\ref{Aaa}) contributes the last term.
Since the theory with energy momentum tensor (\ref{Aaa}) and (\ref{Aab})
has still
the same operator content and multiplicities $\n_{{i^*}a}{}^b
=n_{ib}{}^a$ as before,we may write
\be
Z_{b|a}=\tr e^{-T H_{ba}}= e^{2\pi k {L\over T}\sum_\za \nu_\za^2  }\
\sum_i \n_{{i^*}a}{}^b \chi_i(q,\nu )
\ee
with $q=e^{-\pi{T\over L}}$.
 We then use the modular transformation of unspecialized
characters (see~\cite{Kac}, page 264):
\be
\chi_i(q,\nu)=e^{2i\pi k \sum {\tilde z_\za^2\over 2\tilde \tau}}
\sum_j S_{ij} \chi_j(\tilde q, \nu \tilde \tau)
\ee
together with the linear independence of the
$\chi_i(q,z)$ to conclude that (\ref{IIi}) is indeed true in full
generality.


\section*{Appendix~B:  \ade\ Diagrams and Intertwiners}
\rnc{\theequation}{B.\arabic{equation}}\setcounter{equation}{0}


\noindent
In this Appendix we establish notations on \ade\ Dynkin
diagrams and on the associated intertwiners.

Let $G$ be a Dynkin diagram of the \ade\ type with Coxeter number
$g$.  It has $n$ nodes that may be coloured with two colours,
{\it i.e.} its $n\times n$ adjacency matrix $G_{ab}$ connects
only nodes of different colours.  This matrix is symmetric and
it  may thus be diagonalized in an orthonormal basis.  We call
this orthonormal basis $\psii{a}{m}$, it is labelled by the node
$\a$  and the exponent $m$ (see Figure~2 and Table~1).  Hence
\be
\sum_\b G_\a{}^b \psii{b}{\s} = 2\cos{{\pi \s \over g}} \,
\psii{a}{\s}\label{Ba}
\ .
\ee
The $\psii{}{}$'s satisfy orthonormality conditions, namely
\bea
\sum_{a}\psii{a}{\s}\psii{a}{s''*}&=&\delta_{\s s''}\qquad
\s,s''\in \Exp(G) \label{IIo}\\
\sum_{\s\in
\Exp}\psii{a}{\s}\psii{b}{\s\,*}&=&\delta_{ab}\label{Bb}
\eea
Because of the 2-colourability of $G$,
one may attach a $\Bbb{Z}_2$ grading $\tau$ to each node $\a$.
One proves that if $\s$ is an exponent, so is
$\sigma(\s)=g-\s$ and the $\psi$'s may be chosen to satisfy
\be
\psii{a}{\sigma(\s)}
 = (-1)^{\tau(a)} \psii{a}{\s} \ . \label{Bc}
\ee
Moreover all graphs having {\it even} exponents, v.i.z. the $A$,
$D_{{\rm odd}}$ and $E_6$ diagrams, have a $\Bbb{Z}_2$
automorphism $\gamma$ acting on their nodes and preserving their
adjacency matrix (i.e. $G_a{}^b=G_{\gamma(a)}{}^{\gamma(b)}$,
this is the natural $\Bbb{Z}_2$ symmetry of these graphs) such
that
\be
\psii{\gamma(a)}{\s}=(-1)^{\tau(\s)} \psii{a}{\s} \ .\label{Bd}
\ee
Finally, one may find in the graph $G$ a distinguished node
labelled $a=1$ such that $\psii{1}{m}>0$ for all $m$.  This
special node is typically an extremal node, i.e. the end of a
branch for the $A$-$D$-$E$ graphs; this is generally the end of a
long leg, but for the $D_{{\rm odd}}$ graphs, for which  we must
choose $1$ as the end point of one of the two short legs.

\medskip
We list hereafter the explicit expressions of eigenvectors
of the various Dynkin diagrams.

The $D_{{g\over 2}+1}$ series are the simplest examples of
orbifold models. Their fundamental graphs can be obtained by
folding the $A_{g-1}$ Dynkin diagram  so that the nodes $
a_i=a_{g-i}\in D_{{g\over 2}+1}\,, i=1,2,3,\dots,
{g\over 2}-1 $ are identified
with the orbit $\{i\}$ of $i$ under the ${\Bbb Z}_2$ automorphism $\sigma$,
$\sigma(i)=g-i$, while the fixed point $i=g/2$ is
resolved into two points $a_{{g\over 2}, \pm}$ on the graph,
denoted $n, n -1$ in Figure~2.  This implies for the adjacency
matrix elements $G_{a_i a_j}=  A_{i\,j}+ A_{i\,\sigma(j)}=
A_{ij}\,,$ for $  i,j\not= {g\over 2}\,,$ and $G_{a_{{g\over2},
\pm} a_j}=A_{{g\over2}\, j}$ and allows us to determine the
eigenvectors $\psi_a^j$ of $G$ in terms of the eigenvectors
$S_{ij}$ of the $A$ adjacency matrix.
To simplify notation we shall use sometimes $\psi_i^j=\psi_{a_i}^j$.

\vskip1cm


\subsection*{{Eigenvectors of the $D_{2l}$ adjacency matrix}}
\bea
\psi_{i}^j &=&\sqrt{2}\, S_{ij} \,, \ \
i,j\not = {g\over 2}\,, \qquad
\psi_{i}^{({g\over 2}, \pm)} = S_{i{{g\over 2}}} \,, \quad
i \not ={g\over 2}\,; \label{orbi} \\
\psi_{{ {g\over 2},\pm}}^j &=& {S_{{g\over 2}j}\over \sqrt{2}}\,,
  \quad j\not = {g\over 2}\,;  \qquad
\psi_{{{g\over 2}, \epsilon}}^{({g\over 2}, \epsilon')}
=S^{ext}_{\{({g\over 2}, \epsilon)\}{\{({g\over 2},
\epsilon')\}}}={1\over 2}\big(S_{{g\over 2}{g\over 2}} +
\epsilon\epsilon'\, i\,\sqrt{(-1)^{l}}\big)\,.
\nonumber
\eea
For  $i$ --odd the orbits $\{i\}$ belong to $\calI^{ext}$ and can
be identified with the subset $T= \{ a_i\,, i=1,3,\dots,  {g\over
2}-2\,, a_{{g\over 2},\pm }\}\,. $ The matrix
$S^{ext}_{\{i\}\{j\}}$ is the extended theory modular matrix.  The
expressions (\ref{orbi}) can be rewritten in the compact form
\be
\psi_a^j =S_{a \{j\}} \, \sqrt{S_{1j} \over
S^{ext}_{\{1\} \{j\}} }\,, \label{universal}
\ee
where $S_{a \{j\}}$ is a rectangular matrix coinciding for
$a\in T$ with $S^{ext}_{a \{j\}}$, while for $a=a_i \not \in
T$, $S_{a_i \{({g\over 2},\pm)\}}=S_{i{{g\over 2}}}=0$ and
$S_{a_i  \{j\}}=\sum_{l\in \{i\}}\, S_{l j}=2  S_{ij}$ for $j\not
=g/2$.
\vskip1cm


\subsection*{{Eigenvectors of the $D_{2l+1}$ adjacency matrix}}
\bea
\psi_{a_i}^j&=&(-1)^{j-1\over 2}\,\sqrt{2}\,S_{ij}\,,\ \
i,j\ne{g\over 2}\,;\quad\psi_{a_i}^{g\over2}=0\,,\ \ i\ne{g\over2}\,;\\
\psi_{{{g\over 2},\pm}}^j &=&(-1)^{j-1\over 2}\, {1\over \sqrt{2}}\,
S_{{g\over 2} j}= {1\over\sqrt{g}} \,, \ \ j\not =  {g\over 2}\,; \quad
\psi_{{{g\over 2},\pm}}^{g\over 2} =\pm {1\over \sqrt{2}}\,.
\nonumber
\eea

The identity node is chosen to coincide with one of the `fork'
nodes $1=a_{{g\over 2},+}$ (denoted by $L$ in Figure~2) so that the
dual Perron--Frobenius eigenvector $\psi_{1}^j=\psi_{{g\over
2},+}^j$ has  positive entries  (while $\psi_{a_1}^{g\over 2}
=0$).  The `fundamental' node $f$ is identified with ${{g\over
2}-1}$, i.e., $G=\hat{N}_{{{g\over 2}-1}}$. Also $a^*=a$ for all
$a$,  while $\gamma(a_{{g\over 2},+})=a_{{g\over 2},-} $.

\vskip1cm
Next we display  the eigenvectors of the exceptional $E_r$ Dynkin
diagrams as a matrix $\{\psi_a^{j}\}$, with the row index $a$
running over the nodes, following the numbering of Figure~2, and
the column index $j$ over the exponents in the same order as in
Table 1. There too, $S_{ij}$ denote
the eigenvectors of the diagonal graph adjacency matrix $A$
with the same Coxeter number.

\medskip

\subsection*{{Eigenvectors of the $E_6$ adjacency matrix}}
\bigskip
\def\psun{\oh\sqrt{{3-\sqrt{3}\over 6}}}
\def\psun{\oh\sqrt{{3-\sqrt{3}\over 6}}}
\def\psde{\oh\sqrt{{3+\sqrt{3}\over 6}}}
\def\psde{\oh\sqrt{{3+\sqrt{3}\over 6}}}
\def\pstr{\oh\sqrt{{3+\sqrt{3}\over 3}}}
\def\pstr{\oh\sqrt{{3+\sqrt{3}\over 3}}}
\def\pssi{\oh\sqrt{{3-\sqrt{3}\over 3}}}
\def\pssi{\oh\sqrt{{3-\sqrt{3}\over 3}}}

\be
(\psi^{j}_a)=\pmatrix{
a & \oh  & b & b & \oh  & a \cr
b & \oh  & a &-a &-\oh  &-b \cr
c &  0   &-d &-d &  0   & c \cr
b &-\oh  & a &-a & \oh  &-b \cr
a &-\oh  & b & b &-\oh  & a \cr
d &  0   &-c & c &  0   &-d \cr
}
\ee
where
$a=\psun \,,b=\psde\,, c=\pstr \,,d=\pssi $ are determined from
$\psi_1^i=   \sqrt{ S_{1 i}\, \sum_{j\in\rho}\,S_{j i} }\,,$
$\psi_6^i=(S_{4 i}+S_{8 i})\sqrt{ S_{1 i}\over
\sum_{j\in\rho}\,S_{j i} }$   for  $\i\in \calE\,,
\rho=\{1\,,7\}\,.$
\vskip1cm

\subsection*{{Eigenvectors of the $E_7$ adjacency matrix}}
\bigskip

\be
(\psi^{j}_a)=\pmatrix{
a & c & b & {1\over\sqrt{3}} & b & c & a\cr
e & f & d &     0            &-d &-f &-e \cr
c & b &-a &-{1\over\sqrt{3}} &-a & b & c  \cr
f &-d &-e &     0            & e & d &-f \cr
{1\over \sqrt {6}} & -{1\over\sqrt{6}} & {1\over \sqrt {6}} & 0 &
{1\over \sqrt {6}} & -{1\over\sqrt{6}} & {1\over \sqrt {6}} \cr
d &-e & f &     0            &-f & e &-d  \cr
b &-a &-c & {1\over\sqrt{3}} &-c &-a & b \cr }
\ee
where $a\,,b\,,c\,,d\,, e\,,f\,$ are determined from
$\psi_1^j= \sqrt{ S_{1j} \sum_{i\in \rho}S_{ij} }$
where $\rho= \{1,9,17\}$, 
and $\psi_2^j= {S_{2j}\over S_{1j}}\, \psi_1^j\,$
(The values in the $5$'th row come from
$\psi_5^j= \sqrt{2}\,S_{6 j}$ for  $j=1,5,7$ .)
Explicitly,
\bea
\begin{array}{ll}
a= [18+12 \sqrt{3} \cos{\pi\over 18}]^{-{1\over 2}}, \qquad\qquad
&d= [12\big(1+\cos{\pi\over 9}\big)]^{-{1\over 2}} \\
b= [18+12 \sqrt{3} \cos{11 \pi\over 18}]^{-{1\over 2}} \qquad\qquad
&e= [12\big(1+\cos{5\pi\over 9}\big)]^{-{1\over 2}} \\
c= [18+12 \sqrt{3} \cos{13 \pi\over 18}]^{-{1\over 2}}, \qquad\qquad
&f= [12\big(1+\cos{7\pi\over 9}\big)]^{-{1\over 2}} \ .
\end{array}
\eea

\vskip1cm

\subsection*{{Eigenvectors of the $E_8$ adjacency matrix}}
\bigskip
\def\xb{\bigg[15+\sqrt{75-30\sqrt{5}}\bigg]^{-{1\over 2}}}
\def\xh{\bigg[15-\sqrt{75-30\sqrt{5}}\bigg]^{-{1\over 2}}}
\def\xg{\bigg[15+\sqrt{75+30\sqrt{5}}\bigg]^{-{1\over 2}}}
\def\xe{\bigg[15-\sqrt{75+30\sqrt{5}}\bigg]^{-{1\over 2}}}
\def\xa{\bigg[{15(3+\sqrt{5})+\sqrt{15(130+58\sqrt{5})}\over 2}\bigg]^{
-{1\over 2}}}
\def\xc{\bigg[{15(3+\sqrt{5})-\sqrt{15(130+58\sqrt{5})}\over 2}\bigg]^{
-{1\over 2}}}
\def\xf{\bigg[{15(3-\sqrt{5})+\sqrt{15(130-58\sqrt{5})}\over 2}\bigg]^{
-{1\over 2}}}
\def\xd{\bigg[{15(3-\sqrt{5})-\sqrt{15(130-58\sqrt{5})}\over 2}\bigg]^{
-{1\over 2}}}

\be
(\psi^{j}_a)=\pmatrix{
a & f & c & d & d & c & f & a \cr
b & e & h & g &-g &-h &-e &-b \cr
c & d &-a &-f &-f &-a & d & c \cr
d & a &-f &-c & c & f &-a &-d \cr
e &-h &-g & b & b &-g &-h & e \cr
f &-c & d &-a & a &-d & c &-f \cr
g &-b & e &-h &-h & e &-b & g \cr
h &-g &-b & e &-e & b & g &-h \cr }
\ee
where $a\,,b\,,c\,,d\,, e\,,f\,,g\,, h$ are determined from
$\psi_1^j= \sqrt{S_{1 j}\, \sum_{i\in\rho}\, S_{ij} }\ $
and $\psi_2^j= {S_{2j}\over S_{1j}}\, \psi_1^j\,$
for  $\j\in \calE\,,$  $ \rho=\{1\,,11\,,19\,,29\}\,.$
Explicitly
\bea
\begin{array}{ll}
a = \xa \qquad\qquad b&= \xb \\
c = \xc \qquad\qquad e&= \xe \\
d = \xd \qquad\qquad g&= \xg \\
f = \xf \qquad\qquad h&= \xh \ .
\end{array}
\eea

\medskip
To such a graph $G$, one then attaches matrices $V_i$ as follows.
The case of reference is the $A_{g-1}$ diagram of same Coxeter
number $g$ as $G$.  For this $A$ graph, both the nodes and the
exponents take all integer values in $\{1, \cdots g-1\}$.  The
$\psii{}{}$'s are then nothing other than the entries of the
(symmetric, unitary) matrix $S$ of modular transformations of
characters of the affine algebra $\slh_2$ at level $g-2$
\be
\psii{i}{(A)\,i'}=S_{i i'}=\sqrt{{2\over g}} \sin{\pi ii'\over
g}\ , \label{Bf}
\ee
in terms of which the fusion coefficients
$ N_{i_1 i_2}{}^{i_3} $ may be expressed through Verlinde formula.
Note also that
\be
N_{g-i\, i_1}{}^{g-i_2}=N_{ii_1}{}^{i_2} \label{Bg}
\ee
because of property (\ref{Bc}) applied to $\psi^{(A)}=\hS$.
\medskip

We now return to the graph $G$ of Coxeter number $g$.
The fused adjacency matrices
$V_i$ with $i=1,\ldots,g-1$ are $n\times n$ matrices
defined recursively by the $sl(2)$ fusion algebra
\be
V_i=V_2\, V_{i-1} -V_{i-2}\qquad 2<i\le g \label{Bh}
\ee
and subject to the initial conditions $V_1=I$ and $V_2=G$.
(One may see that $V_g=0$). The matrices $V_i$
are symmetric and mutually commuting with entries given by a
Verlinde-type formula
\be
V_{ia}{}^b=(V_i)_\a{}^b
=\sum_{m\in \Exp(G)}{S_{im}\over
S_{1 m}}\;\psii{a}{m}\psii{b}{m\, *} \ . \label{Bi}
\ee
Regarded as $(g-1)\times n$ rectangular matrices, for $\a$ fixed,
the $V_{ia}{}^b$ intertwine the $A$ and $G$ adjacency matrices
\be
\sum_{i'} A_i{}^{i'} V_{i'a}{}^b = \sum_{b'} V_{ia}{}^{b'}
G_{b'}{}^b\ . \label{Bj}
\ee
Regarded as $n\times n$ matrices, the $V_i$ satisfy not only their
defining relation (\ref{Bh}) but also the whole $\slh(2)$ fusion
algebra  (\ref{IIif})
\be
V_{i_1} V_{i_2}=\sum_{i_3} N_{i_1i_2}{}^{i_3} V_{i_3}\ .
\label{Bk}
\ee

{}From their recursive definition and initial conditions, it
follows that the entries of the $V$ matrices are {\it integers}.
What is not obvious is that these entries are {\it non-negative }
integers.  This follows either from a direct inspection or from
an elegant group theoretic argument due to Dorey~\cite{Dorey}.
We refer the reader to~\cite{DFZ1,PZh} for the explicit
expressions of these intertwiners.  As a consequence of the
existence of the  automorphism $\gamma$ defined above (\ref{Bd}),
a $\Bbb{Z}_2$ symmetry on  $A, D_{{\rm odd}}, E_6$ graphs and the
identity for $D_{{\rm even}}, E_7,E_8$, one has
\be
V_{g-s\, a}{}^{\gamma(b)}=V_{sa}{}^b \ .\label{Bl}
\ee

Using (\ref{Bh}) and (\ref{IIzc}) ({\it i.e.} in the notations of
this Appendix, $V_i \hN_a=\sum_b V_{ia}{}^b \hN_b$) one can
express the graph algebra $\hat{N}_a$ matrices for all but the
$D_{\rm even}$ cases as polynomials of $V$'s with integer
coefficients This in particular ensures that they have integer
matrix elements.  For  $D_{\rm even}$   one of the extended
fusion algebra generators $N^{ext}_{\{{g\over 2}, \pm\}}$ has to
be added since $V_{g\over 2}= N^{ext}_{\{{g\over 2},
+\}}+N^{ext}_{\{{g\over 2}, -\}}$ while $N_i=V_i=V_{g-i}$ for
$i=1,2,...,{g\over 2}-1\,.$ For the three exceptional cases $E_r$
we have $N_i=V_i\,, i=1,2,...r-3\,,$ $N_{r-1}=V_{r-1}-V_{r-3}\,,$
$N_{r-2}=V_{r}-V_{r-4}\,,$ $N_{r}=V_{r-2} +V_{r-4}-V_{r}\,,$
which translates into relations between the eigenvalues and given
$\psi_1^j$ allows to express any $\psi_a^j$ in terms of the
modular matrix $S$ elements.

\medskip
Over recent years, these matrices have  made repeated
appearances in a variety of problems. Originally introduced in
the discussion of local height probabilities in lattice
models~\cite{Pa} and of boundary partition functions~\cite{SaBa,
DFZ1} (see below), they have also appeared in the following contexts:
\renewcommand{\theenumi}{(\roman{enumi})}
\begin{enumerate}
\item The ``cells'' or intertwiners of Boltzmann weights of
height models~\cite{DFZ1, PZh}.
\item The decomposition of the representation of the Temperley-Lieb
algebra on the space of paths from $\a$ to $\b$ on graph $G$ onto
the irreducible ones on the paths from 1 to $s$ on graph
$A_{g-1}$ \cite{PaSa} according to
$R^{(G)}_\a{}^b =\oplus_s V_{sa}{}^b R^{(A)}_1{}^s$.
\item The counting of ``essential paths'' on graphs~\cite{Ocn}; see
also recent mathematical work by Xu, B\"ockenhauer and Evans
\cite{Xu, BE}.
\item The expression of the blocks of the partition function
(\ref{IIa}) as (\ref{subset}, \ref{parent}), see Section 3.3.
\omit{
\item For an appropriate choice of the node 
$f$, the integers
$V_{sf}{}^b$ are the coefficients of the Kostant polynomials, in
the decomposion of representations of $SU(2)$ of dimension $s$
onto irreducible representations $\b$ of the finite group
associated with $G$  in the McKay correspondence~\cite{McK, Ko}
($f$ is the node connected to the node representing the
identity representation and deleted in passing from the affine
Dynkin diagram to the ordinary one).
}
\item
The $sl(2)$ intertwiners  appear in the computation
of  the multiplicities $m_{s'}^b$ 
of an irreducible representation $b$  of the finite group, 
associated with $G$  in the McKay correspondence~\cite{McK, Ko},
in the $SU(2)$ representations of dimension $s'$~\cite{DFZ0}.
Namely the coefficients of the
 Kostant polynomials in the generating function
$F_b$ of these multiplicities are given for  a non trivial $b$
by $\sum_c\, G_{0c} V_{s c}^b$, where $G_{ab}$ is the adjacency matrix
of the affine Dynkin diagram and $a=0$ is the affine node deleted
in passing from the affine
Dynkin diagram to the ordinary one. The proof of this fact
is reduced to the recursive relation (\ref{Bh}).
\item These same entries seem to appear ubiquituously in the
description of $S$-matrices of affine Toda
theories~\cite{BradenCDS} and in the description of the
excitation spectrum of integrable lattice models~\cite{MCO,BaS,Suz}.
\end{enumerate}

\section*{Appendix~C:
Uniqueness of the Boundary Conditions of Minimal Models}
\rnc{\theequation}{C.\arabic{equation}}\setcounter{equation}{0}

\subsection*{C.1. Matrices with spectrum $\gamma <2$}
We first recall general results on symmetric matrices
with non-negative integer entries and with eigenvalues between
$-2$ and $2$.

 It is a standard result that symmetric matrices
with non-negative integer entries and eigenvalues $\gamma \in
]-2, 2[$ may be classified. A lemma of Kronecker
asserts that the eigenvalues are of the form $2 \cos {p_i\pi\over h_i}$
for integers $p_i$ and $h_i$ and for the largest one(s), $p=1$.
One may regard any such matrix
as the adjacency matrix of a graph. Irreducible matrices
correspond to connected graphs, and by an abuse of language
one may call a matrix bicolourable if the graph has that property.
One proves~\cite{GHJ}
 that any irreducible bicolourable symmetric
matrix with spectrum in $]-2, 2[$ is the adjacency matrix
of one of the simply laced Dynkin diagrams of type \ade.

If one relaxes the assumption of bicolourability, with any symmetric
non-bicolourable irreducible
matrix $G$ one may associate a bicolourable symmetric matrix with a block form
$G'=\pmatrix{0&G\cr G&0\cr}$. The corresponding graph is irreducible
and  has a $\Bbb{Z}_2$ symmetry that exchanges the two colours.
 Any eigenvalue $\gamma$ of $G$ gives rise to
two eigenvalues $\pm \gamma$ for $G'$ and one thus concludes that $G'$
is of \ade\ type, and its irreducibility forces $G'=A_{2p}$.
Its $\Bbb Z_2$ quotient
$G$ is what we call the tadpole graph $T_p=A_{2p}/\Bbb Z_2$.

 Finally if one relaxes the assumption of irreducibility, one
concludes that any matrix (with non-negative entries and spectrum
between $-2$ and $2$) is the direct sum of \ade\ or tadpole graphs
$$ G=\oplus G_i \qquad G_i\ {\rm of\ } \ade\ {\rm or\ tadpole\ type }$$
and this decomposition is unique, up to the permutation
of factors.  The uniqueness  may be easily
proved by induction on the number of terms or on the dimension of the matrix:
Given a matrix $G$, one first identifies its largest eigenvalue,
of the form $\gamma_1=2 \cos {\pi\over h_1}$. By the previous
statement, there is an \ade\ or tadpole graph $G_1$ with Coxeter number
$h_1$ and exponents $m_i$,
such that all its eigenvalues $2 \cos {m_j\over h_1}$ appear in the spectrum
of $G$. Thus $G=G_1\oplus G''$, and one may apply on $G''$ the
induction hypothesis. If $\gamma_1$ has multiplicity 1, this suffices to
establish the uniqueness of the decomposition (up to permutations),
while the case where $\gamma_1$ has non-trivial multiplicity is
also easily dealt with.
The uniqueness of this decomposition implies a property used several
times in the text, namely that the spectrum (between $-2$ and
$2$) determines the form of the matrix up to a permutation of
its rows and columns.

\medskip

\noindent

\subsection*{C.2. Representatives of $n_{12}$ and $n_{21}$}

 We now return to minimal models.

\noindent{\bf Explicit form of $n_{12}$ and $n_{21}$}\\
It is convenient to work in a basis different from that used
in~(\ref{jbsduty}). In the basis $r_1=1,\cdots, p$, $a\in G$,
the second term in (\ref{Minf}) does not contribute to $n_{12}$
since $N_{2p\,r_1}{}^{r_2}=\delta_{r_1,2p+1-r_2}=0$ for $1\le r_1,r_2\le p$.
Thus $n_{12}= {I}_p \otimes V_2={I}_p \otimes G$,
where ${I}_p$ is the $p$ dimensional unit matrix
\be
n_{12}={I}_p \otimes G =\pmatrix{G  & & & \cr
                  & G & & \cr
                  &  & \ddots & \cr
                  & & & G\cr} \ . \label{nonetwo}
\ee
As for $n_{21}$,  in the same basis, the second term of (\ref{Minf})
receives a contribution only from $r_1=r_2=p$, namely $(N_{2p-1})_p{}^p=1$,
while $V_{g-1}=\Gamma$, the matrix that realizes the
 automorphism $\gamma$:
\be
\Gamma_a{}^b=\delta_{a\gamma(b)} \ .
\ee
Thus one finds that
$$ n_{21} =\pmatrix{0  & I_n & & & \cr
		   I_n&  0 &I_n & & \cr
		       & \ddots & \ddots & I_n & \cr
		         && I_n & \Gamma\cr } \ .$$
After conjugation by a block-diagonal matrix  
with  $\Gamma$ and $I_n$ in alternating positions, which leaves
the form (\ref{nonetwo}) of $n_{12}$ unchanged,
$n_{21}$ may be recast in the form
\be
n_{21} = \pmatrix{0 & \Gamma && \cr
	  		\Gamma & 0 &\Gamma & \cr
			 & \ddots& \ddots & \Gamma \cr
			 & & \Gamma& \Gamma\cr} = T_p \otimes \Gamma \ ,
\label{ntwoone}
\ee
in terms of the tadpole $T_p$ adjacency matrix.
All the other $n_{rs}$ are obtained as universal polynomials
of the two matrices $n_{12}$ and $n_{21}$.


\medskip
\noindent{\bf Uniqueness of the form of $n_{12}$ and $n_{21}$}\\
Conversely, suppose we only know that the representation $n_{rs}$
has a spectrum specified by the set of exponents $\calE$.
We want to prove that there exists a basis in which
$n_{12}$ and $n_{21}$ take the forms (\ref{nonetwo})  and
(\ref{ntwoone}).

We first make use of the property that the set
$\Exp(G)$ is stable modulo the Coxeter number $g$ of $G$
under multiplication by any integer coprime to
$g$ and is also stable under the reflection $s\to g-s$.
We then find that the spectrum of
$n_{12}$ is made of $p$ copies of $\Exp(G)$.
As explained above in C.1, this implies that
in some basis
\be
n_{12} = {I}_p \otimes G\ . \label{nunde}
\ee

\def\r{r'}\def\s{s'}

For the other generator $n_{21}$,
one observes first that the set of numbers that appear in
(\ref{Minds}), namely
$\Big\{ 2\cos{\pi g \r\over 2p+1} \Big\}$, $\r=1,3 \cdots, 2p-1$,
is simply the set $\Big\{(-1)^{g+1}
 2\cos{\pi \r'\over 2p+1} \Big\}$,  $\r'=1, 3, \cdots, 2p-1$,
which is $(-1)^{g+1}$ times the spectrum of the tadpole $T_p$.
According to (\ref{Minds}), this has to be multiplied by
$(-1)^{\s}$, as $\s$ runs over the exponents of $G$.
Thus if $g$ is even (which is the general case except when $G=A_{2l}$)
the spectrum of $n_{21}$ is made of as many copies
of that of $T_p$ (resp. $-T_p$) as there are odd (resp. even)
exponents in $G$. For $A_{2l}$ which has as many even as odd
exponents, the same conclusion is still correct!
Finally one notices that these signs are just the eigenvalues
of the $\Gamma$ matrix, and one thus concludes that
\be
n_{21} \sim T_p \otimes \Gamma \ ,
\ee
where the sign $\sim$ means that it holds in some basis
obtained from that of (\ref{nunde}) by a simultaneous permutation
of rows and columns. From this  expression, one can see that
\bea
&&n_{21} \mbox{ has no row or column with more than two $1$s}, \label{ndeun}\\
&& n_{21} \mbox{ has exactly  $n$ rows and columns with one $1$}, \label{xndeun}
\eea
 properties invariant under permutations of rows and columns.

We also know that $n_{21}$
must commute with $n_{12}$. In a basis in which
$n_{12}$ takes the form (\ref{nunde}),
$n_{21}$  may thus be regarded as made of
$n\times n$ blocks that commute with $G$.
We shall combine these facts about $n_{21}$ as follows:
\begin{itemize}
\item A non-vanishing matrix $X$ with elements in $\Bbb{N}$ which
 commutes with $G$ cannot have a row or a column
of zeros. {\it Proof}: let $\psi^1$ be
the Perron-Frobenius eigenvector of $G$,   $GX=XG$ implies that
$X \psi^1$ is an eigenvector of $G$ with the same eigenvalue, hence
proportional to $\psi^1$,
$X\psi^1= c \psi^1$, with $c\ne 0$ since the entries of both
$X$ and of $\psi^1$ are non-negative.
If $X$ had a vanishing row, $X\psi^1$ would have a
vanishing component,
which is impossible for the Perron-Frobenius eigenvector.
If $X$ has a vanishing column, one repeats
the argument with $X^T$.
\item Any matrix $X$  with elements in $\Bbb{N}$ which
 commutes with $G$ and which appears in the block decomposition
of $n_{21}$   cannot have more than one $1$ per row or column.
{\it Proof}: If a matrix $X$ with more than one $1$ in a row (respectively
column) was a block of $n_{21}$,
because of the property~(\ref{ndeun}), all the other blocks on
the left or the right (respectively above or below) of $X$ would have to
have at least one vanishing row (respectively column), which is impossible
by (i) above, or to vanish altogether. In the latter case,
after a possible reshuffling of rows and columns leaving (\ref{nunde})
invariant, one would have
\be
{\rm either}\quad n_{21} =\pmatrix{X & 0 & \cdots & 0\cr
		    0 &  &  & \cr
		    \vdots & & & \cr
		    0 & & & \cr} \quad {\rm or}\quad
n_{21}=    \pmatrix{0 & X & 0 & \cdots \cr
		    X^T & 0 & 0 & \cdots   \cr
		    0 & 0 & & \cr
		    \vdots &\vdots & & \cr}
\ee
 which would lead to a
pair $n_{12},n_{21}$ reducible in the same basis.
\item It follows that the matrices that may appear as blocks
in the decomposition of $n_{21}$ must be matrices with one $1$ on
each row and column, i.e. permutation matrices that commute with
$G$. These permutation matrices are the symmetries of the Dynkin
diagram, and thus are readily listed:
\bea
X=& I ,\ \Gamma  &\hbox{ if } G=A_n, D_{2q+1}, E_6 \\
=& I,\ \Gamma_i, i=1,\cdots, 5 &\hbox{ if } G=D_4 \\
=& I,\ \Gamma' &\hbox{ if } G=D_{2q}, q>2 \\
=& I &\hbox{ if } G=E_7,E_8
\eea
Here $\Gamma_i$ denote the 5 non-trivial permutations of the nodes of the
$D_4$ diagrams, and  the matrix $\Gamma'$  exchanges
the two end points of  $D_{2q}$, $q>2$.
\end{itemize}

 One then  demands that the symmetric matrix $n_{21}$
 made of such blocks is irreducible and satisfies 
(\ref{nunde})-(\ref{xndeun}). This implies that
at most one non-vanishing block appears on the diagonal.
Consistency with the form $n_{21}\sim T_p\otimes \Gamma$
leaves as the only possibility
\be
n_{21}=\pmatrix{0 & X_1 & 0 & \cdots & \cr
	         X_1^T& 0  & X_2 & 0 & \cr
		 0  & X_2^T&\ddots&\ddots& \cr
		  & 0& \ddots  & & X_{p-1}\cr
		 0 & & & X_{p-1}^T & \Gamma \cr}
\ee
where $X_1$, $X_2$, \dots, $X_{p-1}$ are chosen among
the symmetry matrices of $G$.
A final permutation of rows and columns by a block
diagonal matrix ${\rm diag}( Y_1, Y_2, \cdots, Y_{p-1} , I)$
brings $n_{21}$ into the form $n_{21}= T_p \otimes \Gamma$
while leaving the form (\ref{nunde}) of $n_{12}$ unchanged,  provided
$Y_j=\Gamma Y_{j+1}X_{j}^T$, hence $Y_j = \Gamma^{p-j} X^T_{p-1}X^T_{p-2}
\cdots X^T_{j}$. Then both $n_{12}$ and $n_{21}$ have their canonical
forms (\ref{nonetwo}),(\ref{ntwoone}).  Q.E.D

\subsubsection*{Remark}
Although it is not required for the present analysis,
it may be interesting to look at 
the commutant of matrices of \ade\ type.\par
For any $G$ of \ade\ type, with the exception
of $D_{even}$, all eigenvalues are distinct. It follows
that any matrix $X$ that commutes with $G$ may be diagonalized
in the same basis as $G$ and consequently be written as
a polynomial of $G$, i.e. as
a linear combination of $I, G, G^2, \cdots, G^{n-1}$.
In order to look at cases where entries
of $X$ are requested to take values $0$ or $1$ only, and with
constraints on the number
of $1$'s, it is advantageous
to use rather the basis of fused graph matrices: $X$ is a linear combination
of the linearly independent matrices $\hN_1=I, \hN_2=G, \cdots,
\hN_{n}$.  The  $D_{{\rm even}}$  case is slightly more involved, since
the matrices that appear naturally are not independent.

\smallskip

The commutant of an \ade\ matrix is: 
\begin{itemize}
\item 
A linear combination of the graph fusion matrices $\hN_a$ for
$G=A, D_{{\rm odd}}, E_6, E_7, E_8$.
\item 
A linear combination of the $\hN_a$ and of two of the three 
matrices $\Sigma_{ab}$,
$a\ne b=1,3,4$, that exchange two of the three extremal  points of the
$D_4$ graph. 
\item 
A linear combination of the $\hN_a$
 and of the two matrices $X=\Gamma' \hN_{{2q}}$ and $Y=
\hN_{{2q}}\Gamma'$, where the matrix $\Gamma'$ exchanges
the two end points, for $G=D_{2q}$, $q>2$.
\end{itemize}

%
%
%
\section*{Appendix~D:
\rnc{\theequation}{D.\arabic{equation}}\setcounter{equation}{0}
$\widehat{sl(3)}$ Modular Invariants and  Graphs}

\pagestyle{plain}

The WZW $\widehat{sl(3)}$ theories may be discussed along the same
lines as in Sections 2 and 3. Solutions $n_i$ to the Cardy equation
are associated with
graphs, with specific  spectral properties: their eigenvalues
are given by ratios of elements of the modular $S$ matrix
labelled by weights of
the diagonal spectrum ${\cal E}$  of the bulk theory.
Conversely, spectral properties and the fact the $n$'s  form
a representation of the fusion algebra are not restrictive enough to
yield the list of possible bulk spectra, as occurred in
$\widehat{sl(2)}$  (up to the unwanted ``tadpole'' graphs). There are
indeed many solutions, i.e. graphs and representations of the fusion algebra,
that must be discarded as not corresponding to a modular invariant partition
function in the list (Table 2) of Gannon~\cite{Ga}: see~\cite{DFZ1} for
such extra solutions.
We may summarise the salient features of the analysis as follows,
see also the accompanying Tables 2 and 3  and Figures
\ref{ADgraphs} and \ref{Egraphs}.
\begin{itemize}
\item At least one graph (or rather one set of $n$ matrices)
has been identified for each bulk theory, i.e. each modular invariant.
 But it is not known if this list of graphs and $n$'s is exhaustive.
\item  Note that the hypothesis of 3-colourability of the graphs
that looked natural on the basis of the $sl(2)$ case has to
be abandonned if we want to cover all cases. This is manifest on Table 3
where it appears that in some cases (namely $A^{(n)*}$, and  $D^{(n)}$,
$n$ not a multiple of $3$, and $E^{(8)*}$, the set $\cal E$ is {\it not}
invariant under the automorphism $\sigma$ of  (\ref{sigmaa}), as it should
be if the graph was 3-colourable.
\item There are a few pairs or even triplets of isospectral graphs, i.e.
different sets of $n$'s that give distinct solutions of the Cardy equation
for a given bulk theory. These graphs/representations should not only
describe different sets of complete orthonormal sets of
boundary conditions for that bulk theory, but also presumably
be associated with different operator algebras and lattice realisations.
\item In Table 3, which summarises the state of the art, we have
also indicated if the graph is of  type I or type II, following the
discussion of Section 3. Some hybrid cases are also encountered, in which
the $M$ and $\hN$ structure constants are both non-negative, but the $\hN$
algebra has no subalgebra isomorphic to some extended fusion algebra.
\end{itemize}

\nc{\Gs}{\sigma}
\nc{\Gl}{\lambda}
\nc{\GL}{\Lambda}

\subsection*{Notations and footnotes for Tables 2 and  3  }\par

(Shifted) weights of $SU(3)$ $\Gl=(\Gl_1,\Gl_2):= \Gl_1 \GL_1+\Gl_2 \GL_2$,
where $\GL_1$, $\GL_2$ are the fundamental weights
of $SU(3)\,, $ $\lambda^*=(\lambda_2,\lambda_1)\,,$
triality  $\tau(\Gl):= (\Gl_1-1) + 2(\Gl_2-1) \equiv \Gl_1 -\Gl_2
\quad {\rm mod}\ 3$\,.\\
 $Q$ is the set of weights  of triality zero.
\\
Weyl alcove of shifted level, or ``altitude'',  $n:= k+3\,,$\ \

$P_{++}^{(n)}=\{\GL=\Gl_1 \GL_1+\Gl_2 \GL_2 \ | \ \Gl_1,\Gl_2 \ge 1,\
\Gl_1 +\Gl_2 \le n-1 \}\,. \quad $\\
Automorphism $\Gs$ of $P_{++}^{(n)}$
\be
\Gs (\Gl_1,\Gl_2):= (n-\Gl_1-\Gl_2, \Gl_1)\ . \label{sigmaa}
\ee

\renewcommand{\theenumi}{(\alph{enumi})}

\begin{enumerate}
\item One of the two connected parts of the fused Dynkin diagram of
type $A_{n-1}$ (the one that possesses the exponent $(1,1)$).
It looks different depending on whether $n$ is even or odd.
The $M,\hN$ algebras of ${\cal A}^{(n)*}$ are positive,
as they follow simply from
the Verlinde fusion algebra $N$ of $\slh(2)$. If $n$ is odd,
${\cal A}^{(n)*}$ is the connected component of the graph of
adjacency matrix $A_{n-1}^2 -I$
made of the nodes $a$ odd (integer spin in $sl(2)$).  For a
triplet of ``exponents'' $\lambda=(l,l)$, $\mu=(m,m)$
and $\rho=(r,r)$ of ${\cal A}^{(n)*}$,
$M_{\lambda,\mu}{}^{\rho}=N_{l m}{}^r+N_{l m}{}^{n-r}$,
and $\hat N_{ab}{}^c$ is the restriction of the Verlinde
$A_{n-1}$ algebra  to odd $a,b,c$. For even $n$, the $M$ and $\hat N$
algebras of ${\cal A}^{(n)*}={I} +A_{{n\over 2}-1}$ coincide with
the Verlinde algebra of $A_{{n\over 2}-1}$.
\item The  {\it orbifold } of ${\cal A}^{(n)}$, see \cite{Kos}.
\item The  ordinary ${\Bbb Z}_3$ {\it fold } of ${\cal A}^{(n)}$.
\item The unfolded (and 3-colourable) version of  ${\cal A}^{(n)*}$.
Their adjacency matrix is a tensor product by
 the permutation matrix $\sigma_{123}=\mbox{\tiny $\pmatrix{0&1&0\cr
0&0&1\cr 1&0&0
\cr}$}$; their $M$ and $\hat N$ algebras are simply obtained
from those of  ${\cal A}^{(n)*}$, thus also $\ge 0$\par
\item 
The ${\Bbb Z}_3$ fold of  ${\cal E}^{(8)}$. 
\end{enumerate}
\clearpage

%
%
%
%

\begin{figure}[ht]
\vglue-45mm\hglue-45mm
\centering\mbox{\epsfxsize =9in\epsfbox{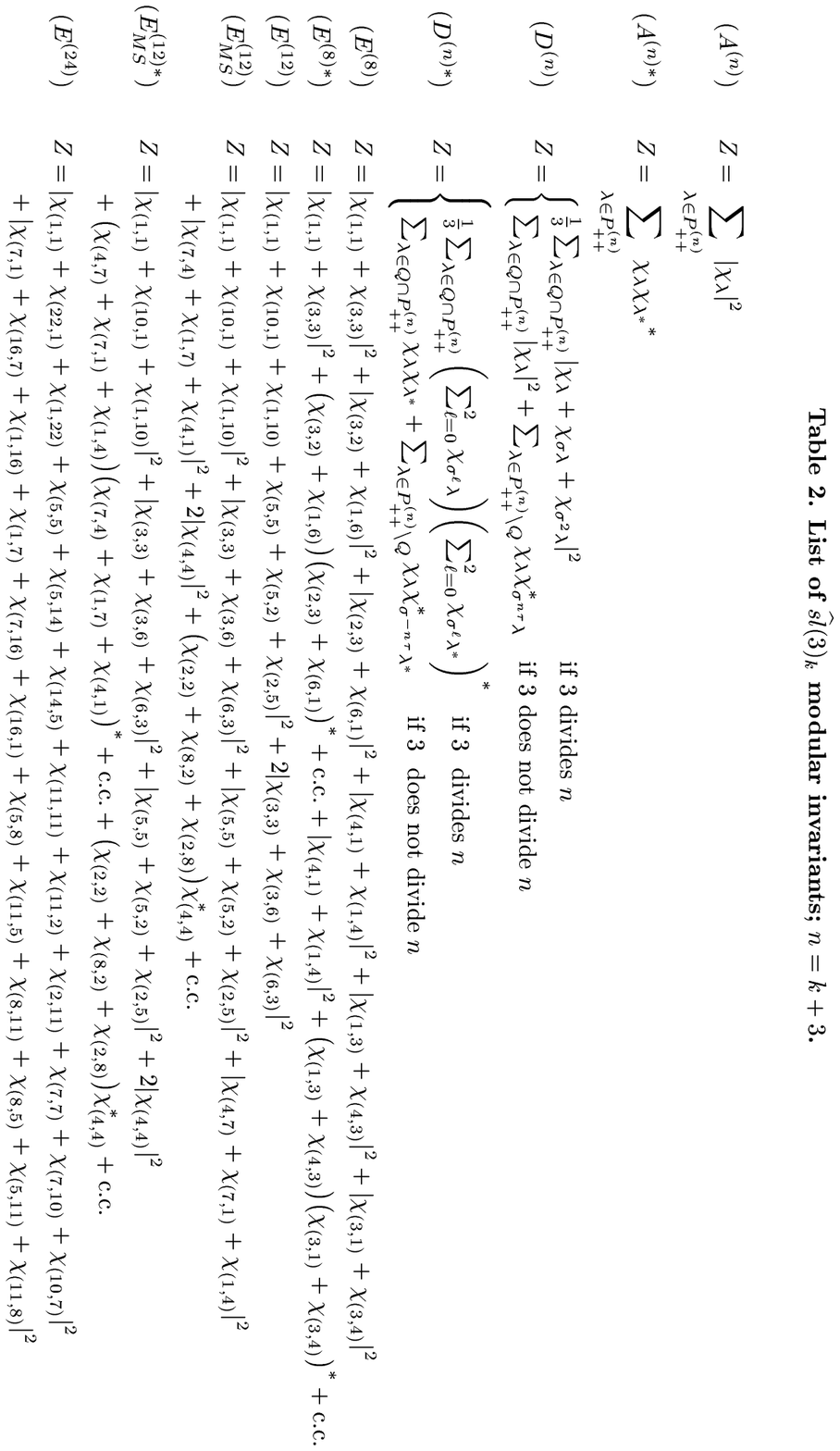}}
\end{figure}

\clearpage
\begin{figure}[ht]
\vglue-45mm
\centerline{\epsfxsize =9in\epsfbox{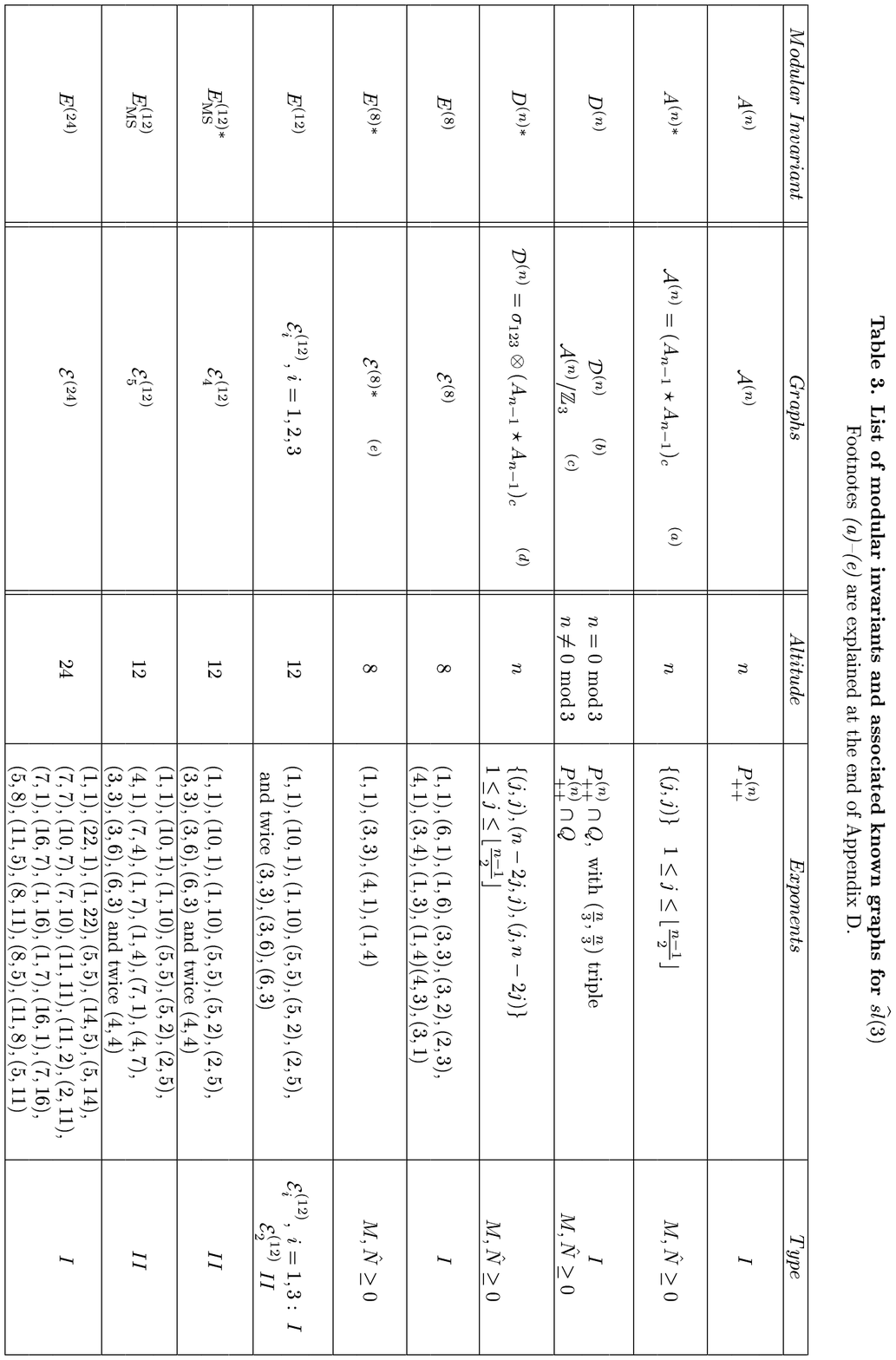}
}
\end{figure}
\pagestyle{empty}
\clearpage
%
%
%
%
\bookfigp{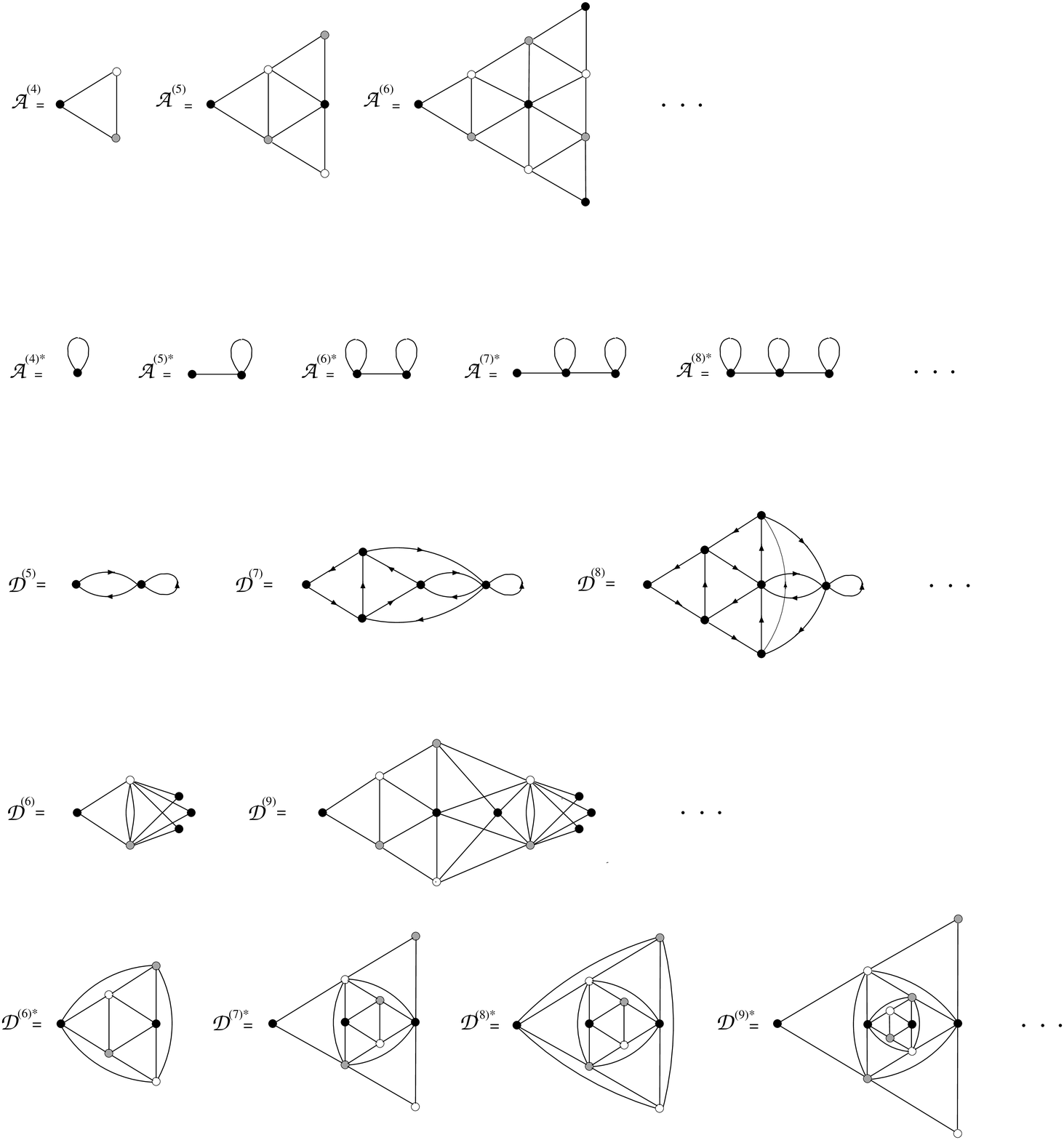}{su3graph1}{ADgraphs}{The known graphs in the case of
 ${\widehat{ sl}}(3)$. Conventions :
(a) For the 3-colourable graphs,
the triality $\tau$ of nodes is indicated  by the colour:
black $\tau=0$, grey $\tau=1$, or white  $\tau=2$; the graph
represents the matrix $n_{21}$ if edges are oriented from
black to grey, or grey to white, etc.
(b) For the non-3-colourable graphs, either the orientation
of all edges (of matrix $n_{21}$, say) is indicated,
(${\cal D}^{(n)}$, $3\not | n$, series,
${\cal E}^{(8)*}$),
or all links are unoriented
(${\cal A}^{(.)*}$ series). }{18}
\pagestyle{plain}
\clearpage
\bookfigp{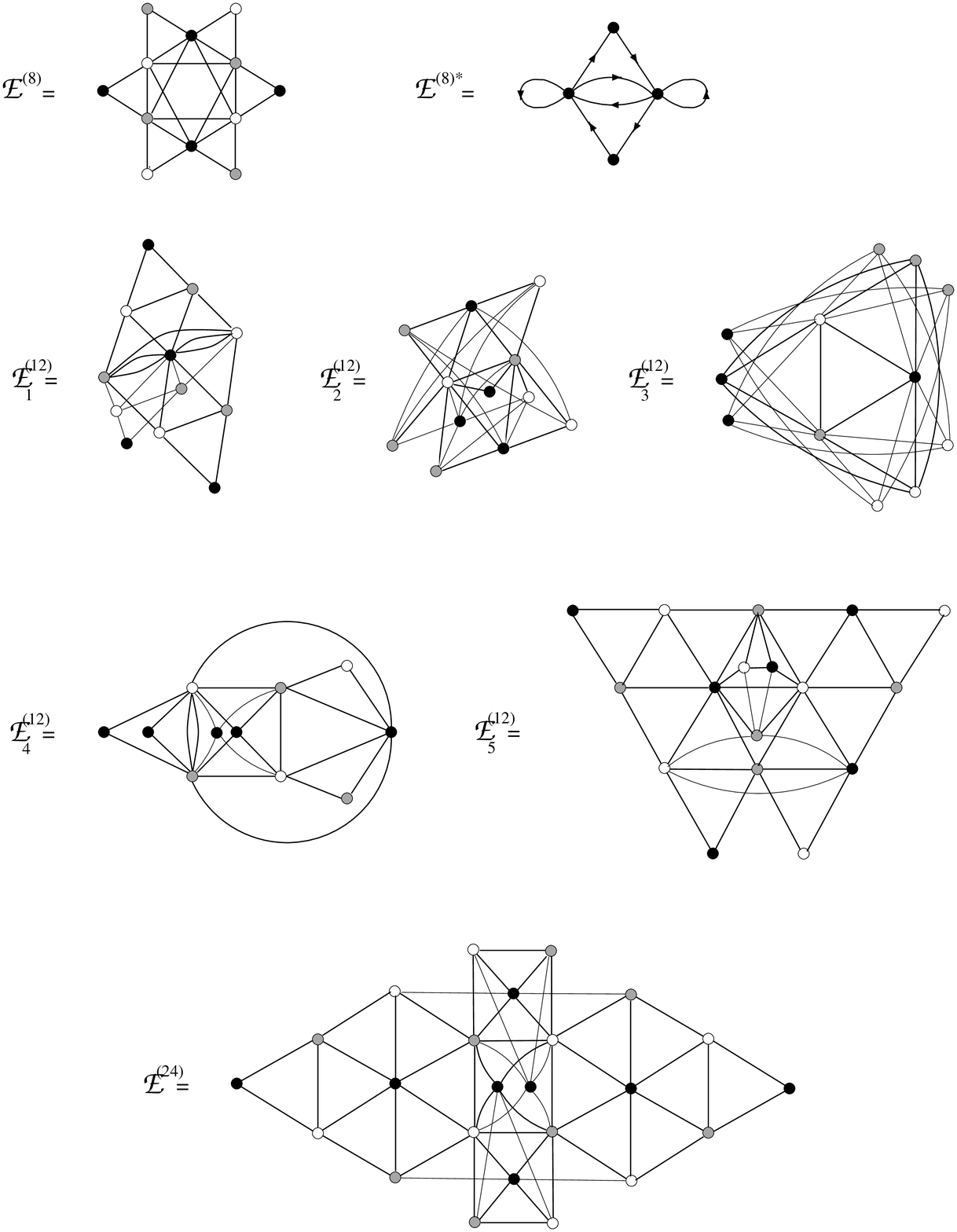}{su3graph2}{Egraphs}{ The known  graphs in the case of
 ${\widehat{ sl}}(3)$, (cont'd). }{18}

\clearpage

\section*{Appendix~E:
Formulae for fusing, braiding and modular matrices}
\rnc{\theequation}{E.\arabic{equation}}\setcounter{equation}{0}


We collect in this appendix some standard formulae for the genus
$0$ and $1$ duality matrices.  The fusing matrices satisfy
several identities implied by the pentagon identity -- they can
be recovered from formulae (\ref{inver}),  (\ref{Va}),
(\ref{tpc})  in the text  making the identification
(\ref{fusid}).

Symmetries:
\be
F_{pq}\left[\matrix{k&j\cr i&l}\right] 
=F_{p^*q}\left[\matrix{j&k\cr l^*&i^*}\right]
=F_{pq^*}\left[\matrix{i^*&l\cr  k^*&j}\right]
=F_{p^*q^*}\left[\matrix{l&i^*\cr j^*&k}\right]\,.
\ee

 Choice of gauge:
\be
 F^{(0)}_{q1}\left[\matrix{i&i^*\cr j&j} \right]_{\za\,
\zb}^{\un_j\, \un_i'} =
 \sqrt{d_q\over d_i d_j }\ \zd_{\zb \,
\sigma_{13}(\za^*)}\label{cgau}\,.
\ee
In the $sl(2)$ case denote by $\sqrt{C_{kj}^q}$
 the  normalisation  of the
CVO in this gauge. Then for the fusion matrix corresponding to
CVO normalised to $1$,  one has
\be
F_{pq}\left[\matrix{k&j\cr i&l}\right]=
\sqrt{C_{kj}^q C_{ql}^i\over C_{kp}^i C_{jl}^p }\
F^{(0)}_{pq}\left[\matrix{k&j\cr i&l}\right]=
\sqrt{C_{kj}^q C_{i^*l}^{q^*} C_{qq^*}^1\over
C_{ki^*}^{p^*} C_{jl}^p C_{pp^*}^1}\
F^{(0)}_{pq}\left[\matrix{k&j\cr i&l}\right] \,, \label{apI}
\ee
or,
\be
C_{ki^*}^{p^*} C_{jl}^p C_{pp^*}^1\
F_{pq}\left[\matrix{k&j\cr i&l}\right]=
C_{kj}^q C_{i^*l}^{q^*} C_{qq^*}^1\
F_{qp}\left[\matrix{k^*&i\cr j&l^*}\right] \,.\label{apII}
\ee
This equation coincides with the quadratic relation resulting
from locality of the physical $4$--point  function in the diagonal
case.
Hence the constants  $C_{kp}^i=C_{(k,k) (p,p)}^{(i,i}\,,$
$C_{kp}^i\,  C_{i i^*}^1 =C_{k i^*}^{p^*}\,  C_{p p^*}^1 $
 can be identified
with the physical OPE structure constants in this case.
For the minimal models  these constants were computed in~\cite{DF};
the matrices $F^{(0)}$ in the gauge (\ref{cgau}) coincide up to
signs with a product of standard $q$-- $6j$ symbols, see
~\cite{KiRe} for the latter.

Braiding matrices:
\bea
B_{pq}\left[\matrix{i&j\cr k&l} \right]_{\zb_1\
\sigma_{23}(\zb_2)}^{\sigma_{23}(\zg_2)\ \zd}
(\epsilon)=e^{\pi i \epsilon (\triangle_k+
\triangle_l-\triangle_p-\triangle_q)}\
F_{pq}\left[\matrix{i&l\cr k&j} \right]_{\zb_1\, \zb_2}^{\zg_2\
\zd}
\eea

The $q$ - analogs of the Racah identity (hexagon identities),
$\epsilon=\pm 1$:
\bea
\sum_q\ F_{mq}\left[\matrix{ i&k \cr j&l}\right]  \
e^{-\pi i \epsilon \triangle_q}\
F_{qp}\left[\matrix{l&i\cr j&k   }\right] =
e^{\pi i \epsilon
(\triangle_m+\triangle_p-\triangle_{l}-\triangle_j-
\triangle_i-\triangle_k) }\
F_{mp}\left[\matrix{i&l\cr j&k}\right]\,.\label{Racah}
\eea

Recall the Moore-Seiberg torus duality identity resulting from a
relation in the modular group of the  torus with two field
insertions, namely, 
$S(j_1,j_2)\, a=b\, S(j_1,j_2)$ where $S(j_1,j_2)$ is the modular
matrix of two-point blocks, expressed in terms of $F$ and $S(p)$,
and $a,b$ are the monodromy transformations moving one of the CVO
around the $a,b$ cycles, \cite{MS1}
\bea
S_{ri}(s) \ \sum_{m}\ e^{2 \pi i
(\triangle_i-\triangle_m)}\
F_{s^*m}\left[\matrix{j_2&i\cr j_1^*&i^*} \right]\
F_{mp}\left[\matrix{j_1&j_2\cr i&i } \right]\
 \nonumber
\\=
\sum_{q}\ S_{qi}(p)\  e^{ \pi i
(\triangle_p-\triangle_{j_1}
-\triangle_{j_2})}\
 F_{sq^*}\left[\matrix{r^*&j_1\cr r^*&j_2 } \right]
F_{rp}\left[\matrix{j_2&j_1\cr q&q} \right]\
 \,. \label{tor}
\eea

Choose $s=1=r$. This implies that $j_1=j_2^*=j^*$ and 
$q=j$, hence
\bea
S_{ji}(p)&&={ S_{1i} \over
F_{1p}\left[\matrix{j&j^*\cr j&j} \right]}\,
\ \sum_{m}\ e^{\pi i
(2\triangle_i+2\triangle_j
-2\triangle_m-\triangle_p)}\
F_{1m}\left[\matrix{j&i\cr j&i^*} \right] \,
F_{mp}\left[\matrix{j^*&j\cr i&i} \right]\  \nonumber\\
&&=
{ S_{1i} \over
F_{1p}\left[\matrix{j&j^*\cr j&j} \right]}\,
\ \sum_{m}\ e^{-\pi i
(2\triangle_i+2\triangle_j
-2\triangle_m)}\
F_{1m}\left[\matrix{j^*&i\cr j^*&i^*} \right] \,
F_{mp}\left[\matrix{j&j^*\cr i&i} \right]\  \label{mtor}\\
&&=
{ S_{1j} \over
F_{p1}\left[\matrix{i^*&i\cr i^*&i^*} \right]}\,
\ \sum_{m}\ e^{ \pi i
(2\triangle_i+2\triangle_j
-2\triangle_m-\triangle_p)}\
F_{p m}\left[\matrix{j&i\cr j&i^*} \right]\
F_{m1}\left[\matrix{j^*&j\cr i&i} \right] \,. \nonumber
\eea
The second equality is obtained reversing the sums in (\ref{tor})
and solving for $S_{ri}(s)$ as above, while the third is obtained
from the transposed version of (\ref{tor}) taking into account
$S(p)^2 = C \, e^{-\pi i \triangle_p}$.
For $p=1$ the formula reproduces the ordinary $S=S(1)$ matrix
\bea
S_{ij} &&={ S_{i1} \over
F_{11}\left[\matrix{j&j^*\cr j&j} \right]}\,
\ \sum_{m}\ e^{2\pi i
(\triangle_i+\triangle_j
-\triangle_m)}\
F_{1m}\left[\matrix{j&i\cr j&i^*} \right]\
F_{m1}\left[\matrix{j^*&j\cr i&i} \right] \nonumber \\
&&=S_{11}\, \sum_{m}\ e^{ 2 \pi i
(\triangle_i+\triangle_j
-\triangle_m)}\ d_m\ N_{ij}{}^m \,. \label{torb}
\eea

\section*{Acknowledgements}
P.A.P. is supported by the Australian Research Council.
V.B.P. acknowledges the hospitality of Arnold Sommerfeld
Institute for Mathematical  Physics, TU Clausthal, the support and
hospitality of Service de Physique Th\'eorique, CEA-Saclay and
the partial support of the Bulgarian National Research Foundation
(contract $\Phi$-643). J.-B.Z.
acknowledges partial support of the EU Training and Mobility of Researchers
Program (Contract FMRX-CT96-0012), which made possible an extremely
profitable stay in SISSA, Trieste.
We acknowledge useful discussions with
M. Bauer, D. Bernard, J. Fuchs, T. Gannon,
A. Honecker, A. Recknagel, P. Ruelle, I.
Runkel,  C. Schweigert and G. Watts.

\end{document}